\documentclass[aps,prd,reprint,superscriptaddress,preprintnumbers]{revtex4-2}

\usepackage{amsmath}
\usepackage{amsfonts}
\usepackage{graphicx}
\usepackage{mathtools}
\usepackage{array}
\usepackage{minted}
\usepackage{verbatim}
\usepackage{siunitx}
\usepackage{algorithmic}
\usepackage{algorithm}
\usepackage{tikz}
\usetikzlibrary{shapes.geometric, shapes.misc, shapes.symbols, decorations.pathreplacing}
\usepackage{outlines}
\usepackage{comment}
\usepackage{varioref}
\usepackage[hidelinks]{hyperref}
\usepackage{cleveref}
\crefname{figure}{Fig.}{Figs.}
\Crefname{figure}{Fig.}{Figs.}
\crefname{table}{Tab.}{Tabs.}
\Crefname{table}{Tab.}{Tabs.}
\crefname{section}{Sec.}{Secs.}
\Crefname{section}{Sec.}{Secs.}
\crefname{appendix}{Appendix}{Appendices}
\Crefname{appendix}{Appendix}{Appendices}
\usepackage{xspace}
\usepackage{printlen}
\usepackage{booktabs}
\usepackage{makecell}
\usepackage{xcolor}
\usepackage{fontawesome5}
\usepackage{comment}
\usepackage[utf8]{inputenc}
\usepackage{chngcntr}
\usepackage{orcidlink}
\usepackage{etoolbox}

\newcommand{\cosmogrid}{\textsc{CosmoGridV1}\xspace}

\newcommand{\ufalcon}{\texttt{UFalcon}\xspace}
\newcommand{\deepsphere}{\texttt{DeepSphere}\xspace}

\newcommand{\pkdgrav}{\texttt{PkdGrav3}\xspace}

\newcommand{\healpix}{\texttt{HEALPix}\xspace}
\newcommand{\healpy}{\texttt{healpy}\xspace}

\newcommand{\pycosmo}{\mbox{\texttt{PyCosmo}}\xspace}
\newcommand{\pyccl}{\texttt{PyCCL}\xspace}

\newcommand{\tensorflow}{\texttt{TensorFlow}\xspace}

\newcommand{\horovod}{\texttt{Horovod}\xspace}
\newcommand{\adam}{\texttt{adam}\xspace}
\newcommand{\emcee}{\texttt{emcee}\xspace}
\newcommand{\flowconductor}{\texttt{FlowConductor}\xspace}

\newcommand{\addgals}{\texttt{ADDGALS}\xspace}

\newcommand{\omatter}{\ensuremath{\Omega_m}\xspace}
\newcommand{\sigeight}{\ensuremath{\sigma_8}\xspace}
\newcommand{\wzero}{\ensuremath{w}\xspace}
\newcommand{\ns}{\ensuremath{n_s}\xspace}
\newcommand{\obary}{\ensuremath{\Omega_b}\xspace}
\newcommand{\hubble}{\ensuremath{H_0}\xspace}
\newcommand{\seight}{\ensuremath{S_8}\xspace}

\newcommand{\mczero}{\ensuremath{M^0_c}\xspace}

\newcommand{\olambda}{\ensuremath{\Omega_\Lambda}\xspace}
\newcommand{\mneutrino}{\ensuremath{m_\nu}\xspace}

\newcommand{\AIA}{\ensuremath{A_{\mathrm{IA}}}\xspace}
\newcommand{\etaIA}{\ensuremath{{\eta_{A_{\mathrm{IA}}}}}\xspace}
\newcommand{\bta}{\ensuremath{b_{\mathrm{TA}}}\xspace}
\newcommand{\zpivot}{\ensuremath{z_\mathrm{pivot}\xspace}}

\newcommand{\bgi}{\ensuremath{b_g^i}\xspace}
\newcommand{\bgone}{\ensuremath{b_g^1}\xspace}
\newcommand{\bgtwo}{\ensuremath{b_g^2}\xspace}
\newcommand{\bgthree}{\ensuremath{b_g^3}\xspace}
\newcommand{\bgfour}{\ensuremath{b_g^4}\xspace}

\newcommand{\bgsource}{\ensuremath{b_{g,s}}\xspace}
\newcommand{\bgsourcei}{\ensuremath{b_{g,s}^i}\xspace}

\newcommand{\mb}{\ensuremath{m_b}\xspace}
\newcommand{\dzsource}{\ensuremath{\Delta z_s}\xspace}
\newcommand{\dzlens}{\ensuremath{\Delta z_l}\xspace}
\newcommand{\sigmazlens}{\ensuremath{\sigma_{z,l}}\xspace}

\newcommand{\dmsource}{\ensuremath{\delta_{m,s}}\xspace}
\newcommand{\dmlens}{\ensuremath{\delta_{m}}\xspace}
\newcommand{\ngsource}{\ensuremath{n_{g,s}}\xspace}
\newcommand{\nglens}{\ensuremath{n_{g}}\xspace}
\newcommand{\nglenshat}{\ensuremath{\hat{n}_{g}}\xspace}
\newcommand{\nsource}{\ensuremath{{\langle n_{g,s}\rangle}}\xspace}
\newcommand{\nsourcei}{\ensuremath{{\langle n_{g,s}^i \rangle}}\xspace}
\newcommand{\nlens}{\ensuremath{{\langle n_{g} \rangle}}\xspace}
\newcommand{\nlensi}{\ensuremath{{\langle n_{g}^i \rangle}}\xspace}

\newcommand{\nside}{\ensuremath{n_\textrm{side}}\xspace}
\newcommand{\npix}{\ensuremath{n_\textrm{pix}}\xspace}
\newcommand{\lbox}{\ensuremath{L_{\mathrm{box}}}\xspace}

\newcommand{\alm}{\ensuremath{a_{\ell m}}\xspace}
\newcommand{\cl}{\ensuremath{C_{\ell}}\xspace}

\newcommand{\lmin}{\ensuremath{\ell_\mathrm{min}}\xspace}
\newcommand{\lmax}{\ensuremath{\ell_\mathrm{max}}\xspace}
\newcommand{\lmaxt}{\ensuremath{\tilde{\ell}_\mathrm{max}}\xspace}

\newcommand{\metacalibration}{\textsc{Metacalibration}\xspace}
\newcommand{\metacal}{\textsc{Metacal}\xspace}
\newcommand{\maglim}{\textsc{Maglim}\xspace}
\newcommand{\sompz}{\textsc{SOMPZ}\xspace}
\newcommand{\buzzard}{\textsc{Buzzard}\xspace}

\newcommand{\balrog}{\texttt{Balrog}\xspace}

\newcommand{\Seight}{\ensuremath{S_8}\xspace}
\newcommand{\wcdm}{\ensuremath{w}CDM\xspace}
\newcommand{\lcdm}{\ensuremath{\Lambda}CDM\xspace}
\newcommand{\x}{\ensuremath{\times}\xspace}
\newcommand{\normal}{\ensuremath{\mathcal{N}}\xspace}
\newcommand{\sobol}{\ensuremath{\mathcal{S}}\xspace}
\newcommand{\latin}{\ensuremath{\mathcal{L}}\xspace}
\newcommand{\FoM}{\ensuremath{\mathrm{FoM}}\xspace}
\newcommand{\FoMSeight}{\ensuremath{\mathrm{FoM}_{\omatter, \Seight}}\xspace}
\newcommand{\FoMwzero}{\ensuremath{\mathrm{FoM}_{\omatter, \wzero}}\xspace}

\newcommand{\R}{\ensuremath{\mathbb{R}}\xspace}
\newcommand{\Z}{\ensuremath{\mathbb{Z}}\xspace}

\definecolor{colorOne}{RGB}{152,68,100}
\definecolor{colorTwo}{RGB}{94,204,171}
\definecolor{colorThree}{RGB}{0,103,138}
\definecolor{colorFour}{RGB}{230,161,118}

\usepackage{natbib}
\defcitealias{fluriFullMathrmCDM2022}{F22}
\defcitealias{kacprzakDeepLSSBreakingParameter2022}{K22}
\defcitealias{kacprzakCosmoGridV1SimulatedCDM2023}{K23}

\begin{document}

\preprint{DES-2025-0952}
\preprint{FERMILAB-PUB-25-0749-PPD}

\title{Dark Energy Survey Year 3 results: Simulation-based \wcdm inference from weak lensing and galaxy clustering maps with deep learning: Analysis design}

\author{A.~Thomsen\orcidlink{0000-0002-0309-9021}\textsuperscript{1}}
\email{athomsen@phys.ethz.ch}
\author{J.~Bucko\textsuperscript{1}}
\author{T.~Kacprzak\textsuperscript{2,3}}
\author{V.~Ajani\textsuperscript{4}}
\author{J.~Fluri\textsuperscript{5}}
\author{A.~Refregier\textsuperscript{1}}
\author{D.~Anbajagane\textsuperscript{6}}
\author{F.~J.~Castander\textsuperscript{7,8}}
\author{A.~Fert\'e\textsuperscript{9}}
\author{M.~Gatti\textsuperscript{6}}
% ...

\collaboration{DES Collaboration}
\thanks{Author affiliations are listed at the end of the paper.}
\begin{abstract}
Data-driven approaches using deep learning are emerging as powerful techniques to extract non-Gaussian information from cosmological large-scale structure.
This work presents the first simulation-based inference (SBI) pipeline that combines weak lensing and galaxy clustering maps in a realistic Dark Energy Survey Year 3 (DES Y3) configuration and serves as preparation for a forthcoming analysis of the survey data.
We develop a scalable forward model based on the \cosmogrid suite of $N$-body simulations to generate over one million self-consistent mock realizations of DES Y3 at the map level.
Leveraging this large dataset, we train deep graph convolutional neural networks on the full survey footprint in spherical geometry to learn low-dimensional features that approximately maximize mutual information with target parameters.
These learned compressions enable neural density estimation of the implicit likelihood via normalizing flows in a ten-dimensional parameter space spanning cosmological \wcdm, intrinsic alignment, and linear galaxy bias parameters, while marginalizing over baryonic, photometric redshift, and shear bias nuisances.
To ensure robustness, we extensively validate our inference pipeline using synthetic observations derived from both systematic contaminations in our forward model and independent \buzzard galaxy catalogs.
Our forecasts yield significant improvements in cosmological parameter constraints, achieving 2--3$\times$ higher figures of merit in the \omatter--\seight plane relative to our implementation of baseline two-point statistics and effectively breaking parameter degeneracies through probe combination.
These results demonstrate the potential of SBI analyses powered by deep learning for upcoming Stage-IV wide-field imaging surveys.
\end{abstract}

%\keywords{}

\maketitle

\section{Introduction}\label{sec:introduction}
The large-scale structure (LSS) of the mass distribution in the Universe encodes a wealth of information about its late-time evolution, offering a unique testbed for cosmological theories.
Two primary observational probes of the LSS are \emph{weak gravitational lensing} and \emph{galaxy clustering}.
Weak lensing directly measures the projected matter density through coherent distortions of galaxy shapes by intermediate mass, while galaxy clustering tracks the distribution of luminous matter, which serves as a biased tracer of the underlying dark matter density.
These probes of the LSS constrain cosmological parameters including the present-day matter density fraction \omatter, the variance of linear density perturbations on the scale of 8 Mpc/h denoted \sigeight, and the dark-energy equation-of-state parameter \wzero (see e.g.~\cite{albrechtReportDarkEnergy2006,schneiderWeakGravitationalLensing2006,kilbingerCosmologyCosmicShear2015,zhanCosmologyLargeSynoptic2018} for reviews).
In addition, these observations are sensitive to astrophysical model parameters including intrinsic alignments and galaxy biasing.
Crucially, combined analyses of weak lensing and galaxy clustering break parameter degeneracies that would otherwise limit the constraining power of each probe individually~\cite{descollaborationDarkEnergySurvey2022,heymansKiDS1000CosmologyMultiprobe2021,krauseDarkEnergySurvey2021,kacprzakDeepLSSBreakingParameter2022}.

In recent years, observing programs targeting the LSS like the Dark Energy Survey (DES)~\footnote{\url{https://www.darkenergysurvey.org/}}, the Kilo-Degree Survey (KiDS)~\footnote{\url{https://kids.strw.leidenuniv.nl/}}, the Subaru Hyper Suprime-Cam (HSC)~\footnote{\url{https://hsc.mtk.nao.ac.jp/ssp/survey/}}, and the DECam Local Volume Exploration survey (DELVE)~\footnote{\url{https://delve-survey.github.io/}} have measured hundreds of millions of galaxy positions and shapes over thousands of square degrees of the sky, enabling cosmological parameter measurements with sub-5\% precision~\cite{descollaborationDarkEnergySurvey2022,heymansKiDS1000CosmologyMultiprobe2021,sugiyamaHyperSuprimeCamYear2023,anbajaganeDarkEnergyCamera2025}.
Successor Stage-IV surveys like the Vera C. Rubin Observatory’s Legacy Survey of Space and Time (LSST)~\footnote{\url{https://rubinobservatory.org/explore/lsst}}, the Euclid mission~\footnote{\url{https://www.euclid-ec.org/public/data/surveys/}}, and the Roman Space Telescope~\footnote{\url{https://science.nasa.gov/mission/roman-space-telescope}} are set to measure orders of magnitude larger datasets, and expected to further increase the measurement precision to the sub-percent level.

At sufficiently large scales, the projected matter density is well described as an isotropic Gaussian random field (GRF)~\cite{peacockCosmologicalPhysics1998}. 
Under this assumption, two-point statistics become sufficient summaries that capture all available statistical information~\cite{bartelmannWeakGravitationalLensing2001}.
This has contributed to their widespread use in cosmological analyses, for example implemented as the correlation function in real space~\cite[e.g.][]{descollaborationDarkEnergySurvey2022b,asgariKiDS1000CosmologyCosmic2021} or the power spectrum in harmonic space~\cite[e.g.][]{douxDarkEnergySurvey2022,asgariKiDS1000CosmologyCosmic2021}.

However, due to non-linear structure formation and baryonic effects, the fields considered in this work only resemble GRFs on the largest scales and contain \emph{non-Gaussian} features at intermediate and small scales~\cite{kacprzakCosmoGridV1SimulatedCDM2023}.
Two-point statistics cannot capture this information, rendering them statistically insufficient.
A growing body of literature therefore explores alternative summary statistics designed to extract the non-Gaussian component (see e.g.~\cite{euclidcollaborationEuclidPreparationXXVIII2023} for an overview in weak lensing) from the \emph{map-level}.
Within DES, examples of weak lensing analyses employing such summary statistics include peak counts~\cite{kacprzakCosmologyConstraintsShear2016,zurcherDarkEnergySurvey2022,jeffreyDarkEnergySurvey2025}, higher-order moments of weak lensing mass maps~\cite{gattiDarkEnergySurvey2020,descollaborationDarkEnergySurvey2022e,descollaborationDarkEnergySurvey2024,darkenergysurveyDarkEnergySurvey2025}, higher order correlation functions~\cite{descollaborationDarkEnergySurvey2022f,gomesCosmologySecondThirdorder2025,gomesDarkEnergySurvey2025}, the cumulative distribution function~\cite{anbajagane3rdMomentPractical2023}, wavelet harmonics~\cite{descollaborationDarkEnergySurvey2024,darkenergysurveyDarkEnergySurvey2025}, scattering transforms~\cite{descollaborationDarkEnergySurvey2024,darkenergysurveyDarkEnergySurvey2025}, and persistent homology~\cite{pratDarkEnergySurvey2025}.
Rather than predefining the summary statistic, an alternative approach parametrizes it as a (typically convolutional) neural network that is trained to automatically extract informative features~\cite{fluriCosmologicalConstraintsNoisy2018,fluriCosmologicalConstraintsDeep2019,fluriCosmologicalParameterEstimation2021,fluriFullMathrmCDM2022,kacprzakDeepLSSBreakingParameter2022,jeffreyLikelihoodfreeInferenceNeural2021,jeffreyDarkEnergySurvey2025,guptaNonGaussianInformationWeak2018,ribliWeakLensingCosmology2019,matillaInterpretingDeepLearning2020,luSimultaneouslyConstrainingCosmology2022,luCosmologicalConstraintsHSC2023,akhmetzhanovaDataCompressionInference2024,lanzieriOptimalNeuralSummarization2025}.
As these methods mature, several have progressed beyond being proofs of concept to yield cosmological constraints from the Dark Energy Survey's first three years (DES Y3) of weak lensing observations~\cite{zurcherDarkEnergySurvey2022,descollaborationDarkEnergySurvey2022e,darkenergysurveyDarkEnergySurvey2025,jeffreyDarkEnergySurvey2025,pratDarkEnergySurvey2025,gomesDarkEnergySurvey2025}.

A key challenge in utilizing these statistics is the general lack of analytical predictions relating them to the cosmological parameters. 
Moreover, the functional form of the \emph{likelihood} is typically unknown. 
Simulation-based inference (SBI) addresses both issues by using large ensembles of simulations across different parameter values to establish the parameter-statistic relationship~\cite{cranmerFrontierSimulationbasedInference2020}. 
Neural density estimation~\cite{papamakariosNeuralDensityEstimation2019,papamakariosNormalizingFlowsProbabilistic2021} can then be used to learn the probability density of interest, in our case the likelihood, directly from the simulated data~\cite{10.1093/mnras/stz1960,papamakariosSequentialNeuralLikelihood2019,alsingMassiveOptimalData2018}.

We present an SBI pipeline that employs map-level neural network summary statistics to constrain cosmology by jointly conditioning on DES Y3 weak lensing and galaxy clustering data. 
This represents the first application of higher-order summary statistics to this probe combination within DES.
In this paper, we validate our methodology on simulations, with the application to the actual DES Y3 observation to be presented in a forthcoming companion paper.

This paper is organized as follows. 
\Cref{sec:survey} provides a brief overview of the blinded DES Y3 source and lens galaxy catalogs. 
\Cref{sec:forward_model} describes our forward model, which transforms the dark matter particle lightcones from the \cosmogrid~\cite{kacprzakCosmoGridV1SimulatedCDM2023} simulation suite into more than one million self-consistent DES Y3 weak lensing and galaxy clustering mocks varying $w$CDM, baryonification, intrinsic alignment, linear galaxy biasing, and observational nuisance parameters.
The SBI methodology is detailed in \cref{sec:simulation_based_inference}, including our map-level compression networks, the two-point statistic baseline, and neural density estimation using normalizing flows.
\Cref{sec:mocks} introduces the \buzzard~\cite{descollaborationDarkEnergySurvey2022d} mock catalogs, which are external to our forward model and serve as an independent validation dataset. 
\Cref{sec:validation} explains how we use these and other mocks to test our pipeline's robustness against model misspecification. 
We present our results in \cref{sec:results}, contrasting our map-level compression with the two-point baseline, as well as comparing the different cosmological probes individually and in combination. 
Finally, we summarize our findings and conclusions in \cref{sec:conclusion} and discuss future prospects for this methodology.

\section{Survey Configuration}\label{sec:survey}

\subsection{Dark Energy Survey Year 3}\label{sec:des}
DES~\cite{thedarkenergysurveycollaborationDarkEnergySurvey2005,darkenergysurveycollaboration:DarkEnergySurvey2016} is an observational program that imaged roughly five thousand square degrees of the southern hemisphere over six years ($2013 - 2019$) providing photometric measurements in the five optical-NIR broadbands \textit{grizY}. 
The measurements were taken with the 570 megapixel Dark Energy Camera~\cite{flaugherDARKENERGYCAMERA2015} mounted on the four meter Blanco telescope at Cerro Tololo Inter-American Observatory in Chile. 
For details on the image processing pipeline, we refer the interested reader to~\cite{morgansonDarkEnergySurvey2018}. 

We utilize the DES Y3 data from the first three years of operations, which have been made publicly available as the DES Data Release 1~\cite{abbottDarkEnergySurvey2018}.
However, in this work, we only employ selected properties of the observed DES Y3 catalogs such as the masking, galaxy number densities, redshift distributions, and calibration properties to conduct a realistic forecast and validate the end-to-end inference pipeline against various synthetic mocks;
no cosmological constraints are derived from the real observations here.
We leave these cosmological results to a later companion paper.
\begin{figure}[!htb]
    \centering
    \includegraphics{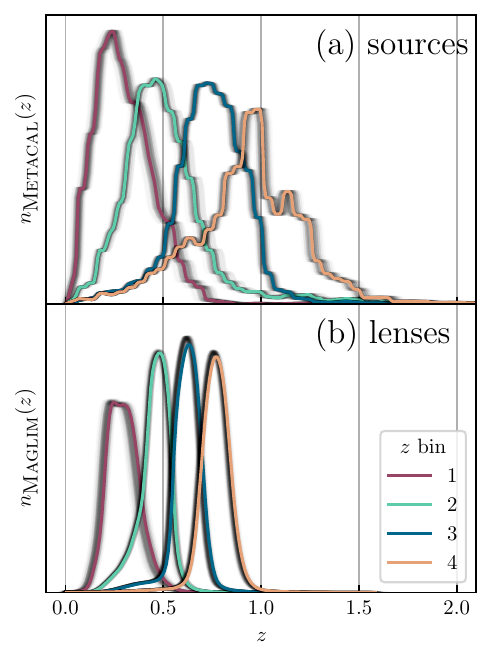}
    \caption{Normalized redshift distributions of (a) the \metacalibration source galaxy sample used for weak lensing and (b) the \maglim lens galaxy sample used for galaxy clustering. The colored lines indicate the base distributions, while the partially transparent overlapping gray lines illustrate the redshift uncertainty via fifty draws from the photo-$z$ distributions parametrized in \cref{eq:z_error_metacal,eq:z_error_maglim} according to \cref{tab:prior_cosmogrid}.}
    \label{fig:redshift_dist}
\end{figure}

\subsection{Source Galaxy Catalog}\label{sec:source_catalog}
The DES Y3 weak lensing shape catalog~\cite{gattiDarkEnergySurvey2021} contains approximately 100 million source galaxies within an area of $\SI{4 \, 143}{\deg \squared}$, yielding a weighted number density of $n_{\rm eff} = 5.59 \, \textrm{gal}/\textrm{arcmin}^2$.

For the catalog, the galactic ellipticities or shears are measured from the observed (multi-band) noisy images using the self-calibrating \metacalibration algorithm~\cite{huffMetacalibrationDirectSelfCalibration2017,sheldonPracticalWeaklensingShear2017}. 
Additional redshift-dependent detection and blending effects identified in~\cite{maccrannDarkEnergySurvey2022} through end-to-end image simulations, which are not accounted for by \metacalibration, can be modeled as a multiplicative bias at the $2-3 \%$ level; see \cref{sec:shear_bias} for our treatment.

The photometric redshifts of the objects in the catalog are determined by the \sompz algorithm~\cite{mylesDarkEnergySurvey2021} and further calibrated in~\cite{gattiDarkEnergySurvey2022,maccrannDarkEnergySurvey2022}.
The galaxies are subdivided into four tomographic bins of roughly equal number density~\cite{mylesDarkEnergySurvey2021}, resulting in redshift distributions we denote as $n_\metacal^i(z)$ for $i \in \{1,2,3,4\}$ and plot in panel (a) of \cref{fig:redshift_dist}.

For this paper, the catalog is utilized only for its mask, redshift distributions, and in the shape noise generation, where randomly sampled and rotated galaxies from the catalog are used to generate noise maps as described in \cref{sec:shape_noise}; 
the original shape catalog is not used.

\subsection{Lens Galaxy Catalog}\label{sec:lens_catalog}
For the lens galaxies, we choose to use the DES Y3 magnitude limited sample~\cite{porredonDarkEnergySurvey2021} denoted as \maglim and employed as fiducial for the DES Y3 3$\times$2pt analysis~\cite{descollaborationDarkEnergySurvey2022a}. 
The sample is characterized by an upper magnitude cut in the \textit{i}-band that depends linearly on the photometric redshift estimate, $z$, from the directional neighborhood fitting (DNF) algorithm~\cite{sevilla-noarbeDarkEnergySurvey2021,devicenteDNFGalaxyPhotometric2016} like $i < 4 z + 18$, which is a selection that has been optimized for \wcdm constrains by balancing the resulting galaxy number density and accuracy of the photometric redshift estimates~\cite{porredonDarkEnergySurvey2021}.
The catalog contains around 10.7 million galaxies grouped into six tomographic bins with redshift distributions $n_\maglim^i(z)$ (see panel (b) in \cref{fig:redshift_dist}), of which we discard the last two following the fiducial analysis in~\cite{descollaborationDarkEnergySurvey2022}.

Just as with the source galaxy sample, we do not use the original lens galaxy catalog in this work;
for the purposes of forecasting and validation, we only consider the catalog's mask, mean galaxy number density, and redshift distributions.
\section{Forward Modeling}\label{sec:forward_model}
In this section, we present our forward model for self-consistent simulated DES Y3-like \emph{weak lensing} and \emph{galaxy clustering} maps.
These maps respectively resemble the source and lens galaxy samples described in \cref{sec:survey}, matching their masking, average number density, and redshift distributions while varying cosmological \wcdm parameters, baryonic feedback effects, intrinsic alignment contributions, linear galaxy bias, and observational nuisances. 

This forward modeling approach offers several advantages.
First, observational systematics such as masking, shear biases, and redshift errors can be directly incorporated despite being difficult to treat analytically even at the two-point level~\cite{linNewModelPredict2015,porqueresBayesianForwardModelling2021,vonwietersheim-kramstaKiDSSBISimulationbasedInference2025}.
Second, within the simulation-based inference framework, map-level forward modeling enables training neural compression networks that potentially capture the full information content of the pixelized fields without relying on handcrafted summary statistics sensitive only to specific features of the data.
The compression networks thus serve as non-Gaussian summary statistics for which no analytical connection to cosmological theory exists; instead, the connection is established through numerical forward simulations.

\begin{figure}[!htb]
    \centering
    \resizebox{\linewidth}{!}{
    \begin{nolinenumbers}
    \begin{tikzpicture}
        % Define styles
        \tikzset{
            % nodes
            map/.style = {
                draw, 
                rectangle, 
                rounded corners, 
                minimum width=3em, 
                minimum height=2em, 
                text centered, 
                text width=3em, 
                % fill=myBlue!30, 
                fill=colorThree!30, 
            },
            partial_map/.style = {
                draw, 
                rectangle, 
                rounded corners, 
                minimum width=3em, 
                minimum height=2em, 
                text centered, 
                text width=3em, 
                % fill=myBlue!10, 
                fill=colorThree!10, 
            },
            operation/.style = {
                draw, 
                rectangle, 
                minimum width=3em, 
                minimum height=2em, 
                text centered,
                text width=3em, 
                font=\scriptsize,
            },
            param/.style = {
                draw, 
                ellipse, 
                % diamond,
                % trapezium,
                % semicircle,
                % kite,
                minimum width=3em, 
                minimum height=2em, 
                text width=2em,
                text centered, 
                rounded corners, 
                % fill=myGreen!30
                fill=colorTwo!30
            },
            nuisance/.style = {
                draw, 
                ellipse, 
                minimum width=3em, 
                minimum height=2em, 
                text centered, 
                text width=2em,
                rounded corners, 
                % fill=myRed!30
                fill=colorOne!30
            },
            % arrows
            myarrow/.style = {thick,->,>=stealth},
        }
        
        % nodes
    
        % y = 4
        \node (params_cosmo) [param, text width=6em] at (2, 4) {
            \renewcommand{\arraystretch}{1.3}
            \begin{tabular}{c}
                $\omatter \; \; \sigeight \; \; \wzero$ \\
            \end{tabular}
        };
        \node (params_nuisance) [nuisance, text width=6em] at (6, 4) {
            \renewcommand{\arraystretch}{1.3}
            \begin{tabular}{c}
                $\ns \; \; \obary \; \; \hubble $ \\
            \end{tabular}
        };

        % y = 2.5
        \node (cosmogrid) [draw, rectangle, text centered, minimum height=4em, text width=10em, anchor=center, fill=gray!30, font=\large] at (4, 2.5) {\cosmogrid};

        % y = 1
        \node (baryonification) [draw, rectangle, text centered, minimum height=2em, text width=6em, anchor=center, font=\scriptsize] at (4, 1) {baryonification};
        \node (params_baryon) [nuisance, text width=3em] at (1, 1) {
            \renewcommand{\arraystretch}{1.3}
            \begin{tabular}{c}
                $\mczero \; \; \nu$ \\
            \end{tabular}
        };

        % y = 0
        \node (projection) [draw, rectangle, text centered, minimum height=2em, text width=6em, anchor=center, font=\scriptsize] at (4, 0) {map projection};
        \node (params_projection) [nuisance, text width=6em] at (7, 0) {
            \renewcommand{\arraystretch}{1.3}
            \begin{tabular}{c}
                $\dzsource^i \; \; \dzlens^i \; \; \sigmazlens^i$
            \end{tabular}
        };

        % y = -2
        \node (k_sig_in) [map] at (0, -2) {$\kappa_\mathrm{WL}$};
        \node (k_ia_in) [map] at (2, -2) {$\kappa_\mathrm{IA}$};
        \node (d_wl_in) [map] at (4, -2) {\dmsource};
        \node (d_gc_in) [map] at (8, -2) {\dmlens};
    
        % y = -4
        \node (sig_ks) [operation] at (0, -4) {inverse Kaiser-Squires};
        \node (ia_ks) [operation] at (2, -4) {inverse Kaiser-Squires};
        \node (d_g_wl) [map] at (4, -4) {\ngsource};
        \node (params_bias_wl) [nuisance] at (6, -4) {\bgsourcei};
    
        % y = -6
        \node (g_sig) [map] at (0, -6) {$\gamma_\mathrm{WL}$};
        \node (g_ia) [map] at (2, -6) {$\gamma_\mathrm{IA}$};
        \node (g_noise) [map] at (4, -6) {$\gamma_\mathrm{SN}$};
        \node (g_data) [map,fill=colorFour!30] at (6, -6) {$\tilde{\gamma}_\metacal$};
        \node (gc_mask) [operation] at (8, -6) {masking};
    
        % y = -8
        \node (sig_mask) [operation] at (0, -8) {masking \& B-mode removal \& Kaiser-Squires};
        \node (ia_mask) [operation] at (2, -8) {masking \& B-mode removal \& Kaiser-Squires};
        \node (noise_mask) [operation] at (4, -8) {masking \& B-mode removal \& Kaiser-Squires};
        \node (params_bias_gc) [param] at (6, -8) {\bgi};
        \node (d_g_gc) [partial_map] at (8, -8) {\nglenshat};
        
        % y = -10
        \node (k_sig) [partial_map] at (0, -10) {$\kappa_\mathrm{WL}$};
        \node (k_ia) [partial_map] at (2, -10) {$\kappa_\mathrm{IA}$};
        \node (k_noise) [partial_map] at (4, -10) {$\kappa_\mathrm{SN}$};
        \node (poisson_gc) [operation] at (8, -10) {Poisson};

        % y = -12
        \node (k) [partial_map] at (2, -12) {$\kappa$};
        \node (params_ia) [param] at (6, -10.3) {\AIA \\ \etaIA \\ \bta};
        \node (params_shear) [nuisance] at (6, -12) {$m_b^i$};
        \node (d) [partial_map] at (8, -12) {\nglens};

        % y = -13
        \node at (2, -13) [font=\large] {Weak Lensing};
        \node at (8, -13) [font=\large] {Clustering};
    
        % lines
        \draw[dashed, black] (7, -2) -- (7, -12);
        \draw[dashed, black] (5, -2) -- (5, -12);
          
        % connections
        \draw[myarrow] (params_cosmo) -- (cosmogrid);
        \draw[myarrow] (params_nuisance) -- (cosmogrid);

        \draw[myarrow] (cosmogrid) -- (baryonification);
        \draw[myarrow] (params_baryon) -- (baryonification);

        \draw[myarrow] (baryonification) -- (projection);
        \draw[myarrow] (params_projection) -- (projection);
        
        \draw[myarrow] (projection) -- (k_sig_in);
        \draw[myarrow] (projection) -- (k_ia_in);
        \draw[myarrow] (projection) -- (d_wl_in);
        \draw[myarrow] (projection) -- (d_gc_in);
        
        \draw[myarrow] (k_sig_in) -- (sig_ks);
        \draw[myarrow] (k_ia_in) -- (ia_ks);
        \draw[myarrow] (d_wl_in) -- (d_g_wl);
        \draw[myarrow] (params_bias_wl) -- (d_g_wl);
        
        \draw[myarrow] (sig_ks) -- (g_sig);
        \draw[myarrow] (ia_ks) -- (g_ia);
        \draw[myarrow] (d_g_wl) -- (g_noise);
        \draw[myarrow] (g_data) -- (g_noise);
        \draw[myarrow] (d_gc_in) -- (gc_mask);
        
        \draw[myarrow] (g_sig) -- (sig_mask);
        \draw[myarrow] (g_ia) -- (ia_mask);
        \draw[myarrow] (g_noise) -- (noise_mask);
        \draw[myarrow] (gc_mask) -- (d_g_gc);
        
        \draw[myarrow] (sig_mask) -- (k_sig);
        \draw[myarrow] (ia_mask) -- (k_ia);
        \draw[myarrow] (noise_mask) -- (k_noise);
        \draw[myarrow] (d_g_gc) -- (poisson_gc);
        \draw[myarrow] (params_bias_gc) -- (d_g_gc);
        
        \draw[myarrow] (k_sig) -- (k);
        \draw[myarrow] (k_ia) -- (k);
        \draw[myarrow] (k_noise) -- (k);
        \draw[myarrow] (params_ia) -- (k);
        \draw[myarrow] (params_shear) -- (k);
        \draw[myarrow] (poisson_gc) -- (d);
    \end{tikzpicture}
    \end{nolinenumbers}
    }
    \caption{
        Schematic overview of the processing pipeline we apply to transform (baryonified) particle shells from the \cosmogrid simulations into mock weak lensing and galaxy clustering maps matching selected DES Y3 properties.
        Sharp-cornered boxes represent processing steps, while blue rounded boxes denote (full- or partial-sky) \healpix maps.
        Green and red ellipses indicate constrained and marginalized parameters, respectively.
        The tomographic bin index $i \in \{1,2,3,4\}$ for source (subscript $s$) and lens (subscript $l$) galaxy samples is omitted from map names for simplicity.
        The abbreviation ``WL" indicates weak lensing signal, ``IA" intrinsic alignment, and ``SN" shape noise.
        We denote the scrambled shear catalog of randomly rotated source galaxies used to generate shape noise maps as $\tilde{\gamma}_\metacal$, and omit the $\kappa_\mathrm{IA,TA}$ (see \cref{eq:kappa_sum}) map for clarity.
    }
    \label{fig:pipeline}
\end{figure}
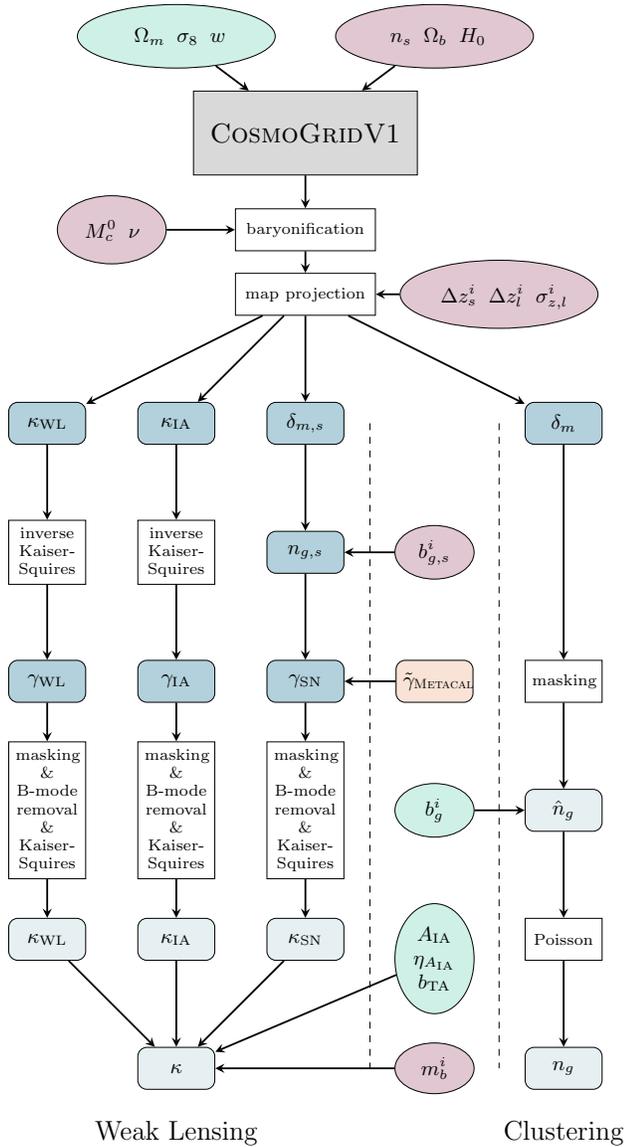

Both training of these compression networks and inference require large numbers of such simulations, motivating our use of the \cosmogrid suite~\cite[hereafter \citetalias{kacprzakCosmoGridV1SimulatedCDM2023}]{kacprzakCosmoGridV1SimulatedCDM2023}, publicly available via~\footnote{\url{http://www.cosmogrid.ai/}}.
A schematic overview of the key post-processing steps applied to these simulations is provided in \cref{fig:pipeline}, with the following subsections following this data flow.

The implementation of the forward model is available on GitHub~\footnote{\href{https://github.com/des-science/multiprobe-simulation-forward-model}{\faGithub\ multiprobe-simulation-forward-model}}.

\subsection{\cosmogrid Simulations}\label{sec:cosmogrid}
The forward model is based on the \cosmogrid simulation suite of flat \wcdm cosmologies.
The dark matter only $N$-body simulations were run with the open-source \pkdgrav code~\cite{potterPKDGRAV3TrillionParticle2017}, which is optimized for hybrid CPU-GPU cluster architectures and has computational cost that scales linearly with the number of particles.

The following subsections describe the suite in more detail.

\subsubsection{Cosmological Parameters}\label{sec:cosmogrid_params}
\begin{table}[htbp]
    \begin{ruledtabular}
        \begin{center}
            \caption{
                Cosmological, astrophysical, and nuisance parameters of the fiducial ($\theta_\mathrm{fid}$) and grid (wide and narrow prior) \cosmogrid subsets.
                The elements of a joint, multivariate Sobol sequence are represented as $\sobol[a, b]$, where the square brackets define the bounding interval.
                Similarly, joint Latin Hypercube Sampling is denoted by $\latin[a, b]$.
                Univariate normal distributions with mean $\mu$ and standard deviation $\sigma$ are designated as $\normal(\mu, \sigma)$.
                We marginalize the cosmological parameters \ns, \obary, \hubble, the baryonification variables, and nuisances below the central double horizontal line.
            }
            \renewcommand{\arraystretch}{1.25}
            \begin{tabular}{l | l l l l}
                $\theta$                                         & $\theta_\mathrm{fid}$ & wide prior                & narrow prior              \\ \hline
                \multicolumn{3}{l}{\textbf{Cosmology}}           &                                                                               \\
                \omatter                                         & 0.26                  & $\sobol [ 0.1, \ 0.5 ]$   & $\sobol[ 0.15, \ 0.45 ]$  \\
                \sigeight                                        & 0.84                  & $\sobol [ 0.4, \ 1.4 ]$   & $\sobol [ 0.5, \  1.3 ]$  \\
                \wzero                                           & $-1$                  & $\sobol [-2, -0.33 ]$     & $\sobol [-1.25, -0.75 ]$  \\
                \ns                                              & 0.9649                & $\sobol [ 0.87, \ 1.07 ]$ & $\sobol [ 0.93, \ 1 ]$    \\
                \obary                                           & 0.0493                & $\sobol [ 0.03, \ 0.06 ]$ & $\sobol [ 0.04, \ 0.05 ]$ \\
                \hubble                                          & 67.3                  & $\sobol [ 64, \ 82 ]$     & $\sobol [ 65,  \ 75 ]$    \\ \hline
                \multicolumn{3}{l}{\textbf{Baryonification}}     &                                                                               \\
                $\log(\mczero)$                                  & 13.82                 & $\sobol [ 12, \ 15 ]$     & \multicolumn{1}{c}{-}     \\
                $\nu$                                            & 0                     & $\sobol [ -2, \ 2 ]$      & \multicolumn{1}{c}{-}     \\ \hline
                \multicolumn{3}{l}{\textbf{Intrinsic Alignment}} &                                                                               \\
                \AIA                                             & 0.5                   & $\latin [ -3, \ 3 ]$      & \multicolumn{1}{c}{-}     \\
                \etaIA                                           & 1.5                   & $\latin [ -4, \ 6 ]$      & \multicolumn{1}{c}{-}     \\
                \bta                                             & 1                     & $\latin [ 0, \ 2 ]$       & \multicolumn{1}{c}{-}     \\ \hline
                \multicolumn{3}{l}{\textbf{Galaxy Biasing}}      &                                                                               \\
                \bgone                                           & 1.34                  & $\latin [ 0.8, 3 ]$       & \multicolumn{1}{c}{-}     \\
                \bgtwo                                           & 1.42                  & $\latin [ 0.8, 3 ]$       & \multicolumn{1}{c}{-}     \\
                \bgthree                                         & 1.50                  & $\latin [ 0.8, 3 ]$       & \multicolumn{1}{c}{-}     \\
                \bgfour                                          & 1.57                  & $\latin [ 0.8, 3 ]$       & \multicolumn{1}{c}{-}     \\ \hline \hline
                \multicolumn{3}{l}{\textbf{Source photo-$z$}}    &                                                                               \\
                $\dzsource^1 \x 10^2$                            & \multicolumn{1}{c}{-} & $\normal(0, 1.8)$         & \multicolumn{1}{c}{-}     \\
                $\dzsource^2 \x 10^2$                            & \multicolumn{1}{c}{-} & $\normal(0, 1.5)$         & \multicolumn{1}{c}{-}     \\
                $\dzsource^3 \x 10^2$                            & \multicolumn{1}{c}{-} & $\normal(0, 1.1)$         & \multicolumn{1}{c}{-}     \\
                $\dzsource^4 \x 10^2$                            & \multicolumn{1}{c}{-} & $\normal(0, 1.7)$         & \multicolumn{1}{c}{-}     \\ \hline
                \multicolumn{3}{l}{\textbf{Source shear bias}}   &                                                                               \\
                $\mb^1 \x 10^2$                                  & \multicolumn{1}{c}{-} & $\normal(-0.6, 0.9)$      & \multicolumn{1}{c}{-}     \\
                $\mb^2 \x 10^2$                                  & \multicolumn{1}{c}{-} & $\normal(-2.0, 0.8)$      & \multicolumn{1}{c}{-}     \\
                $\mb^3 \x 10^2$                                  & \multicolumn{1}{c}{-} & $\normal(-2.4, 0.8)$      & \multicolumn{1}{c}{-}     \\
                $\mb^4 \x 10^2$                                  & \multicolumn{1}{c}{-} & $\normal(-3.7, 0.8)$      & \multicolumn{1}{c}{-}     \\ \hline
                \multicolumn{3}{l}{\textbf{Lens photo-$z$}}      &                                                                               \\
                $\dzlens^1 \x 10^2$                              & \multicolumn{1}{c}{-} & $\normal(-0.9, 0.7)$      & \multicolumn{1}{c}{-}     \\
                $\dzlens^2 \x 10^2$                              & \multicolumn{1}{c}{-} & $\normal(-3.5, 1.1)$      & \multicolumn{1}{c}{-}     \\
                $\dzlens^3 \x 10^2$                              & \multicolumn{1}{c}{-} & $\normal(-0.5, 0.6)$      & \multicolumn{1}{c}{-}     \\
                $\dzlens^4 \x 10^2$                              & \multicolumn{1}{c}{-} & $\normal(-0.7, 0.6)$      & \multicolumn{1}{c}{-}     \\
                $\sigmazlens^1$                                  & \multicolumn{1}{c}{-} & $\normal(0.98, 0.06)$     & \multicolumn{1}{c}{-}     \\
                $\sigmazlens^2$                                  & \multicolumn{1}{c}{-} & $\normal(1.31, 0.09)$     & \multicolumn{1}{c}{-}     \\
                $\sigmazlens^3$                                  & \multicolumn{1}{c}{-} & $\normal(0.87, 0.05)$     & \multicolumn{1}{c}{-}     \\
                $\sigmazlens^4$                                  & \multicolumn{1}{c}{-} & $\normal(0.92, 0.05)$     & \multicolumn{1}{c}{-}     \\
            \end{tabular}
            \label{tab:prior_cosmogrid}
        \end{center}
    \end{ruledtabular}
\end{table}

The simulations comprising the \cosmogrid suite sample the parameter space of \wcdm universes.
A flat universal geometry is maintained throughout by setting the dark energy density parameter to $\olambda = 1 - \omatter$. 
The neutrino mass is fixed to three degenerate species with $\mneutrino = \SI{0.02}{\eV}$ each, giving a total mass sum of $\sum \mneutrino = \SI{0.06}{\eV}$.

While recent results, including those from DES~\cite{collaborationDarkEnergySurvey2025} and the Dark Energy Spectroscopic Instrument (DESI)~\cite{adameDESI2024VI2025,adameDESI2024VII2025,desicollaborationDESIDR2Results2025}, suggest possible time variation in the dark energy equation of state (parametrized by $w_0$ and $w_a$), our analysis employs the \wcdm model with constant \wzero, as the \cosmogrid was designed around this choice.
We defer exploration of dynamical dark energy to future work, which will require generating a new suite of simulations.

The \cosmogrid contains two main sets of simulations, defined by the sampled point(s) in parameter space, which we generally denote as $\theta$:

\paragraph{Fiducial:}
The fiducial cosmological parameters of the \cosmogrid are based on the Planck2018 results~\cite{aghanimPlanck2018Results2020} and are shown in the $\theta_\mathrm{fid}$ column of \cref{tab:prior_cosmogrid}.
There is a total of 200 independent simulation runs at the fiducial cosmology, from which we build $1 \, 000$ semi-independent realizations as described in \cref{sec:shell_permutations}. 

Note that in a traditional likelihood analysis, realizations at the fiducial cosmology are typically used to estimate the summary statistic's covariance matrix. 
However, this is not necessary in the simulation-based inference framework we employ in this work, as an approximation to the likelihood function is learned directly by a normalizing flow (\cref{sec:neural_likelihood_estimation}).

\paragraph{Grid:}
\begin{figure}[!htb]
    \centering
    \includegraphics{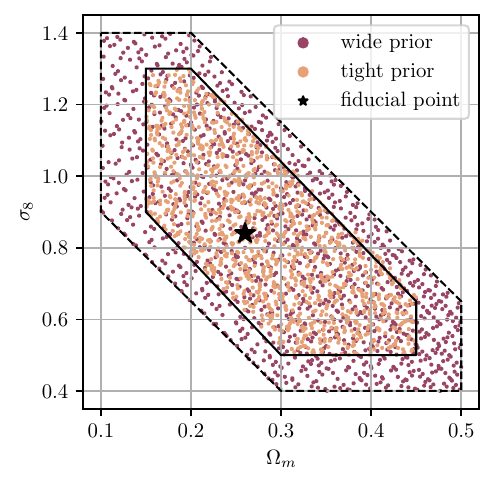}
    \caption{
        Projection of the fiducial and $2 \, 500$ grid cosmologies included in the \cosmogrid to the $\omatter - \sigeight$ plane. 
        Dashed and solid lines represent the wide and narrow priors, respectively, with the corresponding points colored in blue and orange.
        The black star marks the fiducial cosmology.
    }
    \label{fig:Om_s8_prior}
\end{figure}
In simulation-based inference, the grid in parameter space establishes the necessary connection between the cosmological and astrophysical variables making up the vector $\theta$ to be inferred and the (potentially synthetic) observable. 

The parameter grid within the \cosmogrid follows a Sobol sequence~\cite{sobolDistributionPointsCube1967}, which is a quasi-random, low-discrepancy sampling scheme that fills space uniformly.
The dimensionality of the sequence is expandable beyond the six cosmological parameters initially included in the \cosmogrid: \omatter (matter density), \sigeight (matter clustering amplitude), \wzero (dark energy equation of state), \ns (spectral index), \obary (baryon density), and \hubble (Hubble constant).
We use this property to extend the sequence to include baryonification parameters (\cref{sec:cosmogrid_bary}).

All one-dimensional priors over these parameters are uniform and flat, forming a hyperrectangle in aggregate, except for additional restrictions applied in the $\omatter - \sigeight$ (\cref{fig:Om_s8_prior}) and $\omatter - \wzero$ (Appendix A of~\cite[hereafter \citetalias{fluriFullMathrmCDM2022}]{fluriFullMathrmCDM2022}) planes.
The $2 \, 500$ grid points are evenly divided between narrow and wide priors with $1 \, 250$ cosmologies each, where the narrow prior provides higher sampling density in the region of greatest interest.
For the forecasts conducted in this work, we consider the entire parameter range of the wide prior and the narrow prior therefore has no further Bayesian significance. 
The wide and narrow prior parameter ranges are listed in \cref{tab:prior_cosmogrid}.

For every grid point, the \cosmogrid includes seven independent simulation runs stemming from different random initial conditions.
In the later map projection step, we mix these seven runs to yield twenty permutations as outlined in \cref{sec:shell_permutations}.

\subsubsection{Configuration}\label{sec:cosmogrid_config}
Throughout the \cosmogrid, simulation outputs are stored in lightcone format as thin shells of pixel-wise dark matter particle counts in \healpix~\footnote{\url{https://healpix.sourceforge.io/}} maps (detailed in~\cite{gorskiHEALPixFrameworkHighResolution2005}) with resolution $\nside = 2048$, corresponding to an angular scale of approximately 1.72 arcmin.

\begin{table}[htb]
    \begin{ruledtabular}
        \begin{center}
            \caption{
                Summary of the main simulation settings for different subsets of the \cosmogrid.
                The number of distinct cosmologies (or sampled points in parameter space) is denoted by $N_\textrm{cosmos}$, the number of permutations per cosmology by $N_\textrm{perts}$, the non-replicated simulated box size by \lbox, the number of simulated particles by $N_\textrm{part}$, and the number of redshift shells (or stored time steps) by $N_{z\textrm{-shells}}$.
            }
            \renewcommand{\arraystretch}{1.25}
            \begin{tabular}{l l l l l l}
                                                       & $N_\textrm{cosmos}$ & $N_\textrm{perts}$ & $L_\textrm{box}$           & $N_\textrm{part}$   & \makecell{$N_{z\textrm{-shells}}$} \\[-1.5ex]
                                                       &                     &                    & \text{\scriptsize [Mpc/h]} &          &                                    \\ \hline
                \multicolumn{3}{l}{\textbf{Main}}      &                     &                    &                                                                            \\
                Fiducial                               & $1$        & $1\, 000$               & 900                        & $832^3$  & 69                              \\
                Grid                                   & $2 \, 500$                & 20                 & 900                        & $832^3$  & 69                                \\ \hline
                \multicolumn{3}{l}{\textbf{Benchmark}} &                     &                    &                                                                            \\
                base                                   & 1                   & 20                 & 900                        & $832^3$  & 69                                \\
                \lbox                                  & 1                   & 20                 & $2 \, 250$                       & $2 \, 080^3$ & 69                                \\
                $N_\textrm{part}$                              & 1                   & 20                 & 900                        & $2 \, 048^3$ & 69                                \\
                \# $z$-shells                          & 1                   & 20                 & 900                        & $832^3$  & 392                                \\
            \end{tabular}
            \label{tab:cosmogrid_config}
        \end{center}
    \end{ruledtabular}
\end{table}

\Cref{tab:cosmogrid_config} summarizes the main simulation settings.
A distinction is made between:

\paragraph{Main settings:}
The main simulation settings apply to the \emph{fiducial} and \emph{grid} subsets. 
The configuration was informed by trade-offs between accuracy and computational feasibility, aiming to facilitate analyses of Stage-III LSS surveys like the one conducted in this work~\citepalias{kacprzakCosmoGridV1SimulatedCDM2023}.

\paragraph{Benchmark settings:}
The \cosmogrid contains a set of benchmark simulations designed to test and evaluate the robustness of the analysis with respect to the main simulation settings.
These include 
i) larger box sizes at equal particle density to verify the adequacy of the box replication scheme, 
ii) greater numbers of particles to decrease the amount of shot noise, and 
iii) increased numbers of stored redshift shells in the lightcone improving the resolution in the radial direction.
All benchmark simulations were conducted at the fiducial cosmological parameters and share initial conditions with runs performed using the main simulation settings in order to fix cosmic variance in direct comparisons.

\subsubsection{Baryonic Feedback}\label{sec:cosmogrid_bary}
The dark matter particles in \pkdgrav evolve under gravity alone.
However, it has been shown that baryonic feedback can bias cosmological constraints, particularly on small scales~\cite[e.g.][]{vandaalenEffectsGalaxyFormation2011,mccarthyBahamasProjectCalibrated2017,chisariModellingBaryonicFeedback2019,villaescusa-navarroCAMELSProjectCosmology2021}.

The standard DES Y3 modeling strategy~\cite{krauseDarkEnergySurvey2021} addresses this through scale cuts, removing small scales affected by baryonic physics as modeled in the AGN simulation~\cite{vandaalenEffectsGalaxyFormation2011} from the hydrodynamic OverWhelmingly Large Simulations (OWLS) suite~\cite{schayePhysicsDrivingCosmic2010}.
These scale cuts are tuned to ensure posterior shifts with respect to a fiducial uncontaminated data vector remain below $0.3 \sigma$ in the $\omatter - \Seight$ plane.

In this work, we instead incorporate baryonic effects directly into our modeling by post-processing the \cosmogrid lightcone shells with the effective \emph{baryonification model} developed in~\cite{schneiderNewMethodQuantify2015,schneiderQuantifyingBaryonEffects2019,schneiderConstrainingBaryonicFeedback2022}, which displaces the dark matter particles according to a physically motivated prescription.
The interested reader is directed to~\cite{schneiderQuantifyingBaryonEffects2019} for a comprehensive explanation of the original model and to~\citepalias{kacprzakCosmoGridV1SimulatedCDM2023} for details on the slightly modified, shell-level implementation used to baryonify the \cosmogrid.

Following~\citepalias{fluriFullMathrmCDM2022}, we only vary the model parameter $M_c$ defining the mass dependence of the gas profile, as it has been shown to have the biggest impact on cosmology~\cite{giriEmulationBaryonicEffects2021}, and assume a power-law redshift dependence 
\begin{equation*}
    M_c = \mczero \, (1 + z)^\nu
\end{equation*}
in terms of the new model parameters \mczero and $\nu$.
We assign every grid cosmology unique values \mczero and $\nu$ by extending the Sobol sequence with two additional dimensions, which we scale according to the prior ranges given in \cref{tab:prior_cosmogrid}.
Since the \cosmogrid only resolves halos with masses down to approximately $10^{13} M_\odot/h$, values at the lower end of the prior interval result in negligible baryonification.
All other parameters in the baryonification model are fixed to the same values as in~\citepalias{fluriFullMathrmCDM2022}, which are motivated by X-ray observations and listed in Tab.~2 of~\cite{schneiderQuantifyingBaryonEffects2019}.

Throughout this work, we use the baryonified \cosmogrid particle shells unless stated otherwise and marginalize over \mczero and $\nu$.
We assess the impact of baryonic effects at our fiducial scale cuts in \cref{sec:systematic_contamination}.

\subsection{Map Projection}\label{sec:map_projection}
We project the particle shells of the simulated lightcones onto probe maps with the publicly available~\footnote{\url{https://cosmology.ethz.ch/research/software-lab/UFalcon.html}} \ufalcon code~\cite{sgierFastGenerationCovariance2019,sgierFastLightconesCombined2021,reeves122Pt2024}, which has been used in several forecasts~\cite{fluriCosmologicalConstraintsNoisy2018,zurcherCosmologicalForecastNonGaussian2021,kacprzakDeepLSSBreakingParameter2022,zurcherFullWCDMMapbased2023,reeves122Pt2024} and cosmological inferences from real observational data~\cite{fluriCosmologicalConstraintsDeep2019,zurcherDarkEnergySurvey2022,fluriFullMathrmCDM2022,reevesTuningCosmicInstrument2025,reevesMultiprobeConstraintsEarly2025}.
The code employs the Born approximation, which was also assumed in~\cite{descollaborationDarkEnergySurvey2024,darkenergysurveyDarkEnergySurvey2025,jeffreyDarkEnergySurvey2025} and shown to be sufficiently precise for Stage-III surveys like DES in~\cite{petriValidityBornApproximation2017} and the systematics testing in~\cite{fluriCosmologicalConstraintsDeep2019}, an application to KiDS-450 weak lensing data.

As described in more detail in~\citepalias{kacprzakCosmoGridV1SimulatedCDM2023}, each pixel in the projected \healpix probe map $m$ is computed as
\begin{equation}
    m^{\mathrm{pix}} \approx \sum_b W_m \int_{\Delta z_b}\frac{\mathrm{d}z}{E(z)}\delta_{\mathrm{3D}}\left[\frac{c}{H_0}\mathcal{D}(z) \, \hat{n}^\mathrm{pix},z\right],
    \label{eq:projection}
\end{equation}
where the index $b$ runs over the lightcone's redshift shells of thickness $\Delta z_b$, $W_m$ is the kernel associated with a given probe as defined in the following, $\delta_\mathrm{3D}$ is a Dirac delta function, $c$ is the speed of light,
$\mathcal{D}(z)$ is the dimensionless comoving distance, $E(z) \coloneqq \mathrm{d}z/\mathrm{d}\mathcal{D}$ is the dimensionless Hubble parameter, and $\hat{n}^\mathrm{pix}$ is a unit vector pointing to the pixel's center. 

Following~\cite{fluriCosmologicalConstraintsDeep2019,fluriFullMathrmCDM2022,kacprzakDeepLSSBreakingParameter2022}, we define the probe kernels of the weak lensing (WL, first introduced in~\cite{fosalbaOnionUniverseAll2008}) and intrinsic alignment (IA, see \cref{sec:intrinsic_alignment}) signal as
\begin{align}
    % lensing
    W_\mathrm{WL} & = \frac{3}{2}\Omega_\mathrm{m}  \frac{\int_{\Delta z_b}\frac{\mathrm{d}z}{E(z)}\int_z^{z_s}\mathrm{d}z'n(z') \, \frac{\mathcal{D}(z)\mathcal{D}(z,z')}{\mathcal{D}(z')}\frac{1}{a(z)}}{\int_{\Delta z_b}\frac{\mathrm{d}z}{E(z)}\int_{z_0}^{z_s}\mathrm{d}z' \, n(z')}
    \label{eq:kernel_lensing}
    \\[1em]
    % intrinsic alignment
    W_\mathrm{IA} &= - \frac{\int_{\Delta z_b}\mathrm{d}z \, C_1 \rho_\mathrm{crit} \frac{\Omega_m}{D_+(z)} n(z)}{\int_{\Delta z_b} \frac{\mathrm{d}z} {E(z)}\int_{z_0}^{z_s}\mathrm{d}z' \, n(z')},
    \label{eq:kernel_ia}
\end{align}
where $n(z)$ is a normalized redshift distribution, $z_s$ and $z_0$ denote the source and observer redshifts, respectively, $a(z)$ is the scale factor, $C_1 = 5 \x 10^{-14} h^{-2} M_\odot \mathrm{Mpc}^3$ is a normalization constant~\cite{bridleDarkEnergyConstraints2007}, $\rho_\mathrm{crit}$ is the critical density today, and $D_+(z)$ is the normalized linear growth factor with $D_+(0)=1$. 
We evaluate \cref{eq:projection} for these kernels and $n_\metacal(z)$, resulting in the full-sky convergence maps $m = \kappa_\textrm{WL}, \kappa_\textrm{IA}$ in the left part of \cref{fig:pipeline}.

The kernel used to project the full-sky linear matter-density-contrast maps is defined as
\begin{align}
    % clustering
    W_\mathrm{GC} &= \frac{\int_{\Delta z_b}\mathrm{d}z \, n(z)}{\int_{\Delta z_b} \frac{\mathrm{d}z}{E(z)}\int_{z_0}^{z_s}\mathrm{d}z' \, n(z')}.
    \label{eq:kernel_lin_bias}
\end{align}
Insertion into \cref{eq:projection} yields the map $m = \dmlens$ for $n_\maglim(z)$ (used for lens galaxy clustering maps; \cref{fig:pipeline} right, \cref{sec:galaxy_clustering}) and $m = \dmsource$ for $n_\metacal(z)$ (used for shape-noise generation and modeling of intrinsic alignment of the source galaxy sample; \cref{fig:pipeline} left, \cref{sec:shape_noise,sec:intrinsic_alignment}).

During the projection step, we downsample the map resolution to \healpix $\nside = 512$ (corresponding to an angular pixel size of approximately 6.87 arcmin), which 
reduces the, nevertheless substantial, storage and compute requirements of this work. 
The decreased resolution also acts as a low-pass filter, removing small scales. 
However, due to limitations of our physics modeling validated in \cref{sec:scale_cuts}, we apply additional scale cuts to erase further small-scale information as detailed in \cref{sec:smoothing}.

\subsubsection{Shell Permutations}\label{sec:shell_permutations}
To make the $N$-body simulations computationally feasible for the large cosmological volumes necessary at high redshifts, \pkdgrav implements a box replication scheme.
This can introduce unwanted artifacts like discontinuities~\citepalias{kacprzakCosmoGridV1SimulatedCDM2023} and underestimation of cosmic variance on the largest scales~\cite{fluriCosmologicalConstraintsDeep2019}.
To avoid the former effect, we apply the \emph{shell permutation scheme} introduced in~\citepalias{kacprzakCosmoGridV1SimulatedCDM2023} during the map projection step.

As an additional benefit, this procedure increases the number of available pseudo-independent realizations as for a fixed cosmology at a time, simulation boxes stemming from different independent runs are randomly combined. 
We refer the interested reader to Section 4.1 in~\citepalias{kacprzakCosmoGridV1SimulatedCDM2023} for further details.

In this work, the 7 independent runs of the \emph{grid} subset of the \cosmogrid  are mixed to yield 20 permutations per cosmology, while the 200 simulation runs at the \emph{fiducial} cosmology are combined to 1000 distinct permutations of that ordered index set; see the $N_\textrm{perts}$-column in \cref{tab:cosmogrid_config}.

\subsubsection{Redshift Errors}\label{sec:redshift_errors}
Following the analysis in~\cite{descollaborationDarkEnergySurvey2022}, we model the uncertainty associated with the redshift distribution of tomographic bin $i$ for the source and lens galaxy samples using the \emph{shift} parameters $\dzsource^i$ and $\dzlens^i$, respectively, and the \emph{stretch} parameter $\sigmazlens^i$ for the lenses only.
The parameter distributions are assumed to be Gaussian with mean and standard deviation as reported in the \emph{photo-$z$}-rows of 
\cref{tab:prior_cosmogrid}.

We do not aim to constrain the shift and stretch uncertainty parameters; instead, we treat them as nuisances to be marginalized. 
To this end, we employ the same general strategy as~\citepalias{fluriFullMathrmCDM2022} and~\cite{jeffreyDarkEnergySurvey2025}:
First, we draw samples from the prescribed Gaussian distributions.
Then, we project the probe maps using the altered redshift distributions
\begin{equation}
    n_\metacal^i(z) = \hat{n}_\metacal^i  \left(z - \dzsource^i\right)
    \label{eq:z_error_metacal}
\end{equation}
for the source sample and
\begin{equation}
    n_\maglim^i(z) = \frac{1}{\sigmazlens^i} \, \hat{n}_\maglim^i \left( \frac{z - \langle z \rangle - \dzlens^i}{\sigmazlens^i} + \langle z \rangle \right)
    \label{eq:z_error_maglim}
\end{equation}
for the lens sample, where $\hat{n}$ denotes the original, noiseless distribution and $\langle z \rangle$ its mean redshift~\cite{descollaborationDarkEnergySurvey2022a}.
Thus, the redshift uncertainty directly enters the maps we create.

It can be shown~\cite{jeffreySolvingHighdimensionalParameter2020} that this implicit marginalization is mathematically equivalent to the standard integral approach.

\subsubsection{Validation}\label{sec:fm_validation}
The power spectra of the noiseless full-sky probe maps from the \cosmogrid have been validated against theoretical predictions from two independent codes.
In~\citepalias{fluriFullMathrmCDM2022}, comparison was performed with the \pyccl package~\footnote{\href{https://github.com/LSSTDESC/CCL}{\faGithub\ CCL}} described in~\cite{chisariCoreCosmologyLibrary2019} for \cref{eq:kernel_lensing,eq:kernel_ia} using KiDS-1000 redshift distributions.
In~\citepalias{kacprzakCosmoGridV1SimulatedCDM2023}, validation was conducted against the \pycosmo package~\footnote{\url{https://cosmology.ethz.ch/research/software-lab/PyCosmo.html}} introduced in~\cite{refregierPyCosmoIntegratedCosmological2018} for \cref{eq:kernel_lensing,eq:kernel_ia,eq:kernel_lin_bias} using generic Stage-III-like redshift distributions.
Both studies showed agreement within $5\%$, consistent with the minimal error expected from discrepancies between different theory predictions~\cite{euclidcollaborationEuclidPreparationIX2021,tanAssessingTheoreticalUncertainties2023}.

\subsection{Masking and Padding}\label{sec:masking}
We apply realistic masking to our forward-modeled maps to ensure they resemble the DES Y3 observations, which are contained within a complex footprint that excludes numerous intermediate objects, such as stars in the Milky Way.

\begin{figure}[!htb]
    \centering
    \includegraphics[width=\linewidth]{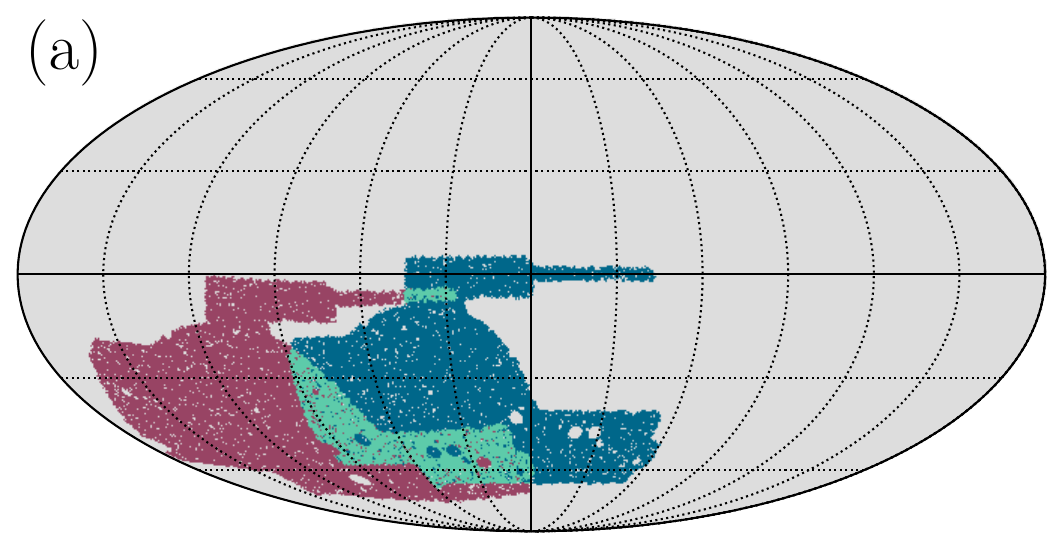}
    \includegraphics[width=\linewidth]{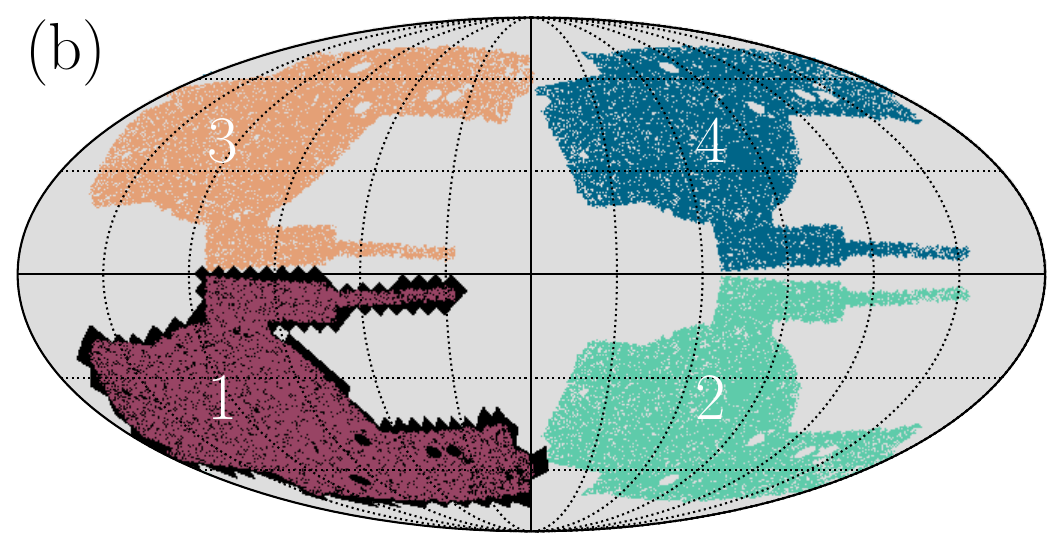}
    \caption{
        Full-sky mollweide projection of (a) how the original DES Y3 footprint (blue) is rotated to a position (red) that allows for (b) four non-overlapping cutouts.
        These distinctly colored patches are related by \healpix symmetries
        such that there is a perfect one-to-one correspondence between the pixels.
        The black padding with zeros along patch 1 is determined by the lowest \nside used within the \deepsphere networks; $\nside = 16$ here.
        It is not part of the original survey area.
    }
    \label{fig:footprint}
\end{figure}
To be able to cut out a total of four independent (except for super-survey modes, which we handle as described in \cref{sec:large_scales}) DES Y3 footprints from our full-sky simulations, we first rotate the fiducial mask from~\cite{descollaborationDarkEnergySurvey2022a} by \SI{-7.16}{\deg} along the $y$-axis and \SI{-69.9}{\deg} along the $z$-axis as illustrated in panel (a) of \cref{fig:footprint}.
At $\nside = 512$, this yields a mask of $4 \, 178$ square degrees. 

From this modified orientation, we apply \healpix octahedral symmetry 
transforms (rotations by \SI{90}{\deg} or mirroring along the equator) to the footprint to obtain three more non-overlapping positions of the DES mask on the celestial sphere.
These transforms maintain a perfect bijection between pixels, preventing the introduction of unwanted artifacts.
Thereby, the four footprints are fully equivalent.
Panel (b) in \cref{fig:footprint} illustrates these additional orientations and the black padding applied to the base footprint.
This zero-padding is required to ensure that the \deepsphere networks~\cite{perraudinDeepSphereEfficientSpherical2019,defferrardDeepSphereGraphbasedSpherical2019} described in \cref{sec:map_level} can internally downsample to a minimal \healpix \nside resolution without error.

As illustrated in \cref{fig:pipeline} and detailed in the following \cref{sec:weak_lensing}, we deliberately apply the masking to the physical fields that are directly accessible observationally (i.e. shear for the weak lensing maps) to ensure that the forward model remains consistent between our \cosmogrid mock maps and (synthetic) observations constructed from a galaxy catalog. 

\subsection{Weak Gravitational Lensing}\label{sec:weak_lensing}
Weak lensing measurements enable the reconstruction of the convergence field, which represents a weighted projection of the intervening matter density distribution between the observer and background source galaxies along the line of sight.
Since these maps directly trace the underlying matter distribution (predominantly dark matter), they are commonly referred to as mass maps~\cite{jeffreyDarkEnergySurvey2021}.

We forward-model DES Y3-like convergence maps $\kappa$ by summing contributions from the noiseless weak lensing signal (WL), intrinsic alignment (IA), and shape noise (SN):
\begin{equation}
    \kappa^i = \kappa_\mathrm{WL}^i + \AIA^i \left(1 + \bta \dmsource \right) \kappa_\mathrm{IA}^i + \kappa_\mathrm{SN}^i,
    \label{eq:kappa_sum}
\end{equation}
where $i$ indexes the redshift bin, \AIA and \bta are free parameters of the intrinsic alignment model, and \dmsource is the matter density contrast map for the source galaxy redshift distribution.
Each field component is detailed in the following subsections.

% details on Kaiser Squires in Section 2.2 of https://arxiv.org/pdf/2206.01450
\subsubsection{Mass-Mapping and B-Mode Removal}\label{sec:mass_mapping}
The convergence field $\kappa$ is not directly observable but can be reconstructed from the noisy shear field $\gamma$ obtained from source galaxy ellipticity measurements.
To ensure consistent processing (see below), we forward-model our weak lensing mass maps through a series of conversions between convergence and shear ($\kappa \rightarrow \gamma \rightarrow \kappa$), as illustrated in the left side of \cref{fig:pipeline}.

In the first step of the forward modeling pipeline, we project full-sky convergence maps $\kappa_\textrm{WL}$ and $\kappa_\textrm{IA}$ from simulated dark matter particle shells as described in \cref{sec:map_projection}. 
We then perform an inverse Kaiser-Squires transform~\cite{kaiserMappingDarkMatter1993,wallisMappingDarkMatter2022} to convert these to shear maps $\gamma_\textrm{WL}$ and $\gamma_\textrm{IA}$, respectively.
This conversion to shear allows us to apply identical subsequent processing steps, including masking, to both the observed $\gamma_\mathrm{obs}$ and forward-modeled maps.

By construction, the $\gamma_\textrm{WL}$ and $\gamma_\textrm{IA}$ obtained from simulated $\kappa$-maps via inverse Kaiser-Squires transform contain only $E$-modes, unlike $\gamma_\textrm{obs}$ and the shape noise map $\gamma_\textrm{SN}$ described in the following section, which may contain $B$-modes.
A $B$-mode mismatch between forward-modeled and observed shear maps can introduce biases in the inference.
Following the \emph{mode removal} procedure in~\cite[\citetalias{fluriFullMathrmCDM2022},][]{fluriCosmologicalConstraintsDeep2019}, we hence remove all potential $B$-modes from the $\gamma$-maps after applying masking.
This ensures that forward-modeled mocks, noise realizations, and real observations all contain the same masking effects and only $E$-modes.

Lastly, to reduce storage requirements by half, we apply a direct Kaiser-Squires transform to the masked spin-2 $\gamma$-maps to convert back to scalar $\kappa$-maps.
The spherical harmonics decomposition involved in this transform is only defined on the full sky, which we address by padding the area outside the survey footprint with zeros.

\subsubsection{Shape Noise}\label{sec:shape_noise}
\begin{figure}[!htb]
    \centering
    \includegraphics[width=\linewidth]{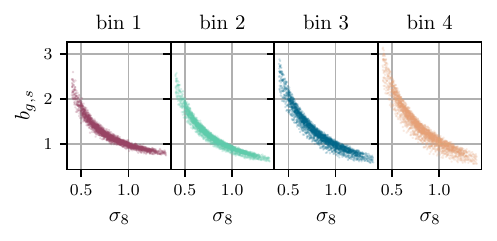}
    \caption{
        Scatterplot depicting the source galaxy bias $\bgsource$ we find for the $2 \, 500$ unique cosmologies of the \cosmogrid by fitting the source galaxy number count histogram to a reference \buzzard simulation of fixed cosmology.
    }
    \label{fig:metacal_bias}
\end{figure}

We generate shape noise stemming from the unknown intrinsic ellipticities of the source galaxies self-consistently using the dark matter density contrast maps \dmsource and the shuffled shear catalog $\tilde{\gamma}_\metacal$ (with positions discarded) of the source galaxies through the following steps:

First, we build the galaxy count map of tomographic bin $i$ assuming a linear galaxy biasing model
\begin{equation*}
    \ngsource^i = \nsourcei \left( 1 + \bgsource^i \, \frac{\dmsource^i - \langle \dmsource^i \rangle}{\langle \dmsource^i \rangle} \right),
\end{equation*}
where \nsource is the mean number of source galaxies per pixel and \bgsource is the linear galaxy clustering bias associated with the source galaxy sample.
We determine this bias by matching the pixel histogram of galaxy counts for each simulated cosmology to a reference \buzzard simulation (\cref{sec:mocks_buzzard}) serving as our synthetic mock observation.
The resulting dependence is shown in \cref{fig:metacal_bias}.
By varying \bgsource per cosmology in this way, the cosmology dependence of the one-point source galaxy distribution is absorbed into the bias parameter, effectively suppressing it in the resulting source galaxy maps.
The source galaxy bias is therefore separately fit to the (synthetic) observational data and fixed in the analysis rather than marginalized.

Given such a map of source galaxy counts, we generate a new catalog of pure shape noise by randomly sampling as many galaxies from the real DES Y3 source galaxy shear catalog $\tilde{\gamma}_\metacal$ as there are galaxies in any given pixel according to \ngsource.
In this step, we discard any positional information from $\tilde{\gamma}_\metacal$ and apply random rotations to the galaxies, such that any weak lensing signal is scrambled.
The \metacalibration inverse variance weights are kept.
Lastly, we obtain shape noise shear maps $\gamma_\textrm{SN}$ by computing the pixel-wise weighted mean of the shear over all contained galaxies.

This approach ensures that each pixel's shape noise level is coupled to its source galaxy count;
pixels with more galaxies experience lower shape noise. 
By linking shape noise to local galaxy densities, we account for the noise component of source galaxy clustering effects, which measurably impact certain map-based statistics in DES Y3 data~\cite{gattiDetectionSignificantImpact2023}.
We neglect the contribution of source clustering to the weak lensing signal itself, as this has been shown to be small compared to the noise component~\cite{gattiDetectionSignificantImpact2023}.

We choose to generate ten shape noise realizations per noiseless signal map at the \emph{fiducial} cosmology, and five realizations each for the cosmologies forming the parameter \emph{grid}.

\subsubsection{Intrinsic Alignment}\label{sec:intrinsic_alignment}
Intrinsic alignment refers to the non-random, correlated orientations of galaxies due to their mutual gravitational interactions or shared formation history, which can bias weak lensing measurements by introducing additional alignments unrelated to the lensing signal~\cite{troxelIntrinsicAlignmentGalaxies2015}.

The kernel in \cref{eq:kernel_ia} constitutes a map-level implementation of the non-linear alignment (NLA) model~\cite{bridleDarkEnergyConstraints2007} and has been validated against other theory predictions in~\cite{fluriFullMathrmCDM2022,kacprzakCosmoGridV1SimulatedCDM2023}.
The NLA model is a special case ($a_2 = b_\textrm{TA} = 0$) of the more general tidal alignment and tidal torquing (TATT) model~\cite{blazekLinearGalaxyAlignments2019} that has been used in the fiducial DES Y3 3$\times$2pt analysis~\cite{descollaborationDarkEnergySurvey2022a}.

Following~\cite[hereafter \citetalias{kacprzakDeepLSSBreakingParameter2022}]{kacprzakDeepLSSBreakingParameter2022}, we modify the standard redshift dependence of the NLA model to the effective prescription
\begin{equation}
    \AIA^i = \AIA \int_z dz \, n^{i}_\metacal(z) \left( \frac{1 + z}{1 + \zpivot}\right)^\etaIA,
    \label{eq:intrinsic_alignment_redshift}
\end{equation}
producing one amplitude per tomographic bin $i$, where $\zpivot = 0.62$ is a pivot redshift taken from~\cite{descollaborationDarkEnergySurvey2022a} and we integrate over the source galaxy redshift distribution $n^{i}_\metacal(z)$.
This formulation preserves the underlying power law dependence of the NLA model, but yields a single value per bin rather than evaluating it at the finer redshift resolution of the individual lightcone shells in \cref{eq:projection}.
We validate this minor modification in \cref{sec:systematic_contamination}.

In addition, we include the \bta term of the TATT model
\begin{equation*}
    \kappa_\mathrm{IA,TA} = \AIA \bta \, \dmsource \kappa_\mathrm{IA}, 
\end{equation*}
which encodes the coupling strength with the local density field~\cite{harnois-derapsCosmicShear2point2021} in \cref{eq:kappa_sum}.
Our intrinsic alignment model therefore represents an intermediate approach between the NLA and TATT models.

For each cosmology or equivalently, point on the Sobol sequence, we jointly sample the free intrinsic alignment parameters \AIA, \etaIA, and \bta using Latin hypercube sampling (LHS) with prior intervals listed in the \emph{Intrinsic Alignment} rows of \cref{tab:prior_cosmogrid}.

Currently, a map-level implementation of the remaining terms of the TATT model is an open research question~\cite{kacprzakCosmoGridV1SimulatedCDM2023} and therefore not included in this work. 
Moreover, our use of the NLA model is motivated by the fiducial DES Y3 3$\times$2pt analysis~\cite{descollaborationDarkEnergySurvey2022a}, which found consistent cosmological constraints between the TATT and NLA models~\cite[Appendix E.4]{descollaborationDarkEnergySurvey2022a} for the unblinded DES Y3 data.

\subsubsection{Shear Bias}\label{sec:shear_bias}
In addition to the multiplicative shear biases corrected by the self-calibrating \metacalibration algorithm~\cite{huffMetacalibrationDirectSelfCalibration2017,sheldonPracticalWeaklensingShear2017}, there are redshift-dependent detection and blending effects found in~\cite{maccrannDarkEnergySurvey2022} that can be modeled as a multiplicative bias on the $2-3 \%$ level.
In line with~\cite{jeffreyDarkEnergySurvey2025}, we account for this uncertainty by sampling the multiplicative bias factor $m_b^i$ from the normal distributions referenced in \cref{tab:prior_cosmogrid} and modify the forward modeled convergence map as
\begin{equation*}
    \kappa^i_\mathrm{WL} = (1 + m_b^i) \, \hat{\kappa}^i_{\mathrm{WL}}
    \label{eq:multiplicative_shear_bias}
\end{equation*}
for each tomographic bin $i$, where $\hat{\kappa}^i_{\mathrm{WL}}$ is the original map lacking the random correction.
This way, the $m_b^i$ are treated as nuisance parameters and marginalized like the redshift errors in \cref{sec:redshift_errors}.

Furthermore, null tests in~\cite{gattiDarkEnergySurvey2020} demonstrated the shear catalog's robustness against additive biases; therefore, we do not include them in this work.

\subsection{Galaxy Clustering}\label{sec:galaxy_clustering}
The galaxy clustering maps \nglens count the number of lens galaxies in each pixel.
To forward-model such maps from the \cosmogrid simulations, we first compute dark matter density contrast maps \dmlens by projecting the particle shells according to the kernel in \cref{eq:kernel_lin_bias}.
We then require a model to relate the matter and galaxy distributions, for which we adopt a linear biasing prescription as defined below.

The adequacy of this approach is validated in \cref{sec:scale_cuts} for the scale cuts described in \cref{sec:smoothing}.

\subsubsection{Linear Bias}\label{sec:linear_bias}
We parametrize the redshift dependence of the linear bias amplitude $b_g$ (for the lens galaxies, distinct from \bgsource for the source sample) by assigning an independent parameter $b_g^i$ to each tomographic redshift bin, avoiding any assumptions of continuous evolution.
Similar to the intrinsic alignment parameters, we include these bias parameters by joint LHS according to our analysis priors in \cref{tab:prior_cosmogrid}.

Given the linear bias, we then build the noiseless galaxy count map of tomographic bin $i$ as
\begin{equation}
    \nglenshat^i = \nlensi \left( 1 + \bgi \, \frac{\dmlens^i - \langle \dmlens^i \rangle}{\langle \dmlens^i \rangle} \right),
    \label{eq:linear_bias}
\end{equation}
where \nlens is the mean number of lens galaxies per pixel.
For large biases in regions of strong underdensity, the expression in \cref{eq:linear_bias} can produce unphysical negative values.
We address this by truncating negative pixels to zero and renormalizing the map to preserve the total galaxy count, an approach detailed in~\cite{hangBiasingGalaxyTrough2025}.

In \cref{sec:scale_cuts}, we validate this simple linear prescription against more sophisticated galaxy clustering models for appropriate scale cuts. 
Specifically, we condition the inference pipeline on external simulations described in \cref{sec:mocks_buzzard}, which employ sub-halo abundance matching and halo occupation distribution models.

\subsubsection{Poisson Noise}\label{sec:poisson_noise}
So far, the maps contain only the pure galaxy clustering signal according to our linear prescription.
To add shot noise, we follow~\citepalias{kacprzakDeepLSSBreakingParameter2022} and replace each pixel value by a draw from an independent Poisson distribution with a mean equal to the noiseless prediction of the number of galaxies:
\begin{equation*}
    \nglens^\mathrm{pix} = \textrm{Poisson}\left[ \nglenshat^\mathrm{pix} \right].
\end{equation*}
Within the pipeline, Poisson noise plays an analogous role to shape noise for weak lensing.

\subsection{Map-Level Smoothing and Scale Cuts}\label{sec:smoothing}
Our simulation-based inference approach uses map-level compression networks that operate directly in real space (\cref{sec:map_level}).
Consequently, we must apply scale cuts to the real-space maps.
This contrasts with summary statistics like the binned power spectrum (\cref{sec:two_point_level}), where scales can be excluded retroactively from the data vector.
\begin{figure*}[!htb]
    \centering
    \includegraphics[width=\linewidth]{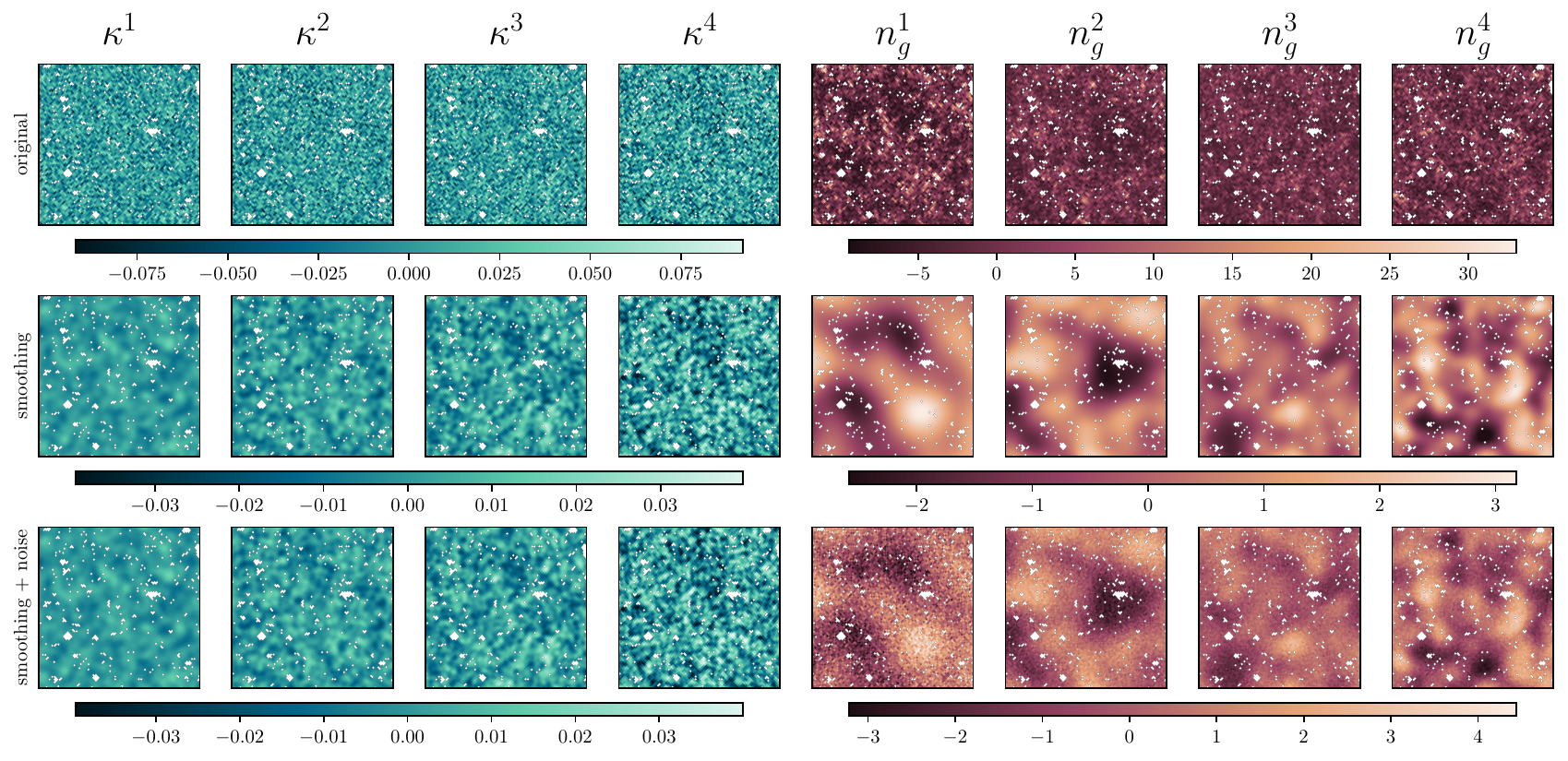}
    \caption{
        Comparison of the original unsmoothed maps (upper row) and the smoothed maps obtained following the steps described in \cref{sec:smoothing} (middle and lower row). 
        The angular smoothing scale is derived from a fixed comoving transverse scale, resulting in varying smoothing between the tomographic redshift bins (columns):
        Bins closer in redshift to the observer (lower index) are smoothed more than those further away.
        The squares show gnomic projections of size $\SI{10}{\deg} \x \SI{10}{deg}$, the colorbars are derived from the 0.001 and 0.999 quantiles, and the masking from \cref{fig:footprint} is visible in white.
    }
    \label{fig:smoothing_maps}
\end{figure*}

\subsubsection{Small Scales}\label{sec:small_scales}
Several assumptions in our forward model (such as the linear bias prescription for galaxy clustering or the effective baryonification) are expected to break down at sufficiently small scales.
To mitigate this source of model misspecification, we follow the standard DES strategy of excluding affected scales from the analysis.

We implement this removal of small-scale information through a two-step process detailed in \cref{sec:smoothing_appendix}.
First, we smooth the maps by convolving them with a Gaussian kernel of scale defined by its standard deviation $\sigma$.
Second, we add a small amount of pixel-wise white noise after smoothing to ensure that small-scale information is removed rather than merely suppressed, yielding an approximate maximal multipole $\tilde{\ell}_\mathrm{max}$ (defined in \cref{fig:smoothing_cls}).

To determine the extent of the smoothing kernel, we follow~\cite{descollaborationDarkEnergySurvey2022a,descollaborationDarkEnergySurvey2022} and fix a comoving transverse scale at the fiducial cosmology. 
From this, we compute distinct smoothing scales $\sigma_\mathrm{min}^i$ per redshift bin $i$, where the varying mean radial distances of the redshift distributions $n^i(z)$ translate the fixed transverse scale into different angular smoothing scales.
Since we keep these angular smoothing scales constant for all cosmologies on the parameter grid, they correspond to slightly varying transverse scales for different cosmologies. 

For weak lensing and galaxy clustering maps respectively, we set our fiducial smoothing scales using comoving transverse scales $R$ of
\begin{equation}
    \begin{aligned}
        R = 8 \, \mathrm{Mpc/h} \, &\Rightarrow \, \sigma_\mathrm{min}^i = [12.5, \, 8.5, \, 6.4, \, 5.4] \, \mathrm{arcmin} \\
        R = 32 \, \mathrm{Mpc/h} \, &\Rightarrow \, \sigma_\mathrm{min}^i = [55.3, \, 37.8, \, 28.9, \, 24.2] \,\mathrm{arcmin},
    \end{aligned}
    \label{eq:smoothing_scales}
\end{equation}
which are determined by the modeling-robustness tests in \cref{sec:scale_cuts} using external simulations.

\subsubsection{Large Scales}\label{sec:large_scales}
We implement a high-pass filter to suppress super-survey modes that could induce unwanted correlations between the four footprints described in \cref{sec:masking}. 
Following~\cite{zurcherDarkEnergySurvey2022}, this filter applies a hard cut in harmonic space, setting $\alm = 0$ for all $\ell < \ell_\textrm{min} = 30$.
The \alm coefficients arise from decomposing the map $m$ into spherical harmonics $Y_{\ell m}$ up to order $\ell_\mathrm{max}$:
\begin{equation}
    m(\theta, \phi) \approx \sum_{\ell=0}^{\ell_\mathrm{max}} \sum_{m=-\ell}^{\ell} a_{\ell m} Y_{\ell m}(\theta, \phi).
    \label{eq:map_spherical_harmonics}
\end{equation}

An example map resulting from the application of the low- and high-pass filters followed by the addition of white noise is compared to the original tomographic map in \cref{fig:smoothing_maps}.
All maps use resolution $\nside = 512$.

\subsection{Dataset Size}\label{sec:number_of_realizations}
In total, we generate $40 \, 000 \, (= \textrm{$1 \, 000$ permutations} \x \textrm{4 footprint cut-outs} \x \textrm{10 noise realizations})$ semi-independent realizations at the \emph{fiducial} cosmology and $400 \, (= \textrm{20 permutations} \x \textrm{4 footprint cut-outs} \x \textrm{5 noise realizations})$ per \emph{grid} cosmology, yielding $1 \, 000 \, 000 \, (= 400 \x \textrm{$2 \, 500$ cosmologies})$ grid maps overall.
This results in a total dataset size of more than 15 TB.

\section{Simulation-Based Inference}\label{sec:simulation_based_inference}
Bayes' Theorem~\cite{bayesLIIEssaySolving1763} relates the \emph{posterior} $p(\theta | x)$ to the \emph{likelihood} $p(x | \theta)$ and \emph{prior} $p(\theta)$ as
\begin{equation}
    p(\theta | x) \propto p(x | \theta) \, p(\theta),
    \label{eq:bayes}
\end{equation}
where $\theta$ denotes the model parameters and $x$ the data, neglecting the constant Bayesian evidence $p(x)$ in the denominator.

Traditional cosmological parameter inference relies on analytical or semi-analytical models to connect theoretical predictions with observational data~\cite[e.g.][]{krauseDarkEnergySurvey2021}. 
However, the increasing complexity of cosmological and observational models can render these likelihoods intractable or computationally prohibitive. 
In weak lensing and galaxy clustering cosmology, such intractability commonly arises from the inapplicability of the central limit theorem~\cite{hallNonGaussianLikelihoodWeak2022,oehlExactNonGaussianWeak2025}, non-linear structure formation at late times~\cite{linNonGaussianityWeakLensing2020,hallNonGaussianLikelihoodWeak2022}, realistic measurement systematics~\cite{sellentinSkewedWeakLensing2018,sellentinInsufficiencyArbitrarilyPrecise2018,taylorCosmicShearInference2019}, and summary statistics such as the map-level compression networks employed in this work, for which even the likelihood's expectation value cannot be predicted analytically and must instead be estimated from simulations.

In such scenarios, SBI (also known as implicit or likelihood-free inference) offers a powerful alternative by deriving posterior constraints $p(\theta | x_\mathrm{obs})$ conditioned on an observation $x_\mathrm{obs}$ without the need for explicit likelihood evaluation~\cite{cranmerFrontierSimulationbasedInference2020}.

A common strategy for SBI is splitting the problem into two consecutive steps: 
First, a compression function is constructed to map the data $x$ to a lower-dimensional space, facilitating subsequent density estimation.
In this work, this mapping is parametrized as an artificial neural network taking the forward-modeled maps or power spectra as input.
Second, a conditional probability distribution --- such as the likelihood or posterior --- relating the compressed data to the parameters $\theta$ is learned from an ensemble of simulations using neural density estimation.
Here, the density estimator is implemented as a normalizing flow and approximates the unknown likelihood.
This completes the framework, allowing amortized inference of the desired posterior constraints conditioned on any (mock) observation.

In \cref{sec:neural_compression}, we introduce the optimization objective that allows us to train neural networks to implement informative compression functions.
\Cref{sec:map_level} details how we train map-level compression networks, which have access to the whole information content of the forward-modeled maps in principle.
Following previous works~\cite[\citetalias{fluriFullMathrmCDM2022},][]{guptaNonGaussianInformationWeak2018,fluriCosmologicalConstraintsNoisy2018,ribliWeakLensingCosmology2019,fluriCosmologicalConstraintsDeep2019,jeffreyLikelihoodfreeInferenceNeural2021,matillaInterpretingDeepLearning2020,kacprzakDeepLSSBreakingParameter2022,luCosmologicalConstraintsHSC2023,jeffreyDarkEnergySurvey2025}, we present angular power spectra as a second-order baseline to compare to the map level in \cref{sec:two_point_level}.
Consistently following a simulation-based approach, we compute these angular power spectra from our forward-modeled maps.

\subsection{Neural Compression}\label{sec:neural_compression}
The forward-modeled maps in this work contain millions of pixels, and although the power spectrum detailed in \cref{sec:two_point_level} constitutes a summary statistic of the maps by itself, the resulting data vector still has over a thousand entries.
With such high dimensionality, estimating the likelihood function becomes practically intractable due to the curse of dimensionality.
This motivates the introduction of a \emph{compression function}
\begin{equation*}
    \begin{aligned}
        S: \R^{d_x} &\rightarrow \R^{d_s} \\
        x &\mapsto S(x)        
    \end{aligned}
    \label{eq:compression}
\end{equation*}
that maps from the high-dimensional input space of maps or power spectra to a much lower dimensionality $d_s \ll d_x$ where density estimation is feasible.
In our case, $d_s$ is of the same order as the number of constrained parameters.
The output of the compression function is also referred to as a summary statistic.

Ideally, this compression preserves all information about the parameter vector $\theta$ contained in the input $x$.
Such a compression is called sufficient (in the Bayesian sense~\cite{bernardoBayesianTheory2001}), meaning the posterior distribution remains unchanged when conditioned on the compression rather than the full input:
\begin{equation}
    p(\theta | x) = p(\theta | S(x)).
    \label{eq:sufficiency}
\end{equation}
While no theoretical guarantee of sufficiency exists for our setting, this is not strictly necessary. 
Suboptimal compressions yield inflated or overly conservative posterior contours rather than biased ones, since the same compression function is applied consistently to both forward-modeled and observational data~\cite{jeffreyDarkEnergySurvey2025}.
Moreover, studies on simplified weak lensing models have empirically demonstrated that neural compressions $S_\varphi$ with trainable parameters $\varphi$ can achieve sufficiency in practice, provided the network operates at the map level and is optimized with respect to an appropriate loss function like the one discussed below~\cite{fluriCosmologicalParameterEstimation2021,lanzieriOptimalNeuralSummarization2025}.

The implementation is available on GitHub~\footnote{\href{https://github.com/des-science/y3-deep-lss}{\faGithub\ y3-deep-lss}}.

\subsubsection{Optimization Objective}\label{sec:loss}
Alongside the architecture, the choice of optimization objective and procedure determines the quality of the compression;
specifically, how much information the compressed outputs retain about the parameters of interest in relation to \cref{eq:sufficiency}.

The simplest loss functions that produce manifestly informative compressions within our setting are the mean squared error~\cite[e.g.][]{luCosmologicalConstraintsHSC2023,jeffreyDarkEnergySurvey2025} 
and mean absolute error~\cite[e.g.][]{guptaNonGaussianInformationWeak2018,ribliWeakLensingCosmology2019,matillaInterpretingDeepLearning2020}.
These objectives train the network to respectively predict the mean or median of the one-dimensional marginal posterior of the individual components $\theta_i$ of the parameter vector, which does not fully characterize the multivariate posterior distribution in general and can therefore lead to suboptimal compression~\cite{jeffreyLikelihoodfreeInferenceNeural2021,lanzieriOptimalNeuralSummarization2025}.

Hence, we employ the information theoretical objective dubbed variational mutual information maximization (VMIM, introduced to the field of cosmology in~\cite{jeffreyLikelihoodfreeInferenceNeural2021}).
Note that in this case, a single network is trained to find a compression that is jointly informative on the whole parameter vector $\theta$.

Under this objective, the mutual information
\begin{equation}
    \begin{aligned}
        I(s, \theta) &= D_{KL}\left(p(s, \theta) \parallel p(s) p(\theta)\right) \\
        &= \mathbb{E}_{p(s, \theta)} [\log p(\theta | s)] - H(\theta)
    \end{aligned}
    \label{eq:mutual_info}
\end{equation}
between the output of the trainable compression $s \coloneqq S_\varphi(x)$ and the parameters $\theta$ is maximized, where $D_{KL}$ is the Kullback-Leibler (KL) divergence~\cite{kullbackInformationSufficiency1951}, and $H$ the Shannon entropy~\cite{shannonMathematicalTheoryCommunication1948}.
This expression is intractable because the exact posterior $p(\theta | s)$ is unknown.
However, there exist a number of tractable lower bounds to \cref{eq:mutual_info} that can instead serve as optimization targets \cite[see e.g.][]{chenNeuralApproximateSufficient2020}.
Following~\cite{jeffreyLikelihoodfreeInferenceNeural2021,lanzieriOptimalNeuralSummarization2025}, and based on our empirical finding that this approach outperforms alternative estimators, we employ the variational lower bound~\cite{barberIMAlgorithmVariational2003}
\begin{equation*}
    I(s, \theta) \geq \mathbb{E}_{p(s, \theta)} \left[ \log q(\theta | s; \phi) \right] - H(\theta),
\end{equation*}
where we have introduced the variational distribution $q(\theta | s; \phi)$ with trainable parameters $\phi$ to approximate the unknown true posterior $p(\theta | s)$.
The approximate learning objective then becomes
\begin{equation*}
    \arg\max_{\varphi, \phi} \, \mathbb{E}_{p(s, \theta)} \left[ \log q(\theta | s; \phi) \right],
\end{equation*}
where we have discarded the constant entropy term.
This yields the loss function
\begin{equation}
    L_\mathrm{VMIM}(x, \theta; \varphi, \phi) = - \log q(\theta | S_\varphi(x); \phi),
    \label{eq:loss_vmim}
\end{equation}
which we estimate from mini-batches following standard practice.

In our multi-probe setup, the parameter vector $\theta = (\omatter, \sigeight, \wzero, \AIA, \etaIA, \bta, \bgone - \bgfour)$ comprises all ten parameters to be constrained (see the green ellipses in \cref{fig:pipeline}), while marginalized nuisance parameters do not enter the loss.
For simplicity, we implement the density estimator $q_\phi$ as a Gaussian mixture model. 

Note that after training, we discard the variational distribution $q(\theta | s; \phi)$ and retain only the compression network $S_{\varphi^*}$ with optimal parameters $\varphi^*$.
We discard the approximate posterior $q_\phi$ for two reasons:
First, the prior implicit in the parameter space sampling by the \cosmogrid suite consists of a \emph{wide} and \emph{tight} Sobol sequence (see \cref{fig:Om_s8_prior}). 
When learning the posterior $q_\phi$ directly as in \cref{eq:loss_vmim} (neural posterior estimation~\cite{cranmerFrontierSimulationbasedInference2020,papamakariosFastEfreeInference2016,lueckmannFlexibleStatisticalInference2017,greenbergAutomaticPosteriorTransformation2019}), this non-uniform prior enters the posterior implicitly. 
Learning the likelihood instead (neural likelihood estimation~\cite{papamakariosSequentialNeuralLikelihood2019,lueckmannLikelihoodfreeInferenceEmulator2019,10.1093/mnras/stz1960}) avoids this issue and allows us to use a simpler uniform prior.
Second, dividing the problem into two consecutive steps by fitting a dedicated density estimator for a fixed compression network enables more accurate density estimation in practice.
For more details, see \cref{sec:neural_likelihood_estimation}.

\subsubsection{Map Level}\label{sec:map_level}
The compression networks operating at the map level have, in principle, access to the full information content of the forward-modeled pixelized fields, including non-Gaussian information absent from two-point statistics.
We therefore expect these networks to capture additional information beyond what is encoded in the power spectrum, leading to tighter posterior constraints.
Depending on the scales considered, this additional information can potentially break parameter degeneracies~\citepalias[e.g.][]{kacprzakDeepLSSBreakingParameter2022}.

The input to the networks consists of partial-sky maps within the padded footprint shown in \cref{fig:footprint}.
We treat the two probes and their respective four tomographic redshift bins analogously to color channels (e.g.~RGB) in natural images.
Hence, for the combined-probe analysis, all eight maps (columns in \cref{fig:smoothing_maps}) are used as input, while probe-specific analyses use only the corresponding four $\kappa$ or \nglens maps.
In this setup, all of the tomographic bins are treated symmetrically and can interact freely, allowing the network to capture arbitrary (non-linear) cross-bin and cross-probe interactions between any number of bins. 
This is in contrast to the two-point-level compression, which by construction only operates on cross-maps composed of two tomographic bins at a time (see \cref{eq:cross_cls}).

\paragraph{Architecture:}\label{sec:map_architecture}
All the map-level networks considered in this paper share the same basic structure: feature-extraction layers followed by a regression head, whose design remains unchanged throughout.

\begin{table}[htb]
    \begin{ruledtabular}
        \begin{center}
            \caption{
                Architecture of the fiducial \deepsphere graph convolutional neural network used for map-level compression of combined probes.
                Each row shows the layer type, tensor output shape for a batch of $N_b$ maps, and number of trainable parameters.
                The network processes partial-sky \healpix maps of $\nside = 512$ with $\npix = 458 \, 752$ pixels and eight feature channels corresponding to redshift bins.
                The residual layer is repeated five times.
                The model contains 6.87M trainable parameters in total, with output dimensionality equal to twice the number of (potentially weakly) constrained parameters.
                Networks for individual probe compression share this architecture except for input and output dimensions.
            }
            \begin{tabular}{l l l}
                Layer Type            & Output Shape            & \# Parameters \\ \hline
                Input                 & $(N_b, 458\,752, 8)$    & 0             \\
                Smoothing             & $(N_b, 458\,752, 8)$    & 0             \\
                Pseudo convolution    & $(N_b, 114\,688, 32)$   & 1 056         \\
                Pseudo convolution    & $(N_b, 28\,672, 64)$    & 8 256         \\
                Pseudo convolution    & $(N_b, 7\,168, 128)$    & 32 896        \\
                Chebyshev convolution & $(N_b, 7\,168, 256)$    & 163 840       \\
                Layer-normalization   & $(N_b, 7\,168, 256)$    & 512           \\
                Pseudo convolution    & $(N_b, 1\,792, 256)$    & 262 400       \\
                Chebyshev convolution & $(N_b, 1\,792, 256)$    & 327 680       \\
                Layer-normalization   & $(N_b, 1\,792, 256)$    & 512           \\
                Pseudo convolution    & $(N_b, 448, 256)$       & 262 400       \\
                Residual layer        & $(N_b, 448, 256)$       & 656 896       \\
                                      & \vdots                  &               \\
                Residual layer        & $(N_b, 448, 256)$       & 656 896       \\
                Flatten               & $(N_b, 114\,688)$       & 0             \\
                Layer-normalization   & $(N_b, 114\,688)$       & 229 376       \\
                Fully connected       & $(N_b, 20)$             & 2 293 780     \\
            \end{tabular}
        \label{tab:deepsphere_network}
        \end{center}
    \end{ruledtabular}
\end{table}
Following~\cite[\citetalias{fluriFullMathrmCDM2022},][]{fluriCosmologicalParameterEstimation2021}, our fiducial choice of architecture is the \tensorflow~\footnote{\url{https://www.tensorflow.org/}} implementation of the \deepsphere~\footnote{\href{https://github.com/deepsphere}{\faGithub\ deepsphere}} graph convolutional neural network developed in~\cite{perraudinDeepSphereEfficientSpherical2019,defferrardDeepSphereGraphbasedSpherical2019}.
In this design, the pixels making up the intricate footprint in \cref{fig:footprint} are represented as the nodes of a sparse graph that includes edges between a fixed number of neighbors.

The graph construction avoids projecting the survey data from the curved sphere onto a flat image, which invariably introduces distortions that break the symmetry of the spherical geometry and can hinder the learning of convolutional filters.
For unmasked inputs, that symmetry is nearly preserved by the \deepsphere graph convolutional layers, which are approximately equivariant under rotations.

As in~\cite[\citetalias{fluriFullMathrmCDM2022},][]{fluriCosmologicalParameterEstimation2021} and described in \cref{sec:masking}, we pad the survey footprint with zeros.
This is motivated by our use of the \deepsphere pseudo-convolution layers introduced in~\cite{fluriCosmologicalParameterEstimation2021}, which downsample the \nside of the internal feature maps according to the hierarchy inherent to the \healpix pixelization scheme and require full super-pixels.

The feature extraction part of our fiducial architecture summarized in~\cref{tab:deepsphere_network} is made up of a Gaussian smoothing layer operating in real space, the aforementioned pseudo-convolutional layers reducing the \nside of the internal representation, Chebyshev convolutions with optional residual connections, and layer normalizations~\cite{baLayerNormalization2016}.
Throughout, we use the rectified linear unit (ReLU) activation function.

We opt for a dense regression head of a single layer and dropout~\cite{srivastavaDropoutSimpleWay2014} of rate $0.01$. 
We found this to perform better than an equivariant architecture with a fully convolutional regression head. 
We suspect that this is the case because the masking we apply breaks the spherical symmetry.
The output dimensionality, which is a free choice for the VMIM objective, is fixed to twice the length of the parameter vector $\theta$. 

\paragraph{Training:}\label{sec:map_training}
Besides the forward-modeling of the multi-probe maps, training the map-level compression networks is the most computationally demanding step of the inference pipeline. 
To address this, we utilize the GPU nodes of the Perlmutter cluster~\footnote{\url{https://docs.nersc.gov/systems/perlmutter/architecture/}} at National Energy Research Scientific Computing Center~\footnote{\url{https://www.nersc.gov/}}, each equipped with four NVIDIA A100 GPUs connected via high-speed Nvlink-3.

The individual networks are trained in a fully data-parallel manner, distributing the global batch across multiple GPUs as local batches.
We fix the local batch size $N_b$ to 16 unless stated otherwise.
Then, the global batch size is determined by the number of nodes we utilize.
For our testing, we only distribute the training over a single node's four GPUs using {\tensorflow}’s built-in mirrored strategy. 
For the main runs, we scale the training across 4 nodes using \horovod~\footnote{\url{https://horovod.ai/}}, yielding a global batch size of 256.

For our fiducial training scheme, we use the \adam optimizer~\cite{kingmaAdamMethodStochastic2017} with default momentum parameters and clip the gradients to a global norm of 1.0 to prevent large parameter updates that destabilize training.

Our warm-up schedule linearly increases the learning rate from $10^{-4}$ to $10^{-3}$ over the first 5 thousand steps.
Afterwards, we keep the learning rate constant as we have found a cosine decay schedule to lead to overfitting.
In total, we train the networks for 500 thousand steps, which corresponds to forty epochs and takes approximately 24 wall-hours.

Our validation set is composed of noise realizations that are unseen during training and we did not observe any signs of overfitting in the validation loss.
We attribute this to both the large size of our training set, and the addition of white noise as part of our scale cut implementation, which serves as a regularizer.

\subsubsection{Two-Point Level}\label{sec:two_point_level}
In this work, the angular power spectra \cl serve as a baseline for comparison with the map-level results.
As a second order- or two-point statistic, the power spectra capture only the Gaussian component of the maps, discarding any additional non-Gaussian content.
While the angular power spectrum is a summary statistic of the forward-modeled maps, the dimensionality of the resulting data vector is still too large for direct neural density estimation in our implementation.
Therefore, in analogy to the map-level, we train a neural network to compress the information on $\theta$ contained in the \cl to lower dimensionality.
In summary, operating on the two-point-level can be seen as introducing a fixed, intermediate compression step that lacks trainable parameters and is known to be insufficient in the sense of \cref{eq:sufficiency}.

We define the pseudo $C_\ell$ of tomographic bins $i$ and $j$ using the spherical harmonics coefficients \alm from the map decomposition in \cref{eq:map_spherical_harmonics} as
\begin{equation}
    \cl^{ij} = \frac{1}{2 \ell + 1} \sum_{m=-\ell}^{\ell} \left| \sqrt{a_{\ell m}^i} \sqrt{a_{\ell m}^j} \right|^2,
    \label{eq:cross_cls}
\end{equation}
where cross-probe combinations are included if applicable.
We compute the \alm coefficients from the forward-modeled \healpix maps using the decomposition implemented in the \healpy~\cite{zoncaHealpyEqualArea2019} package.
Thus, we follow a simulation-based approach for the \cl too and do not rely on direct theory predictions, which allows us to easily include all of the systematics going into the map-level analysis and ensures direct comparability between the two approaches.

To form data vectors, we average the \cl within 32 square-root spaced bins between $\ell_\mathrm{min} = 0$ and $\ell_\mathrm{max} = 3 \, \nside$. 
To be consistent with the map-level, the scale cut is applied as a combination of smoothing and white noise as described in \cref{sec:smoothing,sec:smoothing_appendix}, such that the signal is suppressed for some $\tilde{\ell}_\mathrm{max} \leq \ell_\mathrm{max}$.
The fiducial smoothing scales in \cref{eq:smoothing_scales} yield
\begin{equation*}
    \begin{aligned}
        R = 8 \, \mathrm{Mpc/h} \, &\Rightarrow \, \tilde{\ell}_\mathrm{max} = [589, \, 863, \, 1\,159, \, 1382] \\
        R = 32 \, \mathrm{Mpc/h} \, &\Rightarrow \, \tilde{\ell}_\mathrm{max} = [133, \, 195, \, 255, \, 305] 
    \end{aligned}
    \label{eq:l_max}
\end{equation*}
for weak lensing and galaxy clustering respectively.
For comparison, previous simulation-based DES Y3 weak lensing analyses using the Gower Street simulation suite~\cite{jeffreyDarkEnergySurvey2025} employed a hard cut at $\lmax=1024$ for all four \metacal redshift bins~\cite{descollaborationDarkEnergySurvey2024,darkenergysurveyDarkEnergySurvey2025,jeffreyDarkEnergySurvey2025,pratDarkEnergySurvey2025}, while the theory-based DES Y3 harmonic space analysis~\cite{fagaDarkEnergySurvey2025} used scale cuts at $\lmax = [105, 154, 199, 237]$ for galaxy clustering and $\lmax = [139, 204, 264, 315]$ for galaxy-galaxy lensing.

We concatenate the binned \cl along the tomographic axis including cross-$z$ and potentially cross-probe bins with $i \neq j$, resulting in a ($32 \x 10$)-dimensional data vector for the single-probe setting, and a ($32 \x 36$)-dimensional data vector for the probe combination.

\paragraph{Architecture:}\label{sec:two_point_architecture}
Based on the results in Appendix D of~\citepalias{kacprzakDeepLSSBreakingParameter2022} and our own testing, we implement the compression using a fully connected network, as this simple architecture performs on par with more sophisticated ones for this task.

We define the fiducial network as a layer normalization~\cite{baLayerNormalization2016} layer right after the input, followed by two blocks, each containing a dense layer ($1 \, 024$ units and ReLU activation) and dropout (rate of $0.1$) in that order.
Like for the map-level networks, the final dense output layer has a dimensionality of $2 \, \text{dim}(\theta)$.

\paragraph{Training:}\label{sec:two_point_training}
Since the data volume is smaller by more than three orders of magnitude, training the compression networks at the two-point level is far less computationally demanding than at the map level, and can be completed on a single A100 GPU in under half an hour.

For direct comparability with the map-level analysis, we employ the same VMIM training objective here.
Similarly, we train the networks with the \adam optimizer and clip the global gradient norm to 1.0.
We perform 300 thousand training steps at a batch size of $4 \, 096$, which is feasible due to the comparatively small size of the binned \cl-datavector.

Before feeding the \cl into the networks, we take the logarithm of their absolute value to reduce their dynamic range and improve numerical stability. 

\subsection{Neural Likelihood Estimation}\label{sec:neural_likelihood_estimation}
The forward model described in \cref{sec:forward_model} generates samples 
$x \sim p(x | \theta) p(\theta) = p(x, \theta)$, where the vector $\theta$ consists of the cosmological parameters entering the $N$-body simulation, as well as the astrophysical and nuisance parameters incorporated during post-processing (see \cref{tab:prior_cosmogrid,fig:pipeline}). 

In the following, we denote by $\theta$ the parameters to be constrained, while the rest of the parameters are marginalized.
For the \emph{grid} subset of the \cosmogrid, this yields pairs $\{x_i, \theta_i\}_{i=1}^N$ with $N = 1 \, 000 \, 000$.
As discussed in the previous section, the high dimensionality of the data $x$ renders direct density estimation impractical.
Therefore, in the following, we consider $\mathcal{D} \coloneqq \{s_i, \theta_i\}_{i=1}^N$, where $s_i = S_{\varphi^*}(x_i)$ is the output of the fixed compression function resulting from training. 

The main distinction between different SBI methodologies lies in which probability density they approximate from these samples.
In this work, we choose neural likelihood estimation (NLE)~\cite{papamakariosSequentialNeuralLikelihood2019,lueckmannLikelihoodfreeInferenceEmulator2019,10.1093/mnras/stz1960}, which learns the inaccessible density $p(s | \theta)$ from the samples $\mathcal{D}$ using a density estimator $q(s | \theta; \phi)$ with trainable parameters $\phi$.
This way, we avoid making assumptions about the specific functional form of the underlying true likelihood, which is unknown, and we are free to use a different prior than the one implicit in the parameter-space sampling of the \cosmogrid simulations.

Code available on GitHub~\footnote{\href{https://github.com/des-science/multiprobe-simulation-inference}{\faGithub\ multiprobe-simulation-inference}}.

\subsubsection{Normalizing Flows}\label{sec:normalizing_flows}
We implement the neural density estimators using normalizing flows (NF), which model complicated probability distributions by learning a bijection to a simple, for example Gaussian, base distribution.
This bijective mapping is constructed as the composition of multiple discrete layers.
For a pedagogical introduction to NFs, we refer the interested reader to~\cite{papamakariosNormalizingFlowsProbabilistic2021}.

\paragraph{Architecture:}
We use the \flowconductor~\cite{arendtorresFlowConductorConditionalNormalizing2024} package since it offers a wide selection of conditional transformations.

Specifically, we compose the learnable bijection in our NFs from four blocks each made up of a conditional sum-of-sigmoids layer (introduced in Appendix A.1 of~\cite{negriConditionalMatrixFlows2023}) and singular value decomposition layer.
The former parametrize a monotonic, element-wise function as the sum of sigmoid activations and apply it auto-regressively~\cite{papamakariosMaskedAutoregressiveFlow2017}, while the latter allow for interactions between dimensions that are absent from those element-wise operations.

\paragraph{Training:}
In our setting, we only have access to the samples $\mathcal{D}$ and not their underlying probability density.
To approximate this unknown true distribution, we minimize the forward KL divergence between the true and approximate distributions, which yields the objective
\begin{equation}
    \begin{aligned}
        &\arg\min_{\phi} D_{KL} \left( p(s | \theta) \parallel q(s | \theta; \phi) \right) \\
      = &\arg\max_{\phi} \mathbb{E}_{p(s | \theta)} \left[ \log q(s | \theta; \phi) \right],
    \end{aligned}
  \label{eq:objective_nf}
\end{equation}
where $q(s | \theta; \phi)$ is the NF with trainable parameters $\phi$.
By dropping the terms constant with respect to $\phi$ in the second line, we see that this objective reduces to maximum likelihood estimation~\cite{papamakariosNormalizingFlowsProbabilistic2021}.
This defines the loss function
\begin{equation*}
    L_\mathrm{NF}(\phi) = - \log q(s | \theta; \phi),
    \label{eq:loss_nf}
\end{equation*}
which we estimate from mini-batches drawn from $\mathcal{D}$, thereby obtaining a Monte Carlo approximation of the expectation in \cref{eq:objective_nf}.

Due to the low dimensionality of the compressed data $s$, training the NFs is comparatively lightweight and fits on a single GPU.
For our fiducial setup, we use a batch size of $4096$, learning rate of $10^{-4}$, gradient clipping to $1.0$, cosine decay over $200$ epochs, and again the \adam optimizer.

The quality of the approximation $q(s | \theta; \phi) \approx p(s | \theta)$ is assessed with coverage tests in \cref{sec:posterior_coverage}.

\subsubsection{Inference}\label{sec:inference}
To obtain the posterior constraints $p(\theta | s_\mathrm{obs})$ for an observation $s_\mathrm{obs} = S_{\varphi^*}(x_\mathrm{obs})$, we approximate \cref{eq:bayes} as
\begin{equation*}
    \begin{aligned}
        p(\theta | s) &\propto p(\theta) \, p(s | \theta) \\
                      &\approx p(\theta) \, q(s | \theta; \phi^*),
    \end{aligned}
\end{equation*}
and sample using the Markov chain Monte Carlo (MCMC) code \emcee~\cite{foreman-mackeyEmceeMCMCHammer2013} employing $1 \, 024$ walkers performing $2 \, 000$ steps each.

\section{Mock Observations}\label{sec:mocks}
In this section, we describe the construction and key characteristics of synthetic observations used to conduct end-to-end tests of our inference pipeline's robustness against differing modeling choices.

\subsection{\cosmogrid}\label{sec:mocks_cosmogrid}
The \cosmogrid simulation suite includes \emph{benchmark} runs using the alternative $N$-body simulation settings listed in \cref{tab:cosmogrid_config} to assess the adequacy of our fiducial choices for particle number, redshift shell spacing, and replicated box size.
These tests address concerns raised by studies such as~\cite{bayerFieldlevelComparisonRobustness2025}, which demonstrated that varying particle counts can bias SBI when spurious small-scale information is not properly discarded.
Since benchmark runs share identical initial conditions with their fiducial counterparts, cosmic variance is eliminated in direct comparisons at the contour level, isolating the impact of the different simulation settings.
Furthermore, we use the forward model detailed in \cref{sec:forward_model} to construct the benchmark maps, ensuring that observed differences are exclusively attributable to variations in the underlying $N$-body simulations. 

The \cosmogrid also enables direct comparisons between mock observations with and without the baryonic corrections described in \cref{sec:cosmogrid_bary}, and between our treatment of intrinsic alignment amplitude redshift evolution in \cref{eq:intrinsic_alignment_redshift} and the standard NLA model.

\subsection{\buzzard}\label{sec:mocks_buzzard}
Our synthetic \buzzard~\cite{deroseBuzzardFlockDark2019,descollaborationDarkEnergySurvey2022d} DES Y3 mock observations are generated from a separate forward model not based on the \cosmogrid that does not follow the data flow in \cref{fig:pipeline}.
Therefore, these mocks probe potential model misspecification and, as such, play a pivotal role in validating our pipeline. 
The primary validation tests we conduct are cosmological parameter recovery and posterior predictive checks, which enable us to determine robust scale cuts for both the weak lensing and galaxy clustering components of our analysis.

To construct the mocks, we use the \buzzard~{\sc v2.0} suite~\cite{descollaborationDarkEnergySurvey2022d}, which consists of 18 synthetic galaxy catalogs covering the DES Y3 footprint, of which 15 are available to us.
The catalogs are generated from the lightcone output of dark matter-only simulations with halo masses above $\sim 5\times 10^{12} h^{-1} M_\odot$ at $z\leq 0.32$ and $\sim 10^{13} h^{-1}M_\odot$ up to $z\sim 2$, and are populated with galaxies using the \addgals method~\cite{wechslerADDGALSSimulatedSky2022}. 
This results in independent galaxy catalogs with up to $\sim 1.5 \times 10^9$ galaxies.

\addgals employs a hybrid approach combining sub-halo abundance matching (SHAM)~\cite{bernerForwardModellingGalaxy2024} and a halo occupation distribution (HOD)~\cite{asgariHaloModelCosmology2023} for the galaxy-halo connection. 
A machine learning model is trained on a calibrated SHAM model to assign central galaxies, while a probability distribution calibrated on SHAM is used to populate the simulation with subhalos. 
Hence, the procedure employed by \addgals differs fundamentally from the linear bias matter-galaxy connection implemented in our \cosmogrid forward model, which is key to our validation tests.

Additionally, the {\textsc CALCLENS} algorithm \cite{beckerCalclensWeakLensing2013} is used to compute gravitational shear at the position of each galaxy in the catalog.

For every catalog, we produce a single self-consistent tomographic map of synthetic weak lensing and galaxy clustering observations matching the DES Y3 data in terms of masking, redshift distribution, and average galaxy number counts.

Previously, \buzzard~{\textsc v2.0} catalogs were used to validate the DES Y3 3$\times$2pt analysis~\cite{descollaborationDarkEnergySurvey2022d}, while older versions of the catalogs~\cite{deroseBuzzardFlockDark2019} have been used to validate cosmological parameter estimation for DES Y1 \cite{maccrannY1ResultsValidating2018}. 
The underlying cosmological parameters of the \buzzard simulations are $\omatter = 0.286$, $\sigeight = 0.82$, $h = 0.7$, $\ns = 0.96$, and $\obary = 0.046$. 
The simulations do not model baryonic feedback.

\subsubsection{Synthetic Source Galaxy Catalog}\label{sec:mocks_buzzard_source}
To generate weak lensing mock observations from \buzzard galaxy catalogs that resemble the DES Y3 source galaxy sample introduced in \cref{sec:source_catalog}, we assign observational properties such as signal-to-noise ratio, size, and observed colors to individual galaxies in the  catalogs using the \balrog
framework~\cite{everettDarkEnergySurvey2022}, which matches galaxies from the DES Deep Field catalog to their multiple injections into the DES Wide Field image processing pipeline. 

This process involves three steps: 
First, we use the magnitudes of \buzzard galaxies in the {\textit{g}}, {\textit{r}}, {\textit{i}}, and {\textit{z}} bands to identify their closest counterparts in the DES Deep Field catalog. 
Second, we randomly select one Wide Field \balrog injection for each matched Deep Field galaxy and assign its \metacal properties to the corresponding galaxy in the original \buzzard catalog. 
Finally, after applying this procedure to all galaxies in the catalog, we implement the \metacal selection~\cite{gattiDarkEnergySurvey2021}:

\vspace{0.5em}
\noindent
{{\scriptsize
$
\begin{aligned}
    &\texttt{flags == 0} \\
    &\texttt{snr > 10} \\
    &\texttt{snr < 1000} \\
    &\texttt{size\_ratio > 0.5} \quad  \\
    &\texttt{T < 10} \\
    &\texttt{not (T > 2 and snr < 30)} \\
    &\texttt{not (log10(T) < (22.25 - r)/3.5 and sqrt(e\_1**2 + e\_2**2) > 0.8)}\\
    &\texttt{18 < i < 23.5} \\
    &\texttt{15 < (r,z) < 26} \\
    &\texttt{-1.5 < (r-i,z-i) < 4}, \\
\end{aligned}
$
}}
\vspace{0.5em}

\noindent where \texttt{snr} is the signal-to-noise ratio, \texttt{size\_ratio} is the ratio between the size \texttt{T} and the PSF size, and \texttt{e\_1}, \texttt{e\_2} denote the galaxy ellipticity components. 
The cuts yield a larger number of galaxies than the true source galaxy catalog, providing flexibility for further subselection.
We use this degree of freedom to fine-tune the mock catalog to more accurately reproduce the original \metacal redshift distribution by adjusting galaxy counts within thin redshift intervals $\mathrm{d}z$ to match expected values while maintaining a constant total galaxy count per redshift bin.

Based on this processed sample, we can now generate the mock weak lensing shear maps $\gamma_{\rm WL}$.
For the shape noise component $\gamma_{\rm SN}$, we sample the intrinsic ellipticities  $\gamma_{j,\rm SN}$ and weights $w_j$ from the DES Y3 shape noise catalog $\tilde{\gamma}_\metacal$, consistent with the procedure adopted for the \cosmogrid in \cref{sec:shape_noise}.
We choose not to include intrinsic alignment effects into the mocks, equivalent to an intrinsic alignment amplitude \AIA of zero.

The resulting shear values in a given pixel $i$ are weighted by $w_j$, where $j$ indexes the galaxies falling into pixel $i$: 
\begin{equation*}
    \gamma^\text{pix}_i = \frac{\sum_j w_j (\gamma_{j,\rm WL} + \gamma_{j,\rm SN})}{\sum_j w_j}.
\end{equation*}
To ensure consistency with the \cosmogrid mocks and following the procedure outlined in \cref{sec:mass_mapping}, we apply identical masking, remove $B$-modes, and convert the shear maps to convergence maps using the Kaiser-Squires transform, yielding the final tomographic $\kappa_\buzzard$-maps. 

Throughout this paper, we assume the total (per-bin) responses to equal unity in both the \cosmogrid forward model and the construction of the \buzzard mocks.
We will consider realistic response values in the follow-up companion paper dedicated to the analysis of the actual DES Y3 observations.

\begin{table}[htb]
    \begin{ruledtabular}
        \caption{
            Selected properties of the synthetic \metacal sample constructed from a simulated \buzzard galaxy catalog.        
        }
        \renewcommand{\arraystretch}{1.25}
        \begin{tabular}{c c c c c c c}
            Bin & $N_g$ & $\langle z \rangle$ & $n_{\mathrm{eff}}$ & $\sigma_e$ & $\cap n(z)$ & \lmaxt \\
            \hline
            Full & 100\,203\,633 & 0.623 & 5.545 & 0.260 & -- & -- \\
            1 & 24\,940\,369 & 0.32 & 1.463 & 0.244 & 0.97 & 589 \\
            2 & 25\,280\,310 & 0.51 & 1.467 & 0.262 & 0.97 & 863 \\
            3 & 24\,891\,762 & 0.74 & 1.471 & 0.259 & 0.97 & 1159 \\
            4 & 25\,091\,192 & 0.93 & 1.449 & 0.310 & 0.97 & 1382 \\
        \end{tabular}        \label{tab:buzzard_metacal}
    \end{ruledtabular}
\end{table}
\Cref{tab:buzzard_metacal} summarizes the properties of our synthetic \metacal catalogs. 
The column, $N_g$ gives the total number of galaxies in the full sample and each tomographic bin, while $\langle z \rangle$ denotes the mean redshift of the source galaxies. 
We define the effective number density $n_{\rm eff}$ and shape variance $\sigma_e$ following~\cite{gattiDarkEnergySurvey2021}.
The overlap metric $\cap n(z)$ quantifies the similarity between our reconstructed galaxy redshift distributions and those of the DES Y3 \metacal sample, and is computed as the overlap of the normalized $n(z)$ distributions where 0 indicates completely disjoint distributions and 1 indicates identical distributions.

\subsubsection{Synthetic Lens Galaxy Catalog}\label{sec:mocks_buzzard_lens}
In the following, we outline how we post-process the \buzzard galaxy catalogs to generate mock observations resembling the DES Y3 \maglim lens galaxy sample described in \cref{sec:lens_catalog}. 

Starting from the original \buzzard catalogs, we first apply the same \textit{i}-band magnitude cut $i< 4z + 18$ employed in the \maglim sample, where $i$-band magnitudes are available for each galaxy in the \buzzard catalogs and $z$ is the photometric redshift estimated using the DNF algorithm~\cite{sevilla-noarbeDarkEnergySurvey2021,devicenteDNFGalaxyPhotometric2016}. 

Since only one of the \buzzard catalogs includes DNF redshift estimates, we construct the conditional probability distribution $p(z | z_{\rm true})$ of DNF redshifts $z$ given the true redshifts from this catalog $z_{\rm true}$ and sample from this distribution to impute the remaining catalogs. 
This allows us to approximately reconstruct the redshift distributions $n^i_\maglim(z)$ for each tomographic bin of the \maglim catalog $i$. 

As the last step before binning, we fine-tune our catalog to reconstruct the $n_\maglim(z)$ distributions more precisely, following the same approach used for the weak lensing mocks.

\begin{table}[htb]
    \begin{ruledtabular}
        \caption{
            Selected properties of the synthetic \maglim sample constructed from a simulated \buzzard galaxy catalog.        
        }
        \renewcommand{\arraystretch}{1.25}
        \begin{tabular}{c c c c c c c}
            Bin & $N_g$ & Redshift range & $\langle z \rangle$ & $\langle{n_g}\rangle$ & $\cap n(z)$ & \lmaxt \\
            \hline
            Full & 7\,700\,661 & 0.20--0.85 & 0.54 & 0.512 & -- & -- \\
            1 & 2\,256\,056 & 0.20--0.40 & 0.30 & 0.150 & 1.00 & 133 \\
            2 & 1\,609\,242 & 0.40--0.55 & 0.46 & 0.107 & 1.00 & 195 \\
            3 & 1\,639\,381 & 0.55--0.70 & 0.62 & 0.109 & 1.00 & 255 \\
            4 & 2\,195\,982 & 0.70--0.85 & 0.77 & 0.146 & 1.00 & 305 \\
        \end{tabular}
        \label{tab:buzzard_maglim}
    \end{ruledtabular}
\end{table}
\Cref{tab:buzzard_maglim} presents the basic properties of our synthetic \buzzard galaxy clustering maps for comparison with the original DES Y3 \maglim sample.
The table shows the total number of galaxies $N_g$, redshift ranges and mean redshifts $\langle z \rangle$ for each tomographic bin, galaxy number density $n_g$ (per arcmin$^2$), overlap integral $\langle z \rangle$ between our galaxy redshift distributions and those of the DES Y3 \maglim catalog, and the maximum multipole \lmaxt assumed in the galaxy clustering analysis.

For more technical details about the \buzzard mock observations, see~\cite{buckoDarkEnergySurvey}.

\section{Validation}\label{sec:validation}
In this section, we present tests to determine appropriate scale cuts and validate our inference pipeline at these scales using the mock observations described in \cref{sec:mocks}.
The tests are performed for the map-level compression networks (see \cref{sec:map_level}), which are sensitive to non-Gaussian information, thereby providing a more stringent validation setting than the two-point statistic baseline (see \cref{sec:two_point_level}).
We confirmed that the two-point reference satisfies the same validation criteria.

We define the parameter \seight as
\begin{equation*}
    \seight \coloneqq \sigeight (\omatter / 0.3)^{0.5},
\end{equation*}
which breaks the degeneracy between \omatter and \sigeight that is inherent to weak lensing measurements.

\subsection{Scale Cuts}\label{sec:scale_cuts}
Particular approximations in the forward model detailed in \cref{sec:forward_model} are expected to fail at sufficiently small scales, most notably the assumption of linear galaxy biasing in our galaxy clustering maps.
Consequently, implementing appropriate scale cuts that discard parts of the data is essential to ensure unbiased parameter inference.
This strategy is consistent with previous DES Y3 analyses, where small-scale information was excluded~\cite[e.g.][]{descollaborationDarkEnergySurvey2022,descollaborationDarkEnergySurvey2022a,descollaborationDarkEnergySurvey2022b,descollaborationDarkEnergySurvey2022c} to mitigate potential systematic biases from, for example, inadequate modeling of baryonic effects.

For this work, we adopt conservative scale cuts that pass the validation tests presented below.
We do not optimize these cuts to find the least conservative values that would still pass our tests, deferring such optimization to the forthcoming companion paper.

\subsubsection{Recovery of \buzzard Cosmology}\label{sec:recovery_of_buzzard}
As described in \cref{sec:mocks_buzzard}, our DES Y3-like \buzzard weak lensing and galaxy clustering mocks are constructed from HOD-modeled galaxy catalogs.
This contrasts with the \cosmogrid mocks used to train the compression networks and perform SBI, which are generated directly from probe maps at the pixel level (see \cref{sec:map_projection}).
Furthermore, the $N$-body simulations underlying the \buzzard catalogs use a different numerical code with higher-fidelity settings.
Therefore, these mocks introduce systematic modeling discrepancies (of some degree) relative to the \cosmogrid mocks, enabling tests of the inference pipeline's robustness to such misspecification, or equivalently, generalization performance under \emph{covariate shift}.
We aim to demonstrate that these differences in input do not significantly impact the results.
\begin{figure*}[!htb]
    \centering
    \includegraphics[scale=0.75]{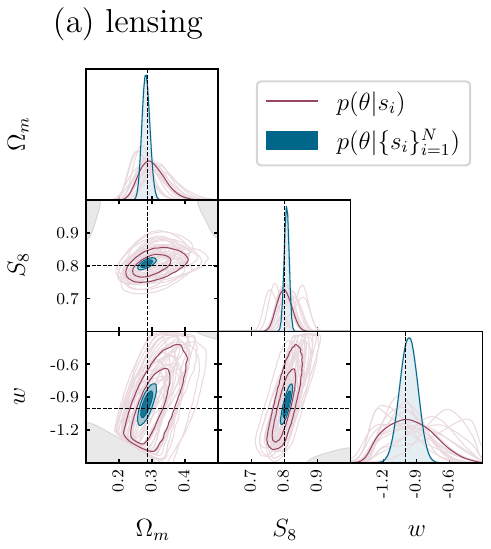}
    \includegraphics[scale=0.75]{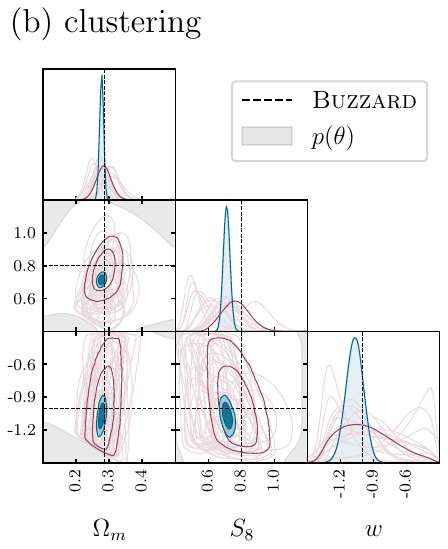}
    \includegraphics[scale=0.75]{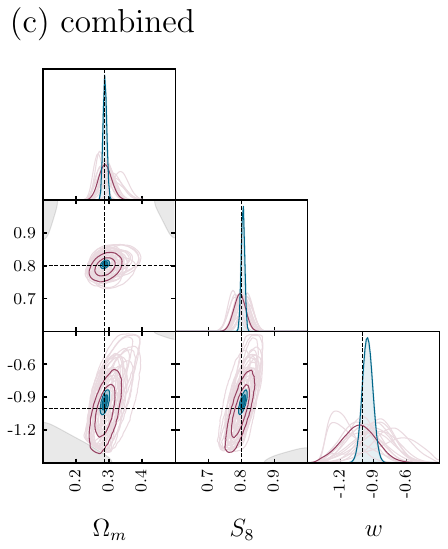}
    \caption{
        Cosmological parameter recovery for individual probes and their combination using an ensemble of $N = 15$ \buzzard mocks $x_i$ with map-level compression $s_i \coloneqq s(x_i)$.
        In all cases, the MAP estimate of the joint posterior $p(\theta|\{s_i\}_{i=1}^N)$ defined in \cref{eq:posterior_product} (solid blue contours) lies within $< 0.5 \sigma$ of a single observation $p(\theta | s_i)$ shifted to align with the ground truth (solid red contour).
        This result demonstrates that potential systematic biases from modeling differences between the \cosmogrid and \buzzard simulations remain below the statistical uncertainty.
        Faint red contours show individual realizations $x_i$, contour levels represent 68th and 95th percentiles, and the gray shaded region indicates where the prior $p(\theta)$ is zero.
    }
    \label{fig:recover_cosmo_buzzard}
\end{figure*}

As a first validation, we test whether the pipeline correctly recovers the known cosmology of the \buzzard simulations.
Following~\cite{maccrannY1ResultsValidating2018}, we evaluate the joint posterior over the ensemble of mocks $x_i$: 
\begin{equation}
    p(\theta|\{s(x_i)\}_{i=1}^N) \propto p(\theta) \prod_{i=1}^N p(s(x_i) | \theta),
    \label{eq:posterior_product}
\end{equation}
where $N = 15$.
This expansion assumes conditional independence between the $x_i$ given $\theta$ and simulates the constraining power of a survey with $N$-fold larger area, thereby reducing the impact of cosmic variance.
The independence assumption is justified because, analogously to our \cosmogrid mocks, we remove potential large-scale correlations between mock observations constructed from disjoint sky regions of the same underlying $N$-body simulation by applying an $\ell_\textrm{min}$-cut as detailed in \cref{sec:large_scales}.
Furthermore, the respective noise realizations are independent by construction. 

We present the results for weak lensing and galaxy clustering individually, and their combination, in \cref{fig:recover_cosmo_buzzard}.
For each configuration, the maximum a posteriori (MAP) estimates of $p(\theta|\{s(x_i)\}_{i=1}^N)$ lie within $0.5 \sigma$ of the true cosmological parameters $\theta = \{\omatter, \Seight, \wzero\}$ in the two-dimensional marginals.
This demonstrates that potential systematic biases from the distributional mismatch between the \cosmogrid and \buzzard simulations are subdominant to statistical uncertainties.
We therefore conclude that for the considered scale cuts, the inference pipeline passes the test.

\subsubsection{Systematic Mock Contamination}\label{sec:systematic_contamination}
This section presents additional tests designed to validate the inference pipeline's robustness to modeling discrepancies using mock observations intentionally contaminated with systematic errors.
Here, the mocks for posterior inference are generated within the \cosmogrid forward model while varying only one selected property at a time.
Shifts in the posterior constraints are therefore directly attributable to the known modeling difference.
We consider tests successful when shifts remain below $0.3\sigma$ in the $\omatter  - \Seight$ plane, the standard criterion for DES Y3~\cite[e.g.][]{descollaborationDarkEnergySurvey2022}.
Note that this differs from the previous cosmology recovery test: 
here, shifts are measured relative to the result obtained from the fiducial setup, not the ground truth.

Following~\cite{descollaborationDarkEnergySurvey2024,jeffreyDarkEnergySurvey2025,pratDarkEnergySurvey2025}, we perform these tests by conditioning on the component-wise mean
\begin{equation}
    \bar{s}_j = \frac{1}{N} \sum_{i=1}^{N} s(x_i)_j
    \label{eq:mean_mock}
\end{equation}
of the compression vector $s(x_i)$, where $j$ indexes vector components and $i$ runs over $N = 80$ pseudo-independent realizations.
This approach is analogous to standard \emph{noise-free} inference.

\begin{figure}[!htb]
    \centering
    \includegraphics[width=\linewidth]{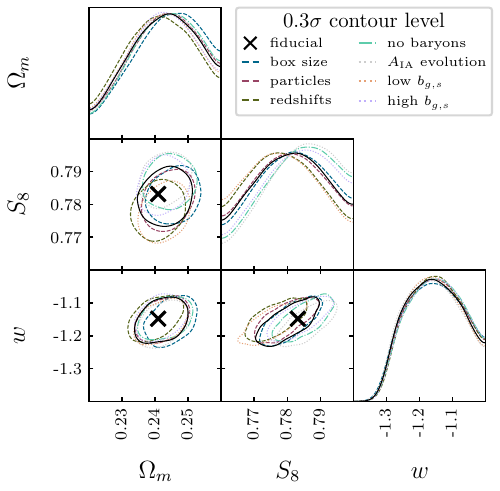}
    \caption{
        Impact of systematic error contamination on cosmological constraints inferred from map-level probe combination.
        The contour lines indicate the $0.3 \sigma$ tolerance for acceptable shifts within the $\omatter - \Seight$ plane.
        All configurations remain within this limit relative to the fiducial MAP (indicated by the cross), demonstrating robustness to alternative modeling.
    }
    \label{fig:validation_contours}
\end{figure}

\Cref{fig:validation_contours} shows the posterior $0.3\sigma$-level contours from the following systematically contaminated mocks for the map-level combined probe analysis:

\paragraph{Simulation Settings:}
The \cosmogrid benchmark mocks (see \cref{sec:cosmogrid_config,sec:mocks_cosmogrid}) validate the fiducial $N$-body simulation setup by increasing the
i) box size, 
ii) particle counts, and 
iii) number of redshift shells.

\paragraph{No Baryonification:}
The baryonification model strength is varied throughout the \cosmogrid according to \cref{sec:cosmogrid_bary}.
Since the associated parameters \mczero and $\nu$ are not expected to be well-constrained at our fiducial scale cuts~\citepalias{fluriFullMathrmCDM2022}, we marginalize over them. 
For this test, we omit baryonification post-processing from the dark matter lightcone, unlike the fiducial mock observation used as reference.

\paragraph{Intrinsic Alignment Amplitude:}
We compare our per-bin parametrization of \AIA redshift evolution in \cref{eq:intrinsic_alignment_redshift} with the standard NLA formulation, where the amplitude varies per redshift shell.

\paragraph{Source Galaxy Bias:}
We test the sensitivity to deviations from the source galaxy bias values introduced in \cref{sec:shape_noise}.
We consider low and high source galaxy clustering biases corresponding to $\pm 2 \sigma$ shifts relative to the distributions in \cref{fig:metacal_bias}: $\bgsource^\mathrm{low} = [1.0, 0.96, 1.0, 1.0]$ and $\bgsource^\mathrm{high} = [1.2, 1.2, 1.3, 1.5]$. 

All configurations produce posterior shifts below $0.3\sigma$ relative to the fiducial baseline, passing our robustness criterion for the combined probes analysis.
Individual probe analyses (not shown) also satisfy this criterion.
The result for the baryonification test indicates that baryonic effects are negligible at our fiducial scale cuts, yet we retain marginalization over the associated parameters as a conservative measure.

\subsubsection{Posterior Predictive Distribution}\label{sec:ppd}
The posterior predictive distribution (PPD)~\cite{gelmanBayesianDataAnalysis}
\begin{equation}
    p_{\text{PPD}}(x^\star | x_{\text{obs}})=
\int p(x^\star | \theta)\,p(\theta | x_{\text{obs}}) \, d\theta
\label{eq:ppd}
\end{equation}
provides an additional diagnostic for detecting potential systematic biases in our inference pipeline by comparing (synthetic) observations $x_\textrm{obs}$ with predicted data $x^\star$ generated from the fitted model.
In this validation test, we employ a \buzzard mock as $x_\textrm{obs}$ while $p(x^\star|\theta)$ represents our \cosmogrid forward model, again creating a scenario where model misspecification may arise.
When the forward model adequately captures the underlying data-generating process, samples $x^\star$ drawn from the PPD should exhibit statistical properties consistent with the observation $x_\mathrm{obs}$. 
Systematic discrepancies between them indicate deficiencies in the forward model, analogous to biased cosmology recovery or significant posterior shifts due to modeling errors, as examined in previous sections.

\begin{figure}[!htb]
    \centering
    \includegraphics[width=\linewidth]{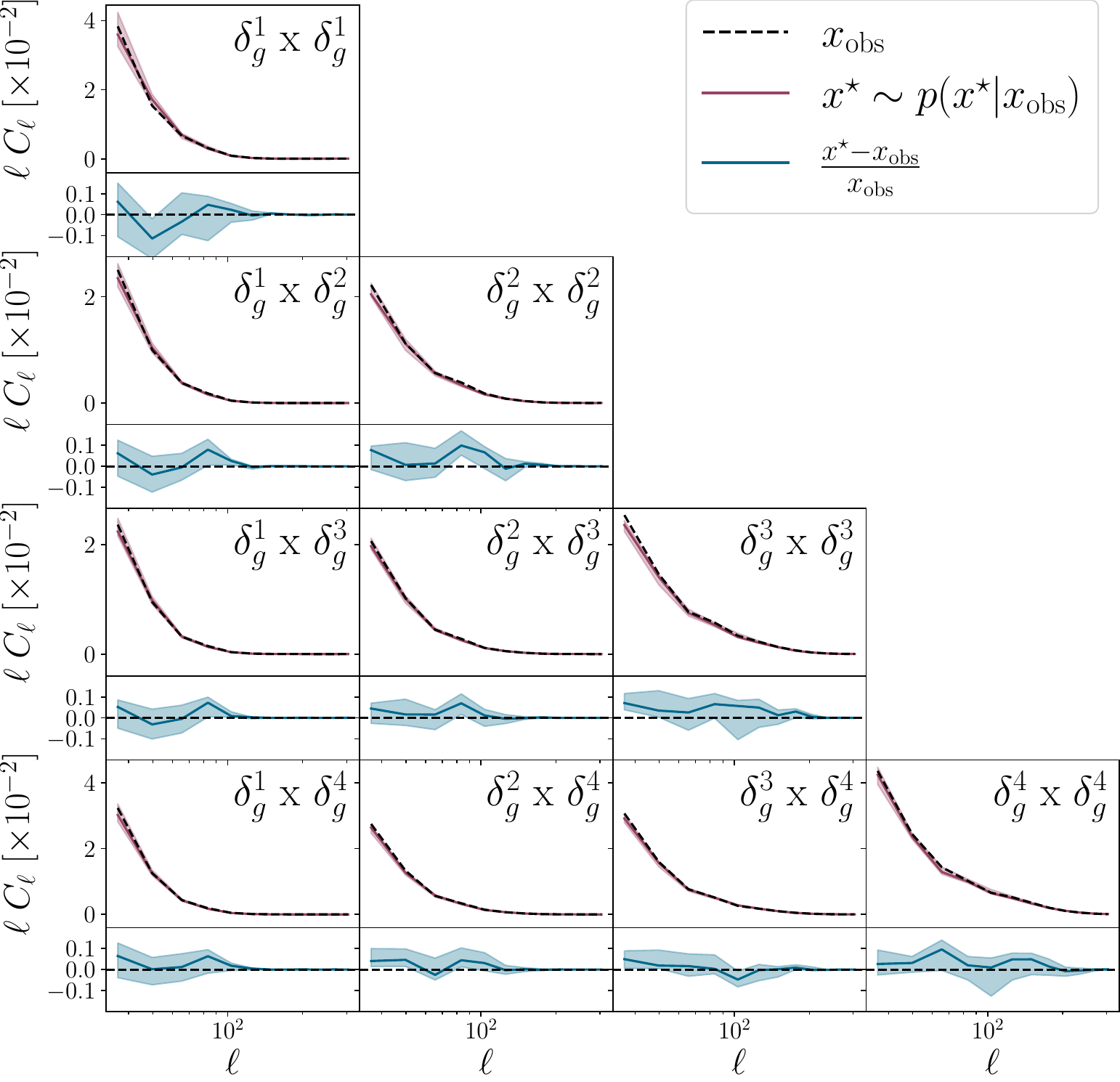}
    \caption{
        Comparison between the power spectrum of one \buzzard simulation (dashed line) to samples from the PPD (see \cref{eq:ppd}) derived from map-level galaxy clustering analysis.
        The solid lines show the mean of the posterior predictive samples, and the shaded regions indicate the 95\%  credible interval.
    }
    \label{fig:ppc}
\end{figure}
We compare the (tomographic cross-) power spectrum of one \buzzard mock with power spectra from the \cosmogrid forward model sampled from the PPD using a posterior $p(\theta | x_\mathrm{obs})$ derived from map-level galaxy clustering in \cref{fig:ppc}.
The predicted and observed power spectra exhibit good agreement with no systematic deviations exceeding the expected statistical fluctuations.
This confirms that our \cosmogrid-based forward model adequately captures the statistical properties of the \buzzard simulations at the relevant scales.
For brevity, we only show the PPD test for map-level galaxy clustering as it most directly tests the adequacy of our linear galaxy biasing assumption.
However, comparable agreement is observed for all other probe configurations.

\subsection{Posterior Coverage Tests}\label{sec:posterior_coverage}
In our SBI framework, we parametrize the intractable likelihood as a normalizing flow and sample from the resulting approximate posterior $p(\theta | s(x))$
using MCMC, where $s(x)$ is the learned compression of a map $x$ (or power spectrum baseline).
For simplicity, we omit $s$ in the following. 

We validate this posterior through empirical coverage tests that measure the calibration of the distribution using held-out simulations. 
Specifically, for a test observation $x_\mathrm{test} \sim p(x | \theta_\mathrm{test})$ drawn from a known point in parameter space $\theta_\mathrm{test}$, a well-calibrated posterior $p(\theta | x_\mathrm{test})$ should produce nominal $(1-\alpha) \in [0,1]$ credible regions containing $\theta_\mathrm{test}$ with frequency $(1-\alpha)$. 
Coverage tests therefore involve repeating the inference procedure across many test observations $x_\mathrm{test}$ to measure this frequency.
Although Bayesian credible regions are not theoretically required to satisfy frequentist coverage properties, demonstrating this correspondence empirically indicates that our inference procedure accurately quantifies parameter uncertainty.

We employ two distinct diagnostics to assess calibration:
The highest posterior density (HPD)~\cite{hermansTrustCrisisSimulationBased2022} test computes the posterior density at the true parameter value $\theta_\mathrm{test}$ and records the fraction $f$ of posterior samples $\tilde{\theta} \sim p(\theta | x_\mathrm{test})$ with lower density. 
The true parameter is counted as lying inside the nominal $(1-\alpha)$ HPD region whenever $f < 1 - \alpha$. 
Aggregating this indicator over many simulations $x_\mathrm{test}$ yields a calibration curve, where deviations below (above) the diagonal indicate over-confident (conservative) posteriors.

The tests of accuracy with random points (TARP)~\cite{lemosSamplingBasedAccuracyTesting2023} method samples random reference points $\theta_\mathrm{ref}$ in parameter space and computes the fraction of posterior samples falling within a ball centered at $\theta_\mathrm{ref}$ and extending to the true value $\theta_\mathrm{test}$, thereby defining a credible region around the reference point.
Repeating this procedure across a large number of reference points and observations $x_\mathrm{test}$ similarly yields a calibration curve that can be compared to the ideal diagonal.
We implement TARP using publicly available code~\footnote{\href{https://github.com/Ciela-Institute/tarp}{\faGithub\ tarp}}.
While TARP provides both necessary and sufficient conditions for well-calibrated posteriors, we include HPD as a complementary diagnostic since we have found it to be more sensitive in most cases.

\begin{figure}[!htb]
    \centering
    \includegraphics[width=0.9\linewidth]{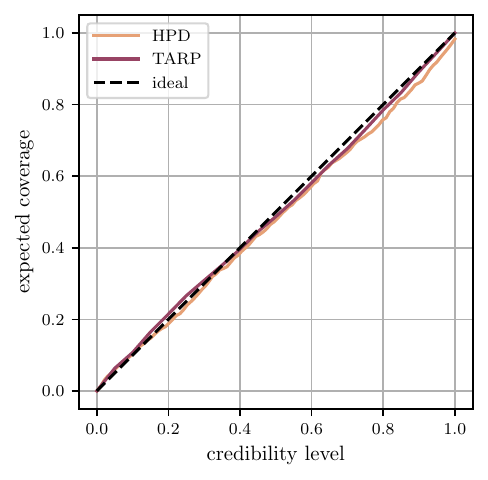}
    \caption{
        Empirical validation of posterior coverage through HPD~\cite{hermansTrustCrisisSimulationBased2022} and TARP~\cite{lemosSamplingBasedAccuracyTesting2023} diagnostics.
        Curves along the diagonal indicate well-calibrated posteriors where credible regions contain true parameters at their nominal frequencies.
        Results shown are for the ten-dimensional posterior from map-level combined probes; all other configurations considered in this work perform similarly.
    }
    \label{fig:nle_coverage}
\end{figure}
We perform both tests in the full 10-dimensional parameter space encompassing cosmological, intrinsic alignment and galaxy biasing parameters (see \cref{fig:pipeline}), using $1 \, 000$ synthetic observations $x_\mathrm{test}$ randomly drawn from held-out \cosmogrid mocks not seen during training of the compression network or density estimator.
The results, presented in \cref{fig:nle_coverage}, demonstrate accurate calibration.

\section{Results on Mock Observations}\label{sec:results}
Having passed extensive validation tests for fiducial scale cuts, we now forecast the constraining power of our posterior inference pipeline by conditioning on the mean compression vector as defined in \cref{eq:mean_mock} over the 15 \buzzard mocks.
Application to the measured DES Y3 \metacal and \maglim catalogs, including additional measurement-related systematics modeling and testing, is deferred to a forthcoming companion paper.

We compare performance along two primary axes: first, compression networks operating at the map level versus those using power spectra (our two-point statistic baseline) as input; second, individual probes of weak lensing and galaxy clustering versus their combination.
We include additional figures in \cref{sec:additional_figures}. 

\subsection{Map- vs. Two-Point Level}\label{sec:map_vs_two_point}
As explained in \cref{sec:neural_compression}, two-point statistics like the power spectrum only capture Gaussian information.
However, the forward-modeled weak lensing and galaxy clustering maps contain additional non-Gaussian information, which is only accessible to the map-level compression networks within our analysis pipeline.
Therefore, comparing posterior constraints from map-level and power spectrum analyses provides a lower bound on the non-Gaussian information content of the maps, as extracted by the map-level compression networks.

We quantify the constraining power in the two-dimensional $X-Y$ parameter plane using the figure of merit (FoM) of the marginal distribution~\cite{albrechtReportDarkEnergy2006}:
\begin{equation}
    \FoM_{X,Y} \coloneqq \left(\operatorname{det}\left(\mathrm{Cov}_{X,Y}\right)\right)^{-0.5}.
    \label{eq:fom}
\end{equation}
The results are presented in Table~\ref{tab:1d_results} and discussed below.
\begin{table*}[htb]
    \begin{ruledtabular}
    \caption{
        Marginalized standard deviations $\sigma(\cdot)$ and two–parameter figures of merit (FoM, see \cref{eq:fom}) for different analysis configurations.
        For map-level statistics, the fractional improvement relative to the power spectrum baseline is shown in parentheses, with positive percentages indicating tighter constraints.
    }        
    \renewcommand{\arraystretch}{1.25}
        \begin{tabular}{l l c c c c c c c}
            Probe & Statistic &
            \shortstack{$\sigma(\omatter)$\\$[\times100]$} &
            \shortstack{$\sigma(\Seight)$\\$[\times100]$} &
            \shortstack{$\sigma(\wzero)$\\$[\times100]$} &
            \shortstack{$\sigma(\AIA)$\\$[\times100]$} &
            \shortstack{$\sigma(\bgone)$\\$[\times100]$} &
            \shortstack{$\mathrm{FoM}_{\omatter,\Seight}$\\\phantom{$[]$}} &
            \shortstack{$\mathrm{FoM}_{\omatter,\wzero}$\\\phantom{$[]$}} \\ \hline
            % ---------- Lensing ----------
            Lensing & $C_\ell$ baseline & 5.1 & 2.9 & 21 & 20 & -- & 686 & 92 \\ &
            \shortstack{map-level\\\phantom{(+)}} &
            \shortstack{3.9\\\textcolor{gray}{\emph{(+32\,\%)}}} &
            \shortstack{2.2\\\textcolor{gray}{\emph{(+30\,\%)}}} &
            \shortstack{21\\\textcolor{gray}{\emph{(+0\,\%)}}} &
            \shortstack{21\\\textcolor{gray}{\emph{(–4\,\%)}}} &
            \shortstack{--\\\phantom{+}} &
            \shortstack{1340\\\textcolor{gray}{\emph{(+95\,\%)}}} &
            \shortstack{158\\\textcolor{gray}{\emph{(+71\,\%)}}} \\ \hline
            % ---------- Clustering ----------
            Clustering & $C_\ell$ baseline & 3.2 & 16 & 28 & -- & 39 & 198 & 121 \\ &
            \shortstack{map-level\\\phantom{(+)}} &
            \shortstack{2.3\\\textcolor{gray}{\emph{(+35\,\%)}}} &
            \shortstack{7.9\\\textcolor{gray}{\emph{(+105\,\%)}}} &
            \shortstack{24\\\textcolor{gray}{\emph{(+17\,\%)}}} &
            \shortstack{--\\\phantom{+}} &
            \shortstack{16\\\textcolor{gray}{\emph{(+147\,\%)}}} &
            \shortstack{572\\\textcolor{gray}{\emph{(+189\,\%)}}} &
            \shortstack{186\\\textcolor{gray}{\emph{(+54\,\%)}}} \\ \hline
            % ---------- Combined ----------
            Combined & $C_\ell$ baseline & 3.1 & 2.7 & 24 & 14 & 13 & 1224 & 142 \\ &
            \shortstack{map-level\\\phantom{(+)}} &
            \shortstack{2.0\\\textcolor{gray}{\emph{(+54\,\%)}}} &
            \shortstack{1.8\\\textcolor{gray}{\emph{(+52\,\%)}}} &
            \shortstack{16\\\textcolor{gray}{\emph{(+52\,\%)}}} &
            \shortstack{13\\\textcolor{gray}{\emph{(+4\,\%)}}} &
            \shortstack{6.9\\\textcolor{gray}{\emph{(+89\,\%)}}} &
            \shortstack{2969\\\textcolor{gray}{\emph{(+143\,\%)}}} &
            \shortstack{389\\\textcolor{gray}{\emph{(+175\,\%)}}} \\
        \end{tabular}
        \label{tab:1d_results}
    \end{ruledtabular}
\end{table*}

\begin{figure*}[!htb]
    \includegraphics[scale=0.8]{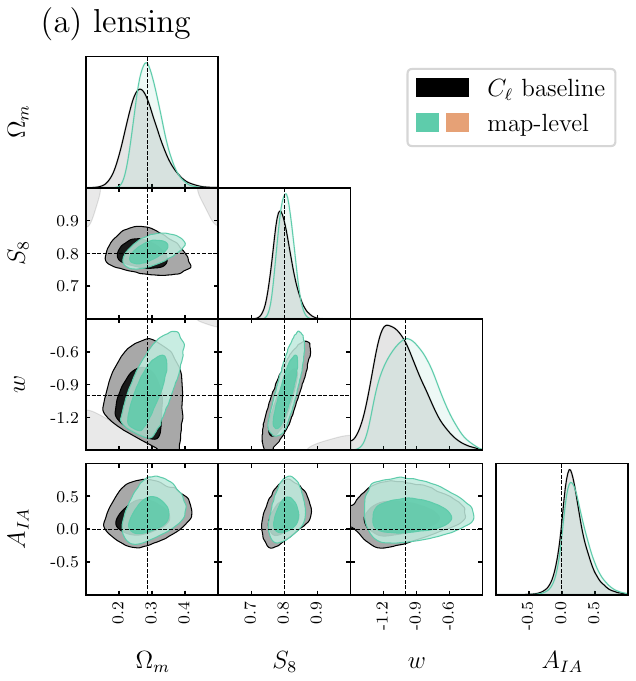}
    \hspace{10pt}
    \includegraphics[scale=0.8]{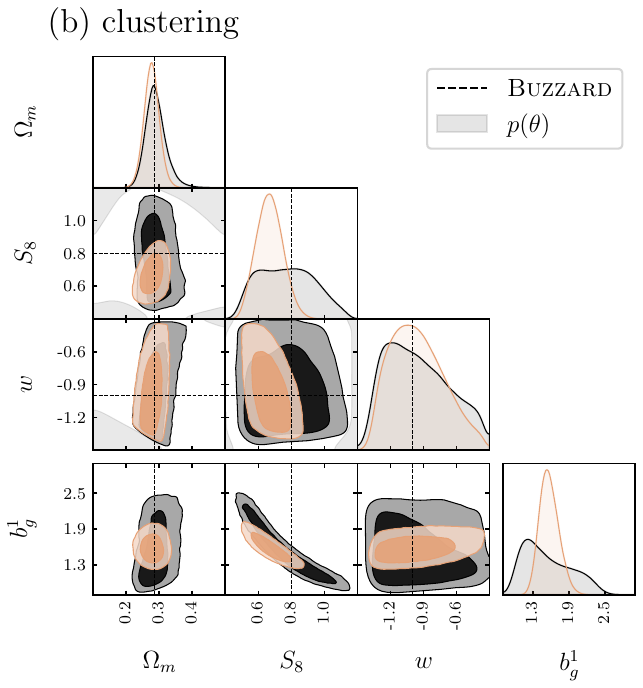}
    \caption{
        Comparison of posterior contours between map-level analysis and power spectrum baseline for fiducial scale cuts.
        The posterior is conditioned on the mean compression of the \buzzard ensemble.
        For weak lensing, we exclude the weakly constrained intrinsic alignment parameters \etaIA and \bta.
        For galaxy clustering, we show only \bgone as the remaining bias parameters $\bgtwo - \bgfour$ behave similarly.
        Complete parameter plots are provided in \cref{sec:additional_figures}.
    }
    \label{fig:maps_vs_cls_individual}
\end{figure*}

\begin{figure*}[!htb]
    \centering
    \includegraphics[width=0.8\linewidth]{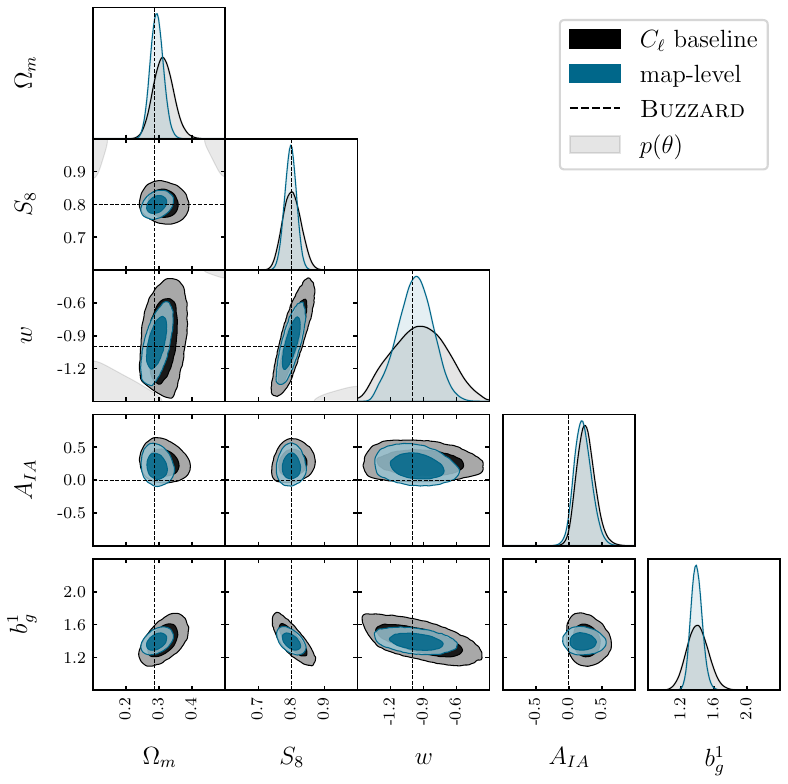}
    \caption{
        Like \cref{fig:maps_vs_cls_individual}, but for combined probes of weak lensing and galaxy clustering.
    }
    \label{fig:maps_vs_cls_combined}
\end{figure*}

\subsubsection{Weak Lensing}
For weak gravitational lensing as an individual probe, we compare two-dimensional posterior marginals in panel (a) of \cref{fig:maps_vs_cls_individual}.
We include the three constrained cosmological parameters $\{\omatter, \Seight, \wzero\}$, and the intrinsic alignment amplitude \AIA in the plot, while excluding the weakly constrained \etaIA and \bta parameters for clarity; the complete parameter space is shown in \cref{fig:maps_vs_cls_lensing_full}.
Map-level networks yield significantly improved constraints compared to the power spectrum baseline, with a 95\% increase (i.e.~1.95$\times$ improvement) in \FoMSeight.
This increase is consistent with previous simulation-based DES Y3 weak lensing analyses employing higher-order summary statistics~\cite{zurcherDarkEnergySurvey2022,darkenergysurveyDarkEnergySurvey2025,jeffreyDarkEnergySurvey2025,pratDarkEnergySurvey2025}.

In contrast, intrinsic alignment constraints show essentially no improvement. 
This result differs from \citepalias{kacprzakDeepLSSBreakingParameter2022}, which found drastic enhancement in intrinsic alignment constraints for deep learning-based summaries of weak lensing mass maps compared to conventional two-point statistics.
However, the forward models in \citepalias{kacprzakDeepLSSBreakingParameter2022} and this work differ substantially. 
We employ a three-parameter intrinsic alignment model compared to the vanilla NLA model assumed by \citepalias{kacprzakDeepLSSBreakingParameter2022}, constrain \wcdm instead of \lcdm cosmology, and adopt a more realistic Stage III survey-like setup including additional nuisances like redshift errors, multiplicative shear bias, and baryonification.
Furthermore, we apply more conservative scale cuts, which reduces the expected gain from higher-order relative to Gaussian statistics.

We demonstrate in \cref{sec:ia_projection} that the marginal posterior contours of \AIA are not centered on the true value of zero due to projection effects.
In \cref{sec:ia_cosmogrid}, we present posterior results on \cosmogrid internal mock observations with $\AIA \neq 0$ to show accurate parameter recovery in the presence of intrinsic alignments.

\subsubsection{Galaxy Clustering}
Analogously, for standalone galaxy clustering, we compare posterior results for power spectrum and map-based analyses in panel (b) of \cref{fig:maps_vs_cls_individual}. 
For conciseness, we include only a single representative bias parameter \bgone and refer to \cref{fig:maps_vs_cls_clustering_full} for the complete parameter space. 

The map-level approach partially breaks the degeneracy between linear bias parameters and clustering amplitude \sigeight (or \Seight), leading to substantially improved constraints on these parameters despite the conservative scale cuts employed.
The improved constraint on \Seight translates to a 189\% increase in the \FoM in the $\omatter - \Seight$ plane.
For one-dimensional marginal constraints, we find approximately two-fold reductions in uncertainty for both \Seight and the $\bgone - \bgfour$ parameters, with modest gains for the dark energy parameter \wzero.

To motivate the observed improvement in \Seight and galaxy bias constraints for the map-level analysis, we draw a qualitative analogy with spectroscopic galaxy surveys.
Effective field theory (EFT) analyses of the Baryon Oscillation Spectroscopic Survey (BOSS)~\cite{dawsonBARYONOSCILLATIONSPECTROSCOPIC2012} and Dark Energy Spectroscopic Instrument (DESI)~\cite{deyOverviewDESILegacy2019} have demonstrated that combining the bispectrum with the power spectrum significantly improves measurement precision for EFT galaxy bias parameters relative to power spectrum-only analyses, particularly for quadratic terms~\cite{philcoxBOSSDR12Fullshape2022,philcoxCosmologyRedshiftspaceGalaxy2022,chudaykinReanalyzingDESIDR12025}.

Considering the EFT linear galaxy bias $b_1$ (related to but distinct from our $b_g$ parameter), this improvement arises from different degeneracy directions between the power spectrum and bispectrum:
As derived in~\cite{philcoxCosmologyRedshiftspaceGalaxy2022}, the power spectrum scales quadratically with $b_1$, while the bispectrum exhibits cubic scaling.
These distinct scaling relations create different degeneracy directions relative to \sigeight, enabling improved parameter constraints when both statistics are combined.

Since the bispectrum contribution remains statistically significant at our fiducial smoothing scale of 32 Mpc/h for galaxy clustering~\cite[Fig.~1 in][]{philcoxCosmologyRedshiftspaceGalaxy2022}, and our map-level compression captures the beyond-Gaussian information content of the bispectrum in principle, we hypothesize that a similar degeneracy-breaking mechanism underlies the improvements observed in our analysis.
A quantitative assessment of this plausible analogy is beyond the scope of this work. 

\subsubsection{Combined Probes}
The main result of this work is the first forecast within DES using combined galaxy clustering and weak lensing maps.
For the combined probe analysis, we compare contours from two-point and map-level statistics in \cref{fig:maps_vs_cls_combined} using the same parameters as in \cref{fig:maps_vs_cls_individual}.
The complete parameter space is shown in \cref{fig:maps_vs_cls_combined_full}.

Map-based summaries yield significant improvement in constraints on cosmological parameters, reducing marginal standard deviations $\sigma(\cdot)$ by approximately $50\%$ for \omatter, \Seight, and \wzero. 
Interestingly, while map-level inference enhances constraints on linear bias parameters (with nearly twofold reduction in one-dimensional marginal uncertainties), the intrinsic alignment constraint remains comparable between two-point and map-level analyses, consistent with our weak lensing-only results.
For the overall \FoMSeight, we forecast a 143\% improvement due to the extracted non-Gaussian information. 

\begin{figure}[!htb]
    \centering
    \includegraphics[width=\linewidth]{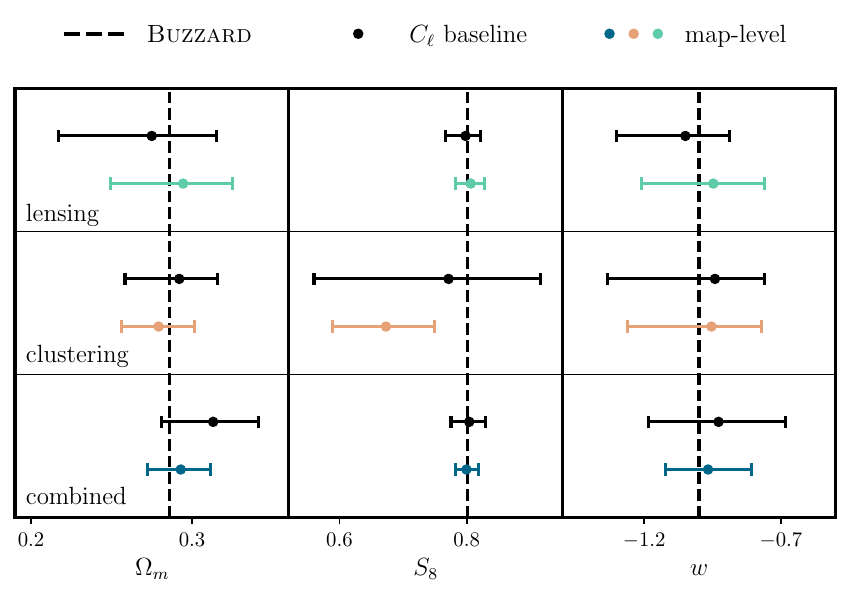}
    \caption{
        Comparison of one dimensional 68\% credible intervals for cosmological parameters across different analysis configurations. 
        Dots indicate posterior means.
    }
    \label{fig:1d_results}
\end{figure}

We summarize these findings as one-dimensional marginals in \cref{fig:1d_results}.

\subsection{Probe Comparison}\label{sec:probe_comparison}
\begin{figure*}[!htb]
    \centering
    \includegraphics[width=\linewidth]{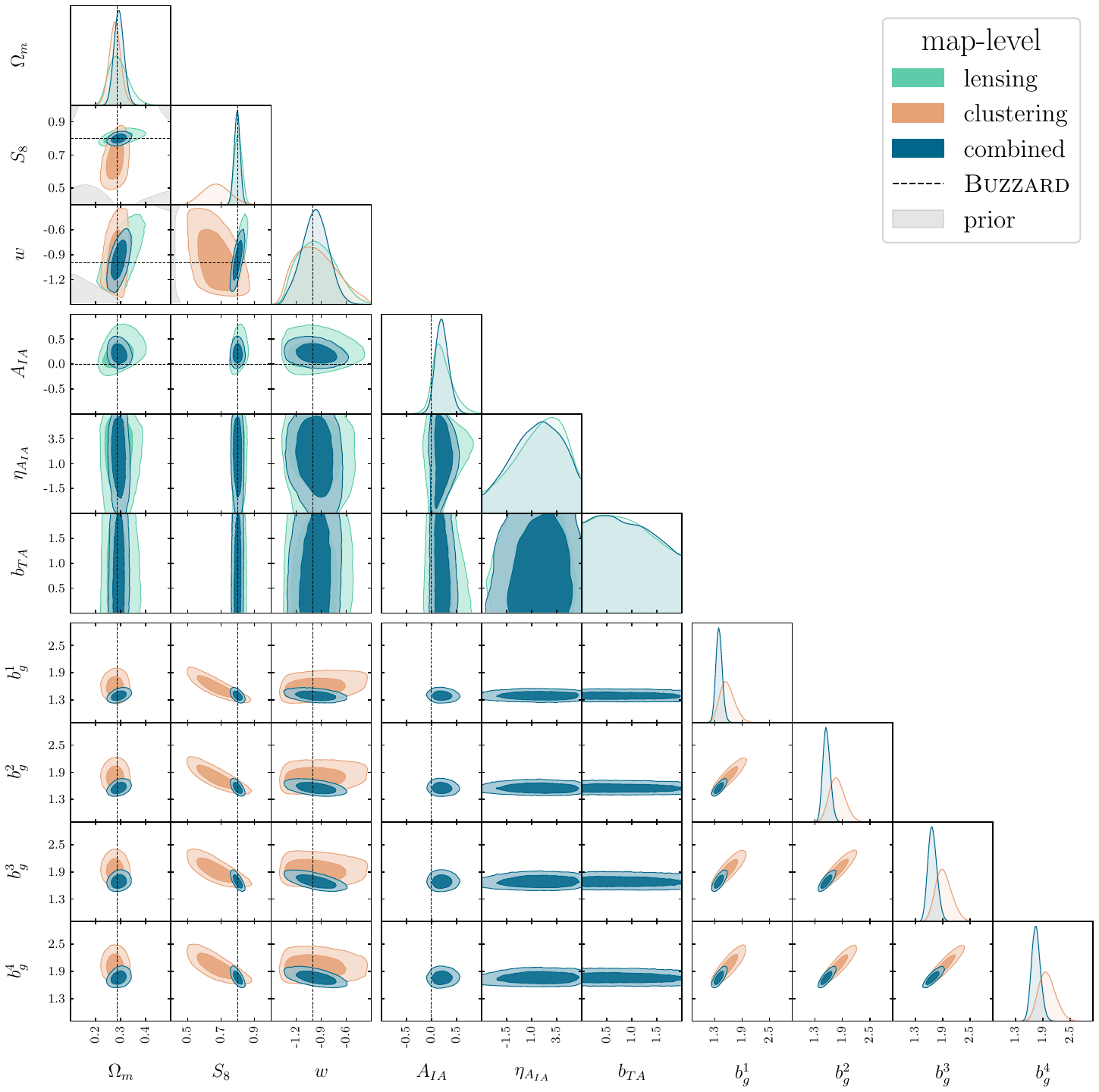}
    \caption{
        Comparison of posterior distributions resulting from individual or combined probes for map-level analysis using fiducial scale cuts.
        The plot shows the full parameter space including cosmological, intrinsic alignment and galaxy biasing parameters.
        The synthetic observation is the ensemble mean over the \buzzard mocks, with ground truth values indicated by dashed lines.
    }
    \label{fig:probe_comparison_maps}
\end{figure*}

In this section, we discuss our map-level results from the perspective of probe combination.
Each probe configuration employs an independent compression network trained on tomographic weak lensing maps $\kappa$, galaxy clustering maps \nglens, or their joint combination.
\Cref{fig:probe_comparison_maps} compares posterior constraints from individual and combined probes across the full parameter space, while \cref{fig:probe_comparison_cls} presents equivalent results for the two-point statistic baseline.

As expected, the higher sensitivity of weak lensing to the clustering parameter \Seight helps breaking the degeneracy between the clustering bias parameters $\bgone - \bgfour$ and \Seight, tightening constraints on these parameters by more than a factor of two. 
Relatedly, galaxy clustering provides additional information via larger sensitivity to \omatter, yielding approximately twofold reduction in the \FoMSeight compared to weak lensing alone.

We also find substantial improvements in the $\omatter - \wzero$ plane due to the different degeneracy directions of these parameters for weak lensing and galaxy clustering. 
This difference is significantly more pronounced for map-level inference than for the power spectrum baseline:
While \FoMwzero increases by approximately 17\% when combining galaxy clustering with weak lensing compared to galaxy clustering alone for power spectra, the corresponding gain for map-based analysis reaches 109\%.

\section{Conclusion}\label{sec:conclusion}
In this paper, we present the first pipeline for simulation-based cosmological inference using neural compressions of combined weak lensing and galaxy clustering maps for a realistic Stage-III survey setup.
We forecast its constraining power and provide detailed validation on simulations in preparation for analysis of the observed DES Y3 data in a forthcoming companion paper.

The main conclusions of this work can be summarized as follows:
\begin{itemize}
    \item We develop a scalable forward model based on the \cosmogrid simulation suite, generating one million combined weak lensing and galaxy clustering DES Y3 mock observations, requiring approximately 15 TB of storage.
    Of these maps, $800 \, 000$ train the neural compression networks and the remaining $200 \, 000$ enable neural density estimation of the unknown likelihood.
    This framework allows for robust SBI in a ten-dimensional parameter space.
    \item In addition to cosmological \wcdm parameters, the forward model varies baryonification, intrinsic alignment, and linear galaxy biasing parameters. Furthermore, we marginalize over photometric redshift uncertainties and multiplicative source galaxy shear biases.
    \item The graph convolutional neural networks we employ to compress the maps operate on the full survey footprint in spherical geometry, avoiding patchification and projection effects.
    We train them jointly on all constrained parameters using a theoretically motivated mutual information loss.
    \item As part of the validation effort, we construct 15 independent synthetic observations from \buzzard galaxy catalogs. 
    We use these potentially out-of-distribution mocks to determine appropriate scale cuts for both galaxy clustering and weak lensing, ensuring robust cosmology recovery and successful posterior predictive checks.
    \item We carry out additional systematics checks, including tests for alternative source clustering bias values and underlying $N$-body simulation settings.
    For all of these, we confirm that resulting posterior shifts in the $\omatter - \Seight$ plane remain below $0.3\sigma$.
    \item Our map-level compression networks are highly effective at extracting non-Gaussian information from the weak lensing and galaxy clustering fields for the validated scale cuts.    
    We report substantial improvements in cosmological parameter constraints compared to our baseline two-point statistics, with increases in the figure of merit of up to 189\% (i.e.~2.89$\times$) depending on the specific probe and parameters, corresponding to narrower posteriors.
    For galaxy clustering alone, the map-level analysis partially breaks the \sigeight--linear bias degeneracy, which is further reduced when combining with weak lensing as expected.
    We find no significant gain in constraints on intrinsic alignment parameters.
\end{itemize}

In this work, we did not systematically optimize scale cuts to identify the least conservative configuration satisfying our validation criteria. 
Instead, we validated a deliberately conservative choice. 
Consequently, our results represent a lower bound on the constraining power of the methodology, with scope for further improvement through scale optimization in future analyses.

The potential of the inference framework we introduce extends beyond the implemented forward model.
For example, while we include baryonic effects and marginalize over associated parameters, their impact is negligible at our current scale cuts.
Since the baryonification model remains valid at smaller scales~\cite{zhouMaplevelBaryonificationUnified2025}, higher resolution weak lensing analyses should be feasible with the existing simulations.
For galaxy clustering, our present modeling is primarily limited by the assumption of simple linear biasing for the matter-galaxy connection.
Although our neural compression extracts non-Gaussian information from galaxy clustering maps even even at conservative smoothing scales, more sophisticated clustering models could further enhance the results.
Possible improvements include incorporating higher-order bias terms or, given access to higher-fidelity halo catalogs, a SHAM-like galaxy-halo connection~\cite{bernerFastForwardModelling2024,fischbacherSHAMOTRapidSubhalo2025} in combination with a galaxy population model~\cite{fischbacherGalSBIPhenomenologicalGalaxy2025}.
However, the current \cosmogrid suite does not support these approaches, which will be addressed in future versions.

The Stage-IV photometric galaxy surveys LSST~\cite{collaborationLSSTScienceBook2009} and Euclid~\cite{laureijsEuclidDefinitionStudy2011} have recently entered operations and will provide data of substantially higher volume and quality than Stage-III surveys like DES. 
While significant challenges in modeling and simulations for Stage-IV data analysis remain, the forward model and inference pipeline we develop provide a solid foundation for SBI analyses of next-generation large-scale structure observations.

\section*{Contribution Statement}
AT: Led the analysis; implemented the forward modeling, neural compression, and simulation-based inference pipelines; and served as primary author of the manuscript.
JB: Generated the \buzzard mocks, co-led the validation, and advised the project.
TK: Co-conceptualized the project, co-led and provided the \cosmogrid simulation suite, performed the map projection and baryonification, and advised the project.
VA: Helped conceptualize the forward model, interpret results, and provided manuscript feedback.
JF: Co-led the \cosmogrid simulation suite and advised the pipeline development, contributing a precursor implementation.
AR: Co-conceptualized the project and advised on all aspects.

AF, DA, FC: DES internal reviewers.
MG, NJ: DES beyond-2pt analysis team members. 
% https://arxiv.org/abs/2105.13547
JD, RW, MB, ER, SP, NM, AA, JM: Production of \buzzard simulations.
JM, AA, AA, CS, SE, JD, JM, DG, GB, MT, SD, AC, NM, BY, MR: Production of the DES redshift distribution.
MG, ES, AA, MB, MT, AC, CD, NM, ANA, IH, DG, GB, MJ, LFS, AF, JM, RPR, RC, CC, SP, IT, JP, JEP, CS: Production of the DES shape catalog. 
MJ, GB, AA, CD, PFL, KB, IH, MG, AR: Production of the DES PSF. 
SE, BY, NK, EH, YZ: Production of the DES Balrog. 
% https://arxiv.org/abs/2011.03411
AP, MC, PF, JE: Production of the DES lens galaxy catalog.
% https://arxiv.org/abs/2105.13540
MR, NW, JE, MC: Production of lens galaxy systematics maps.

The remaining authors have made contributions to this paper that include, but are not limited to, the construction of DECam and other aspects of collecting the data; data processing and calibration; developing broadly used methods, codes, and simulations; running the pipelines and validation tests; and promoting the science analysis.

\section*{Acknowledgements}
AT thanks Silvan Fischbacher, Alexander Reeves, Veronika Oehl, Tilman Tröster, and Dominik Zürcher for helpful discussions on cosmology, and Marcello Negri and Aurelien Lucchi for valuable input on deep learning.
We also thank the reviewers for their insightful comments and helpful feedback. 

We acknowledge the use of the following software packages: \texttt{numpy}~\cite{harrisArrayProgrammingNumPy2020}, \texttt{scipy}~\cite{virtanenSciPy10Fundamental2020}, \texttt{TensorFlow}~\cite{tensorflow2015-whitepaper}, \texttt{horovod}~\cite{sergeevHorovodFastEasy2018}, \texttt{PyTorch}~\cite{paszkePyTorchImperativeStyle2019},  \texttt{FlowConductor}~\cite{negriConditionalMatrixFlows2023,torresLagrangianFlowNetworks2023}, \texttt{wandb}, \texttt{healpy}~\cite{zoncaHealpyEqualArea2019,gorskiHEALPixFrameworkHighResolution2005}, \texttt{emcee}~\cite{foreman-mackeyEmceeMCMCHammer2013}, and \texttt{h5py}~\cite{collette_python_hdf5_2014}. 
Jobarrays were submitted with \texttt{esub-epipe}~\cite{zurcherCosmologicalForecastNonGaussian2021,zurcherDarkEnergySurvey2022,zurcherFullWCDMMapbased2023}, and plots were created using \texttt{matplotlib}~\cite{hunterMatplotlib2DGraphics2007} and \texttt{trianglechain}~\cite{kacprzakDeepLSSBreakingParameter2022,fischbacherRedshiftRequirementsCosmic2023}.

This work was supported in part at ETH Zurich by the research grant 200021\_192243 from the Swiss National Science Foundation.

This work was supported by the Alexander von Humboldt Foundation (Tomasz Kacprzak, Humboldt Experienced Research Fellowship).

% see https://www.nersc.gov/users/become-a-nersc-user/acknowledgement
This research used resources of the National Energy Research Scientific Computing Center (NERSC), a Department of Energy User Facility using NERSC awards HEP-ERCAP 0031464 and DDR-ERCAP 0034644 (AI4Sci@NERSC).

% see https://cdcvs.fnal.gov/redmine/projects/pubboard/wiki/Ack
Funding for the DES Projects has been provided by the U.S. Department of Energy, the U.S. National Science Foundation, the Ministry of Science and Education of Spain, 
the Science and Technology Facilities Council of the United Kingdom, the Higher Education Funding Council for England, the National Center for Supercomputing 
Applications at the University of Illinois at Urbana-Champaign, the Kavli Institute of Cosmological Physics at the University of Chicago, 
the Center for Cosmology and Astro-Particle Physics at the Ohio State University,
the Mitchell Institute for Fundamental Physics and Astronomy at Texas A\&M University, Financiadora de Estudos e Projetos, 
Funda{\c c}{\~a}o Carlos Chagas Filho de Amparo {\`a} Pesquisa do Estado do Rio de Janeiro, Conselho Nacional de Desenvolvimento Cient{\'i}fico e Tecnol{\'o}gico and 
the Minist{\'e}rio da Ci{\^e}ncia, Tecnologia e Inova{\c c}{\~a}o, the Deutsche Forschungsgemeinschaft and the Collaborating Institutions in the Dark Energy Survey. 

The Collaborating Institutions are Argonne National Laboratory, the University of California at Santa Cruz, the University of Cambridge, Centro de Investigaciones Energ{\'e}ticas, 
Medioambientales y Tecnol{\'o}gicas-Madrid, the University of Chicago, University College London, the DES-Brazil Consortium, the University of Edinburgh, 
the Eidgen{\"o}ssische Technische Hochschule (ETH) Z{\"u}rich, 
Fermi National Accelerator Laboratory, the University of Illinois at Urbana-Champaign, the Institut de Ci{\`e}ncies de l'Espai (IEEC/CSIC), 
the Institut de F{\'i}sica d'Altes Energies, Lawrence Berkeley National Laboratory, the Ludwig-Maximilians Universit{\"a}t M{\"u}nchen and the associated Excellence Cluster Universe, 
the University of Michigan, NSF NOIRLab, the University of Nottingham, The Ohio State University, the University of Pennsylvania, the University of Portsmouth, 
SLAC National Accelerator Laboratory, Stanford University, the University of Sussex, Texas A\&M University, and the OzDES Membership Consortium.

Based in part on observations at NSF Cerro Tololo Inter-American Observatory at NSF NOIRLab (NOIRLab Prop. ID 2012B-0001; PI: J. Frieman), which is managed by the Association of Universities for Research in Astronomy (AURA) under a cooperative agreement with the National Science Foundation.

The DES data management system is supported by the National Science Foundation under Grant Numbers AST-1138766 and AST-1536171.
The DES participants from Spanish institutions are partially supported by MICINN under grants PID2021-123012, PID2021-128989 PID2022-141079, SEV-2016-0588, CEX2020-001058-M and CEX2020-001007-S, some of which include ERDF funds from the European Union. IFAE is partially funded by the CERCA program of the Generalitat de Catalunya.

We  acknowledge support from the Brazilian Instituto Nacional de Ci\^encia
e Tecnologia (INCT) do e-Universo (CNPq grant 465376/2014-2).

This document was prepared by the DES Collaboration using the resources of the Fermi National Accelerator Laboratory (Fermilab), a U.S. Department of Energy, Office of Science, Office of High Energy Physics HEP User Facility. Fermilab is managed by Fermi Forward Discovery Group, LLC, acting under Contract No. 89243024CSC000002.
\section*{Author Affiliations}
{
\scriptsize
\noindent
\textsuperscript{1} Department of Physics, ETH Zurich, Wolfgang-Pauli-Strasse 16, CH-8093 Zurich, Switzerland
\\
\textsuperscript{2} University Observatory Munich, Scheinerstraße 1, D-81679 Munich, Germany
\\
\textsuperscript{3} University of Applied Sciences Northwestern Switzerland FHNW, Bahnhofstrasse 6, 5210 Windisch
\\
\textsuperscript{4} Fondazione LINKS, Via Pier Carlo Boggio 61, 10138 Turin, Italy
\\
\textsuperscript{5} Department of Computer Science, ETH Zurich, 8092 Zurich, Switzerland
\\
\textsuperscript{6} Kavli Institute for Cosmological Physics, University of Chicago, Chicago, IL 60637, USA
\\
\textsuperscript{7} Institut d'Estudis Espacials de Catalunya (IEEC), 08034 Barcelona, Spain
\\
\textsuperscript{8} Institute of Space Sciences (ICE, CSIC),  Campus UAB, Carrer de Can Magrans, s/n,  08193 Barcelona, Spain
\\
\textsuperscript{9} SLAC National Accelerator Laboratory, Menlo Park, CA 94025, USA
\\
\textsuperscript{10} Department of Physics \& Astronomy, University College London, Gower Street, London, WC1E 6BT, UK
\\
\textsuperscript{11} Department of Astrophysical Sciences, Princeton University, Peyton Hall, Princeton, NJ 08544, USA
\\
\textsuperscript{12} Physics Department, 2320 Chamberlin Hall, University of Wisconsin-Madison, 1150 University Avenue Madison, WI  53706-1390
\\
\textsuperscript{13} Argonne National Laboratory, 9700 South Cass Avenue, Lemont, IL 60439, USA
\\
\textsuperscript{14} Department of Physics and Astronomy, University of Pennsylvania, Philadelphia, PA 19104, USA
\\
\textsuperscript{15} Department of Physics, Carnegie Mellon University, Pittsburgh, Pennsylvania 15312, USA
\\
\textsuperscript{16} NSF AI Planning Institute for Physics of the Future, Carnegie Mellon University, Pittsburgh, PA 15213, USA
\\
\textsuperscript{17} Instituto de Astrofisica de Canarias, E-38205 La Laguna, Tenerife, Spain
\\
\textsuperscript{18} Laborat\'orio Interinstitucional de e-Astronomia - LIneA, Av. Pastor Martin Luther King Jr, 126 Del Castilho, Nova Am\'erica Offices, Torre 3000/sala 817 CEP: 20765-000, Brazil
\\
\textsuperscript{19} Universidad de La Laguna, Dpto. Astrofísica, E-38206 La Laguna, Tenerife, Spain
\\
\textsuperscript{20} Department of Astronomy and Astrophysics, University of Chicago, Chicago, IL 60637, USA
\\
\textsuperscript{21} Department of Physics, Duke University Durham, NC 27708, USA
\\
\textsuperscript{22} NASA Goddard Space Flight Center, 8800 Greenbelt Rd, Greenbelt, MD 20771, USA
\\
\textsuperscript{23} Kavli Institute for Particle Astrophysics \& Cosmology, P. O. Box 2450, Stanford University, Stanford, CA 94305, USA
\\
\textsuperscript{24} Lawrence Berkeley National Laboratory, 1 Cyclotron Road, Berkeley, CA 94720, USA
\\
\textsuperscript{25} Fermi National Accelerator Laboratory, P. O. Box 500, Batavia, IL 60510, USA
\\
\textsuperscript{26} Universit\'e Grenoble Alpes, CNRS, LPSC-IN2P3, 38000 Grenoble, France
\\
\textsuperscript{27} Department of Physics and Astronomy, University of Waterloo, 200 University Ave W, Waterloo, ON N2L 3G1, Canada
\\
\textsuperscript{28} California Institute of Technology, 1200 East California Blvd, MC 249-17, Pasadena, CA 91125, USA
\\
\textsuperscript{29} University Observatory, LMU Faculty of Physics, Scheinerstr. 1, 81679 Munich, Germany
\\
\textsuperscript{30} School of Physics and Astronomy, Cardiff University, CF24 3AA, UK
\\
\textsuperscript{31} Jet Propulsion Laboratory, California Institute of Technology, 4800 Oak Grove Dr., Pasadena, CA 91109, USA
\\
\textsuperscript{32} Department of Applied Mathematics and Theoretical Physics, University of Cambridge, Cambridge CB3 0WA, UK
\\
\textsuperscript{33} Instituto de F\'isica Gleb Wataghin, Universidade Estadual de Campinas, 13083-859, Campinas, SP, Brazil
\\
\textsuperscript{34} Centro de Investigaciones Energ\'eticas, Medioambientales y Tecnol\'ogicas (CIEMAT), Madrid, Spain
\\
\textsuperscript{35} Ruhr University Bochum, Faculty of Physics and Astronomy, Astronomical Institute, German Centre for Cosmological Lensing, 44780 Bochum, Germany
\\
\textsuperscript{36} Nordita, KTH Royal Institute of Technology and Stockholm University, Hannes Alfv\'ens v\"ag 12, SE-10691 Stockholm, Sweden
\\
\textsuperscript{37} Department of Physics, University of Genova and INFN, Via Dodecaneso 33, 16146, Genova, Italy
\\
\textsuperscript{38} Centro de Investigaciones Energéticas, Medioambientales y Tecnológicas (CIEMAT), Madrid, Spain
\\
\textsuperscript{39} Jodrell Bank Center for Astrophysics, School of Physics and Astronomy, University of Manchester, Oxford Road, Manchester, M13 9PL, UK
\\
\textsuperscript{40} Brookhaven National Laboratory, Bldg 510, Upton, NY 11973, USA
\\
\textsuperscript{41} Department of Physics and Astronomy, Stony Brook University, Stony Brook, NY 11794, USA
\\
\textsuperscript{42} Institut de Recherche en Astrophysique et Plan\'etologie (IRAP), Universit\'e de Toulouse, CNRS, UPS, CNES, 14 Av. Edouard Belin, 31400 Toulouse, France
\\
\textsuperscript{43} Excellence Cluster Origins, Boltzmannstr.\ 2, 85748 Garching, Germany
\\
\textsuperscript{44} Max Planck Institute for Extraterrestrial Physics, Giessenbachstrasse, 85748 Garching, Germany
\\
\textsuperscript{45} Universit\"ats-Sternwarte, Fakult\"at f\"ur Physik, Ludwig-Maximilians Universit\"at M\"unchen, Scheinerstr. 1, 81679 M\"unchen, Germany
\\
\textsuperscript{46} Berkeley Center for Cosmological Physics, Department of Physics, University of California, Berkeley, CA 94720, US
\\
\textsuperscript{47} Department of Physics, Stanford University, 382 Via Pueblo Mall, Stanford, CA 94305, USA
\\
\textsuperscript{48} Cerro Tololo Inter-American Observatory, NSF's National Optical-Infrared Astronomy Research Laboratory, Casilla 603, La Serena, Chile
\\
\textsuperscript{49} Institute for Astronomy, University of Edinburgh, Edinburgh EH9 3HJ, UK
\\
\textsuperscript{50} Laboratório Interinstitucional de e-Astronomia - LIneA, Av. Pastor Martin Luther King Jr, 126 Del Castilho, Nova América Offices, Torre 3000/sala 817 CEP: 20765-000, Brazil
\\
\textsuperscript{51} INAF-Osservatorio Astronomico di Trieste, via G. B. Tiepolo 11, I-34143 Trieste, Italy
\\
\textsuperscript{52} Physik-Institut, University of Zürich, Winterthurerstrasse 190, CH-8057 Zürich, Switzerland
\\
\textsuperscript{53} Institute of Cosmology and Gravitation, University of Portsmouth, Portsmouth, PO1 3FX, UK
\\
\textsuperscript{54} Department of Physics, Northeastern University, Boston, MA 02115, USA
\\
\textsuperscript{55} School of Mathematics and Physics, University of Queensland,  Brisbane, QLD 4072, Australia
\\
\textsuperscript{56} Institut de F\'{\i}sica d'Altes Energies (IFAE), The Barcelona Institute of Science and Technology, Campus UAB, 08193 Bellaterra (Barcelona) Spain
\\
\textsuperscript{57} Oxford College of Emory University, Oxford, GA 30054, USA
\\
\textsuperscript{58} Hamburger Sternwarte, Universit\"{a}t Hamburg, Gojenbergsweg 112, 21029 Hamburg, Germany
\\
\textsuperscript{59} Department of Physics, IIT Hyderabad, Kandi, Telangana 502285, India
\\
\textsuperscript{60} Instituto de Fisica Teorica UAM/CSIC, Universidad Autonoma de Madrid, 28049 Madrid, Spain
\\
\textsuperscript{61} Santa Cruz Institute for Particle Physics, Santa Cruz, CA 95064, USA
\\
\textsuperscript{62} Center for Cosmology and Astro-Particle Physics, The Ohio State University, Columbus, OH 43210, USA
\\
\textsuperscript{63} Department of Physics, The Ohio State University, Columbus, OH 43210, USA
\\
\textsuperscript{64} Center for Astrophysics $\vert$ Harvard \& Smithsonian, 60 Garden Street, Cambridge, MA 02138, USA
\\
\textsuperscript{65} Australian Astronomical Optics, Macquarie University, North Ryde, NSW 2113, Australia
\\
\textsuperscript{66} Lowell Observatory, 1400 Mars Hill Rd, Flagstaff, AZ 86001, USA
\\
\textsuperscript{67} George P. and Cynthia Woods Mitchell Institute for Fundamental Physics and Astronomy, and Department of Physics and Astronomy, Texas A\&M University, College Station, TX 77843,  USA
\\
\textsuperscript{68} Center for Astrophysical Surveys, National Center for Supercomputing Applications, 1205 West Clark St., Urbana, IL 61801, USA
\\
\textsuperscript{69} Department of Astronomy, University of Illinois at Urbana-Champaign, 1002 W. Green Street, Urbana, IL 61801, USA
\\
\textsuperscript{70} Instituci\'o Catalana de Recerca i Estudis Avan\c{c}ats, E-08010 Barcelona, Spain
\\
\textsuperscript{71} Department of Physics, University of Cincinnati, Cincinnati, Ohio 45221, USA
\\
\textsuperscript{72} Perimeter Institute for Theoretical Physics, 31 Caroline St. North, Waterloo, ON N2L 2Y5, Canada
\\
\textsuperscript{73} Centro de Tecnologia da Informa\c{c}\~ao Renato Archer, Campinas, SP, Brazil - 13069-901, Observat\'orio Nacional, Rio de Janeiro, RJ, Brazil - 20921-400
\\
\textsuperscript{74} Physics Department, Lancaster University, Lancaster, LA1 4YB, UK
\\
\textsuperscript{75} Computer Science and Mathematics Division, Oak Ridge National Laboratory, Oak Ridge, TN 37831
}

\bibliography{references_short}

\clearpage
\appendix
\crefalias{section}{appendix}
\counterwithin{figure}{section}

\section{Map-Level Smoothing and Scale Cuts}\label{sec:smoothing_appendix}
This appendix describes the implementation of the small-scale cuts discussed in \cref{sec:small_scales} and presents validation of the method.

\subsection{Implementation Details}\label{sec:smoothing_implementation}
\begin{figure}[!htb]
    \centering
    \includegraphics[width=\linewidth]{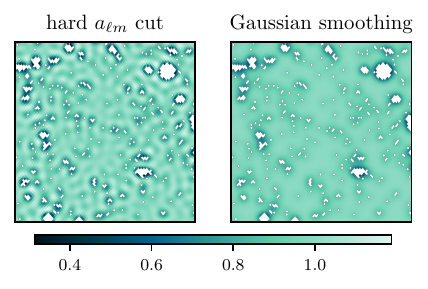}
    \caption{
        Comparison between the artifacts introduced by a hard cut in harmonic space and Gaussian smoothing in real space.
    }
    \label{fig:artifacts}
\end{figure}

We implement the removal of small-scale information in two subsequent steps.
\paragraph{Gaussian Smoothing:}
Because the survey footprint depicted in \cref{fig:footprint} covers only part of the sky and has a complicated shape, hard cuts in frequency (or spherical harmonics) space introduce ringing artifacts in real space (see \cref{fig:artifacts}), which can hinder the learning of convolutional filters. 

Therefore, to remove small scales, we smooth the maps by convolving them with a Gaussian kernel, which mitigates this issue.
The smoothing scale is defined by the kernel's standard deviation $\sigma$ or full width at half maximum ($\mathrm{FWHM} = 2 \sqrt{2 \ln 2} \, \sigma$).

\begin{figure}[!htb]
    \centering
    \includegraphics[width=\linewidth]{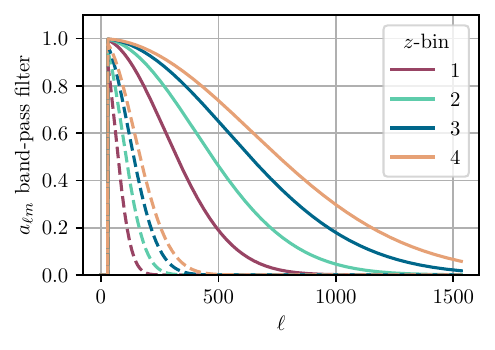}
    \caption{
        Band-pass filters in \alm-space defined by the factor in \cref{eq:low_pass} and a hard $\lmin = 30$ cut for the fiducial smoothing scales used in this work.
        Solid and dashed lines correspond to the redshift bins of the source and lens galaxy sample, respectively.
        }
    \label{fig:alm_filter}
\end{figure}

A useful property of Gaussian smoothing is its closed analytical form in the frequency domain, where smoothing is accomplished by multiplication of the \alm-coefficients from the harmonic decomposition in \cref{eq:map_spherical_harmonics} by the factor
\begin{equation}
    c_\textrm{low-pass}(\ell; \sigma) = \exp \left(-\frac{1}{2} \ell (\ell + 1) \, \sigma ^2 \right),
    \label{eq:low_pass}
\end{equation}
which depends on the multipole order $\ell \in \Z_{>0}$.
As $\ell$ increases, the factor becomes smaller, leading to greater suppression of higher-order modes.
The $c_\textrm{low-pass}$ filters resulting from our fiducial scale cuts in \cref{eq:smoothing_scales} are plotted in \cref{fig:alm_filter}.

Beyond reducing artifacts, another advantage of our implementation of Gaussian smoothing over a hard cut in \alm-space is that we can perform the smoothing operation on the fly, requiring storage of only a single full-resolution version of the dataset.

\paragraph{White Noise:}
Because the Gaussian smoothing kernel (in principle) has infinite support in real and spherical harmonics space, the smoothing operation is theoretically invertible in frequency space by simply dividing the \alm-coefficients through the factor in \cref{eq:low_pass}.
Thus, mathematically there is no loss of information; the small-scale information is merely suppressed, not removed.
While this is not the case in practice due to the finite precision of floating point numbers, we choose not to rely on such numerical noise.
Instead, we address the issue by irreversibly adding white Gaussian noise of a certain scale to the maps after smoothing.

\begin{figure}[!htb]
    \centering
    \includegraphics[width=\linewidth]{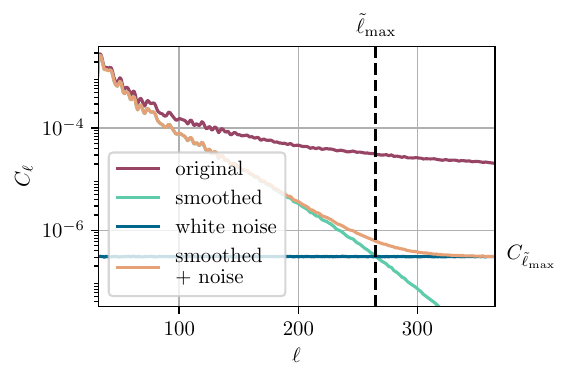}
    \caption{
    Illustration of the \cl-level definition for the amount of white noise added after smoothing. 
    Per construction, the \emph{smoothed} \cl (obtained from convolution with a Gaussian kernel) are suppressed to $1\%$ of the \emph{original} values at \lmaxt.
    The \emph{white noise} (generated from independent Gaussian samples) is calibrated to intersect the smoothed \cl at that value denoted $C_{\tilde{\ell}_\mathrm{max}}$.
    This example depicts the first galaxy clustering redshift bin, which undergoes the most smoothing.
    }
    \label{fig:smoothing_cls}
\end{figure}

We determine this noise level separately for each tomographic bin $i$ with smoothing scale $\sigma_\mathrm{min}^i$ by first finding the smallest spherical harmonic order $\tilde{\ell}_\mathrm{max}^i$ for which \cref{eq:low_pass} is below a threshold of $1 \%$. 
In other words (and omitting the bin-index for improved readability), $\tilde{\ell}_\mathrm{max}$ is the scale at which the smoothing reduces the \alm-coefficients to $1\%$ of their original value.
Then, we take the mean of the angular power spectrum $C_\ell$ (as defined in \cref{sec:two_point_level}) over all realizations at the fiducial cosmology and evaluate the resulting curve at the given $\tilde{\ell}_\mathrm{max}$.
The obtained value $C_{\tilde{\ell}_\mathrm{max}}$ defines the level of white noise we apply.
The different power spectra illustrating the steps are plotted in \cref{fig:smoothing_cls}.

We construct white-noise \healpix maps by drawing pixel-wise samples from independent scalar Gaussians of mean zero and shared standard deviation $\sigma_\textrm{noise}$ (not to be confused with the smoothing scale $\sigma_\mathrm{min}$).
The standard deviation $\sigma_\textrm{noise}$ is chosen such that the approximately constant power spectrum of the white-noise map matches the noise level $C_{\tilde{\ell}_\mathrm{max}}$.
The connection between the white noise power spectrum and the Gaussian distributions is given by
\begin{equation*}
    C_{\tilde{\ell}_\mathrm{max}} = \frac{4 \pi f_\mathrm{sky}}{\npix} \, \sigma_{\mathrm{noise}}^2,
\end{equation*}
where $f_\mathrm{sky}$ is the fraction of the sky occupied by the survey footprint and \npix is the number of pixels contained in the full sky at \nside~\cite{fluriCosmologicalParameterEstimation2021}.

\subsection{Validation}\label{sec:scale_cut_tests}
The real-space smoothing procedure detailed in \cref{sec:smoothing_implementation} allows different trade-offs between the Gaussian smoothing kernel size $\sigma_\mathrm{min}$ and pixel-wise white noise amplitude $\sigma_\mathrm{noise}$.
The same maximal multipole $\tilde{\ell}_\mathrm{max}$ (see \cref{fig:smoothing_cls}) in harmonic space can be achieved by either increasing the smoothing scale while reducing the subsequently added white noise, or vice versa.

\begin{figure}[!htb]
    \centering
    \includegraphics[width=\linewidth]{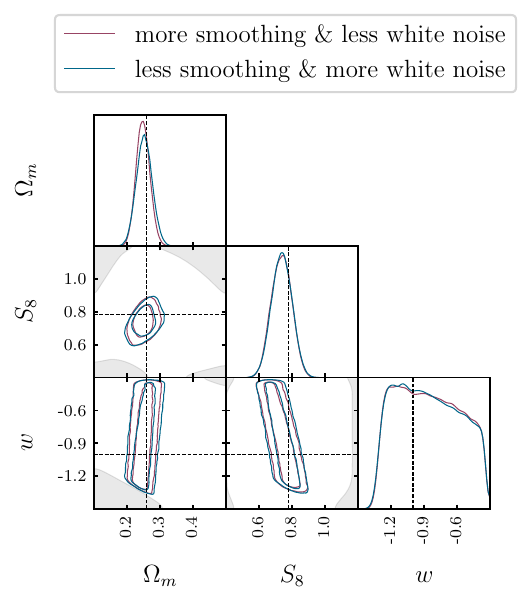}
    \caption{
        Comparison of different trade-offs between Gaussian smoothing scale $\sigma_\mathrm{min}$ and white-noise level $\sigma_\mathrm{noise}$ for fixed multipole $\tilde{\ell}_\mathrm{max}$.
        The marginal contours show no significant differences.
    }
    \label{fig:scale_cut_test}
\end{figure}
We test whether different choices yielding approximately equal multipoles $\tilde{\ell}_\mathrm{max}^i \approx [267, 390, 509, 610]$ significantly impact posterior contours.
\Cref{fig:scale_cut_test} shows results for an example clustering-only analysis evaluated on a fiducial \cosmogrid mock.
The \emph{more smoothing \& less white noise} configuration is derived from a scale of $R = 16 \, \mathrm{Mpc/h}$ with a 1\% threshold in \cref{eq:low_pass}, yielding
\begin{equation*}
    \begin{aligned}
         \sigma_\mathrm{min}^i &= [27.6, \, 18.9, \, 14.5, \, 12.1] \,\mathrm{arcmin} \\
        \sigma_\textrm{noise}^i &= [0.52, \, 0.21, \, 0.15, \, 0.14],
    \end{aligned}
\end{equation*}
while the \emph{less smoothing \& more white noise} setup uses a 10\% threshold, resulting in
\begin{equation*}
    \begin{aligned}
        \sigma_\mathrm{min}^i &= [19.5, \, 13.4, \, 10.2, \,  8.5] \,\mathrm{arcmin} \\
        \sigma_\textrm{noise}^i &= [1.61, \, 0.67, \, 0.47, \, 0.43].
    \end{aligned}
\end{equation*}
The posterior contours show no significant differences.

\section{Additional Figures}\label{sec:additional_figures}

\paragraph{Map- vs. Two-Point Level:}
\begin{figure}
    \centering
    \includegraphics[width=\linewidth]{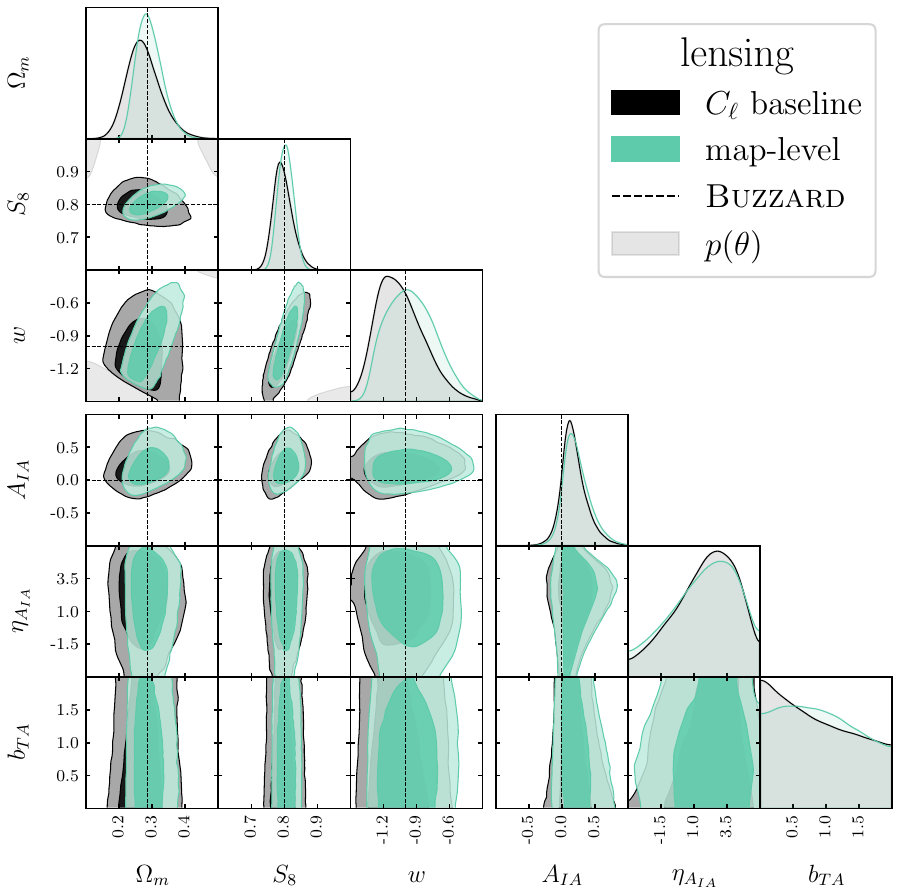}
    \caption{
        Like \cref{fig:maps_vs_cls_individual} lensing, but for all parameters.
    }
    \label{fig:maps_vs_cls_lensing_full}
\end{figure}

\begin{figure}
    \centering
    \includegraphics[width=\linewidth]{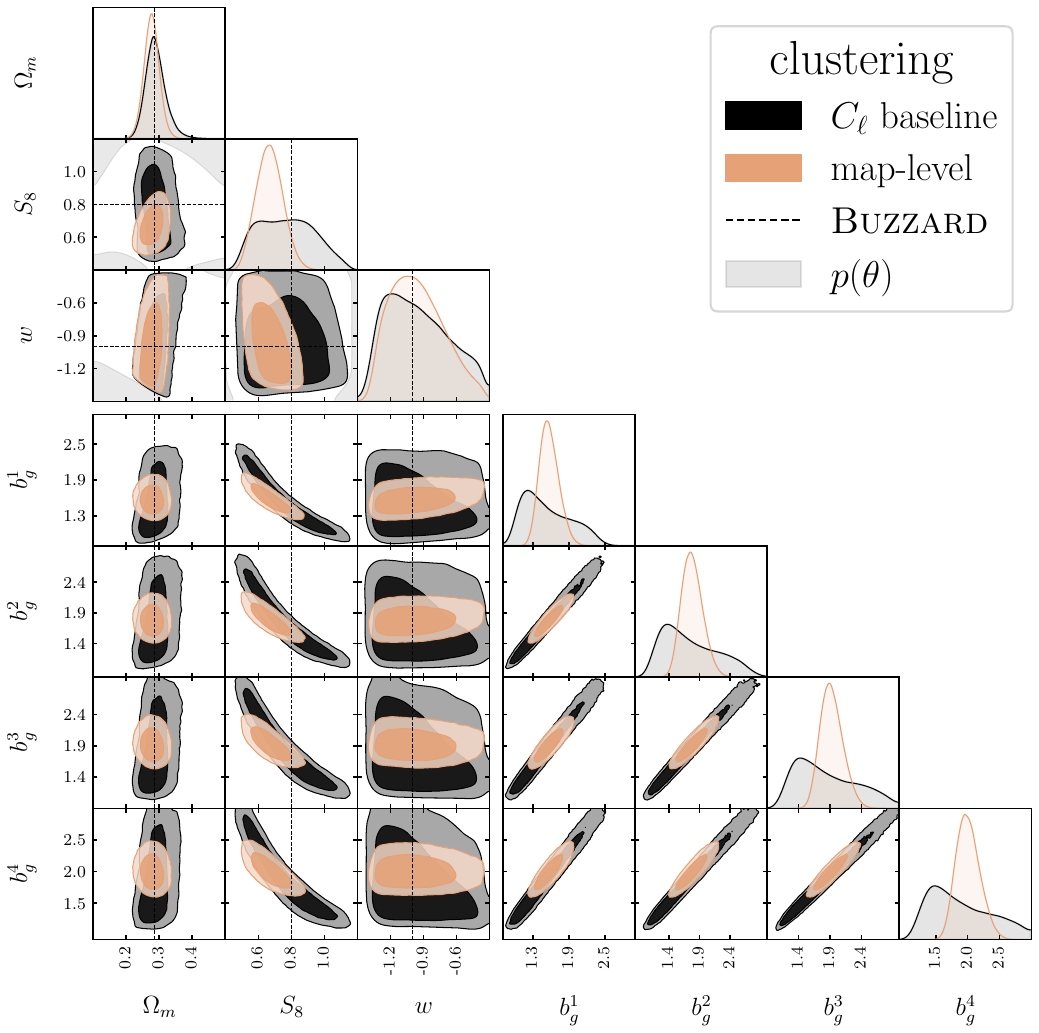}
    \caption{
        Like \cref{fig:maps_vs_cls_individual} clustering, but for all parameters.
    }
    \label{fig:maps_vs_cls_clustering_full}
\end{figure}

\begin{figure}
    \centering
    \includegraphics[width=\linewidth]{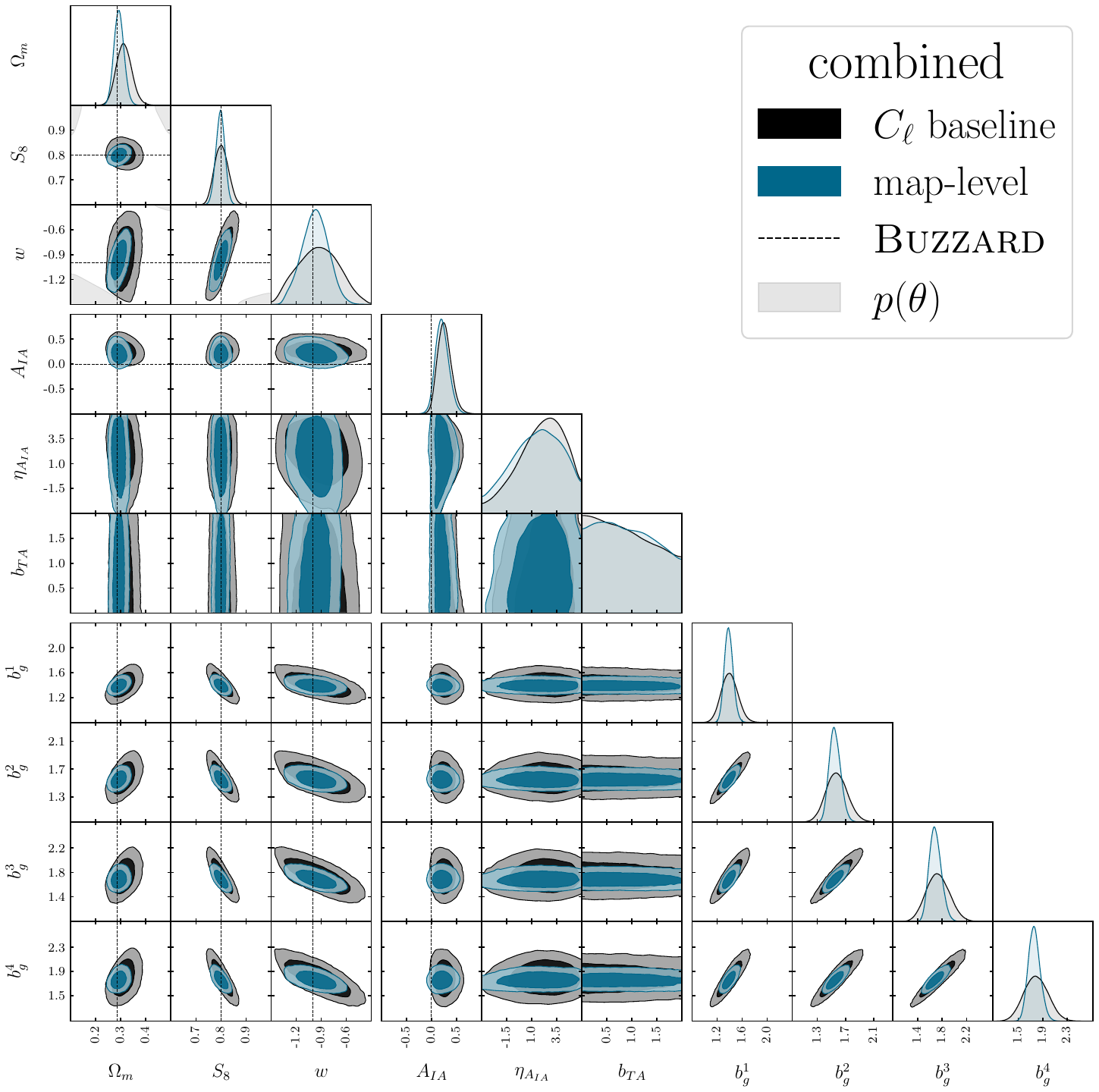}
    \caption{
        Like \cref{fig:maps_vs_cls_combined} combined, but for all parameters.
    }
    \label{fig:maps_vs_cls_combined_full}
\end{figure}

\begin{figure}
    \centering
    \includegraphics[width=\linewidth]{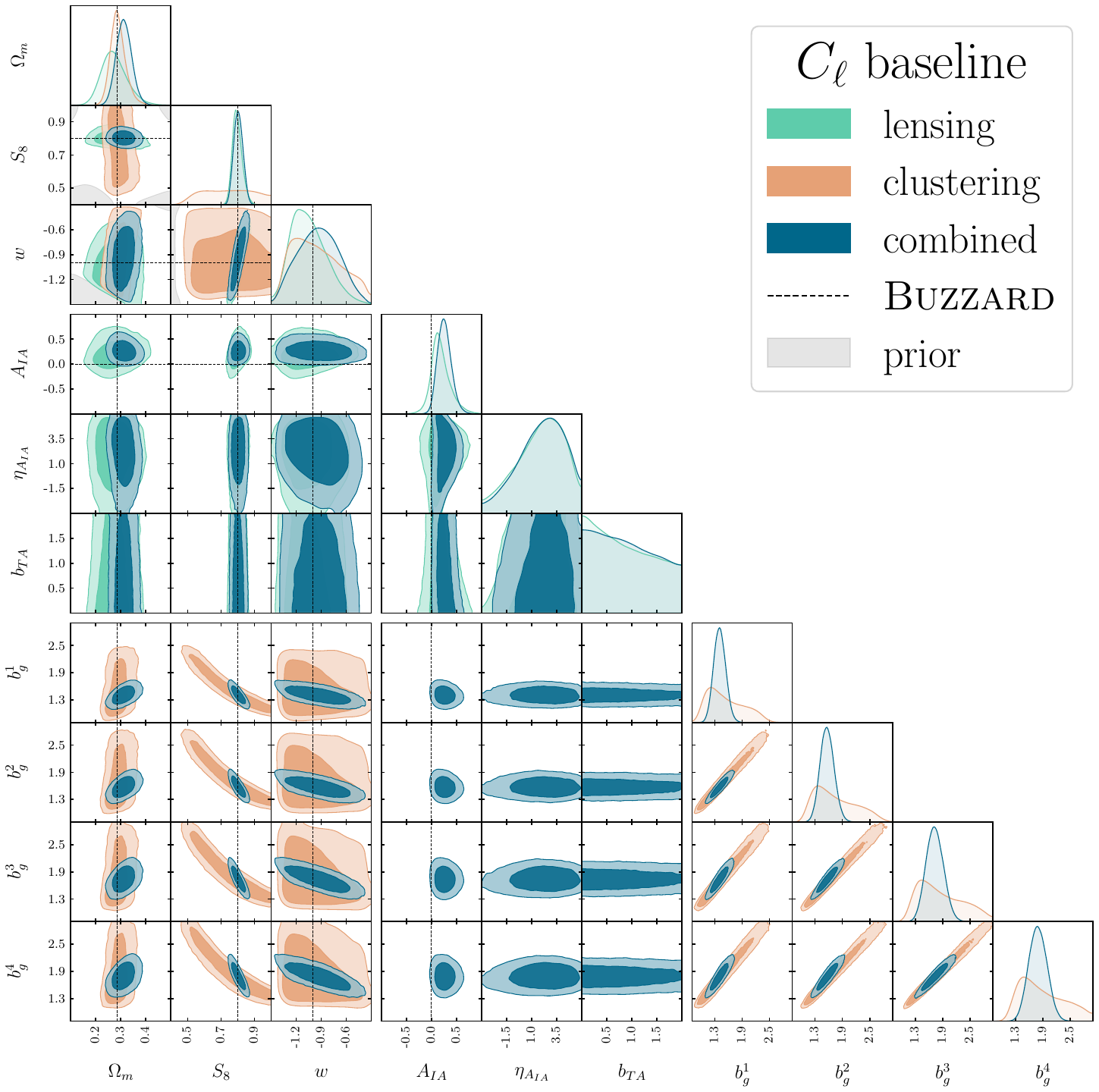}
    \caption{
        Like \cref{fig:probe_comparison_maps}, but for power-spectrum instead of map-level analysis.
    }
    \label{fig:probe_comparison_cls}
\end{figure}

We present the full parameter space versions of the posterior results discussed in \cref{sec:map_vs_two_point} in \cref{fig:maps_vs_cls_lensing_full,fig:maps_vs_cls_clustering_full,fig:maps_vs_cls_combined_full}.
The intrinsic alignment parameters \etaIA, controlling the redshift evolution, and \bta, coupling intrinsic alignment to the local density field, are weakly constrained with no significant difference between the map-level and two-point analyses.
This is expected since the true $\AIA = 0$ for \buzzard, eliminating any dependence on \etaIA and \bta according to \cref{eq:kappa_sum}.
The galaxy clustering biases $\bgone - \bgfour$ show mild redshift-bin dependence.
In all cases, their degeneracy with the \Seight parameter is significantly reduced in the map-level analysis. 
For the combined probes, adding weak lensing to galaxy clustering further reduces this degeneracy.

\paragraph{Probe Comparison:}
\Cref{fig:probe_comparison_cls} compares the different probe configurations for the power spectrum instead of the map-level analysis in \cref{fig:probe_comparison_maps}.

\begin{figure*}[!htb]
    \centering
    \includegraphics[width=\linewidth]{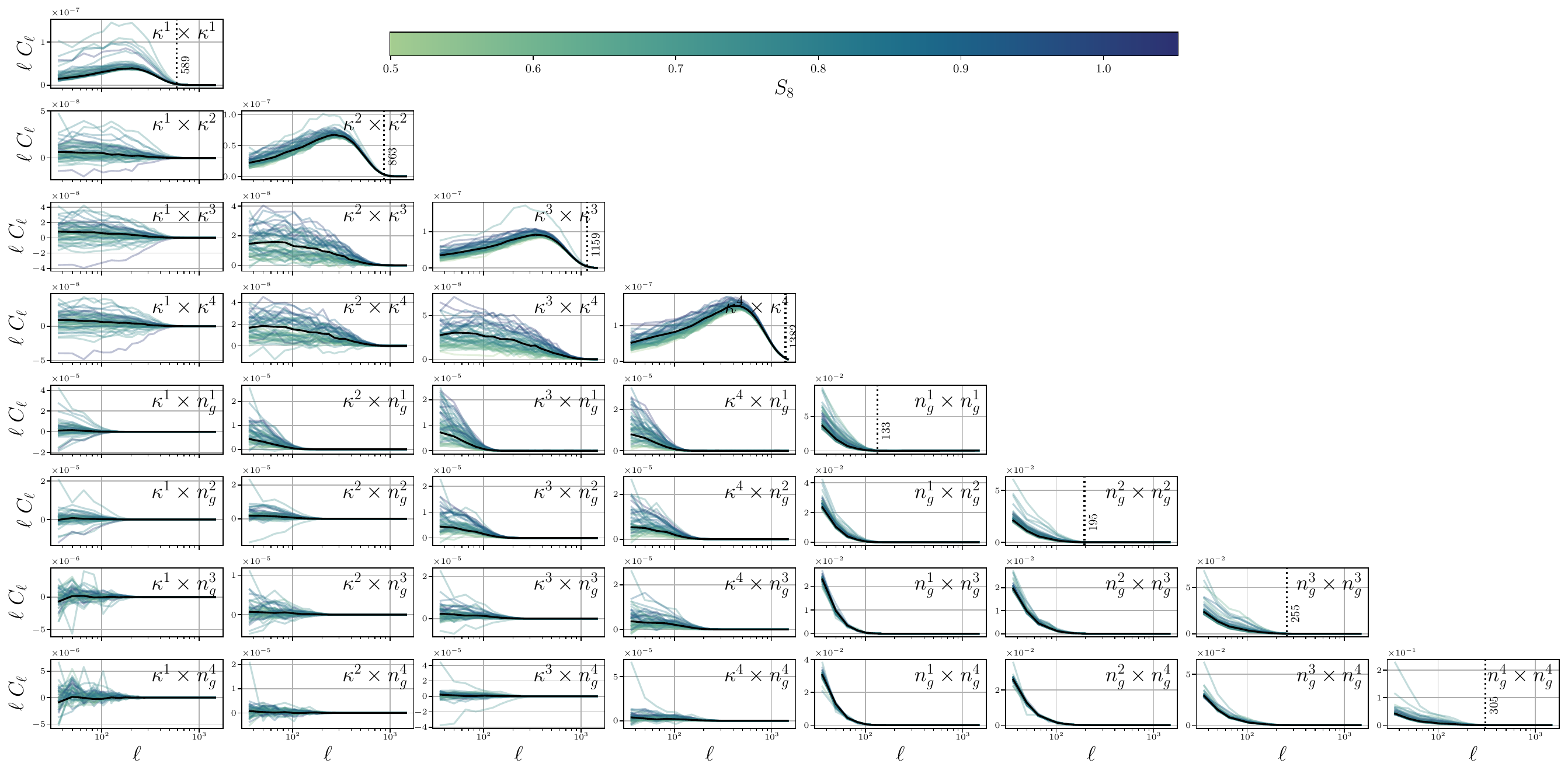}
    \caption{
        Auto, cross redshift bin, and cross probe power spectra. 
        The colored lines correspond to 50 random \cosmogrid grid cosmologies, while the black line is obtained from the mean over the 15 \buzzard realizations.
    }
    \label{fig:power_spectra}
\end{figure*}

\paragraph{Power Spectrum:}
In \cref{fig:power_spectra}, we show all auto- and cross power spectra for weak lensing and galaxy clustering used in the analysis. 
The scale cuts for the power spectra are implemented consistently with the map-level approach described in \cref{sec:scale_cuts}, yielding suppression to near zero for scales smaller than the Gaussian smoothing kernel size.

\section{Intrinsic Alignment}\label{sec:ia_appendix}
This appendix presents additional results regarding the source galaxy intrinsic alignments modeled according to \cref{sec:intrinsic_alignment}.

\subsection{Projection Effects}\label{sec:ia_projection}
Due to projection effects, the posterior contours of the \AIA parameter conditioned on the mean \buzzard mock appear to exhibit non-trivial bias in \cref{fig:maps_vs_cls_individual,fig:maps_vs_cls_combined,fig:probe_comparison_maps}.
\begin{figure}[!htb]
    \centering
    \includegraphics[width=\linewidth]{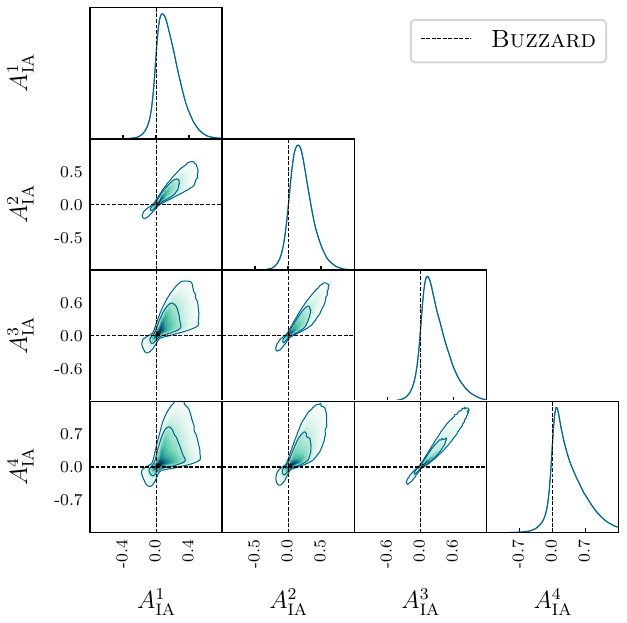}
    \caption{
        Marginal posterior probability densities of the per-bin intrinsic alignment amplitude $\AIA^i$ parametrized by \AIA and \etaIA according to \cref{eq:intrinsic_alignment_redshift}. 
        The analysis configuration matches \cref{fig:maps_vs_cls_individual} (a).
    }
    \label{fig:ia_per_bin}
\end{figure}

We show the posterior densities for the per-bin values $\AIA^i$ computed from \cref{eq:intrinsic_alignment_redshift} in \cref{fig:ia_per_bin}, where $i$ indexes the four redshift bins of the \metacal source galaxy sample.
These two-dimensional marginals demonstrate that the highest posterior probabilities lie close to the true value of zero, consistent with the absence of intrinsic alignment modeling in our \buzzard mocks (see \cref{sec:mocks_buzzard_source}).
We therefore conclude that the shift from zero toward positive \AIA values in \cref{fig:maps_vs_cls_individual,fig:maps_vs_cls_combined,fig:probe_comparison_maps} results from a projection effect arising from the power-law parametrization of $\AIA^i$ in terms of \AIA and \etaIA.

\subsection{\cosmogrid Internal Mocks}\label{sec:ia_cosmogrid}
Since we do not include intrinsic alignments in our \buzzard mocks, we compare different intrinsic alignment amplitudes \AIA using mock observations derived from our weak lensing \cosmogrid forward model detailed in \cref{sec:weak_lensing}.
\begin{figure}[!htb]
    \centering
    \includegraphics[width=\linewidth]{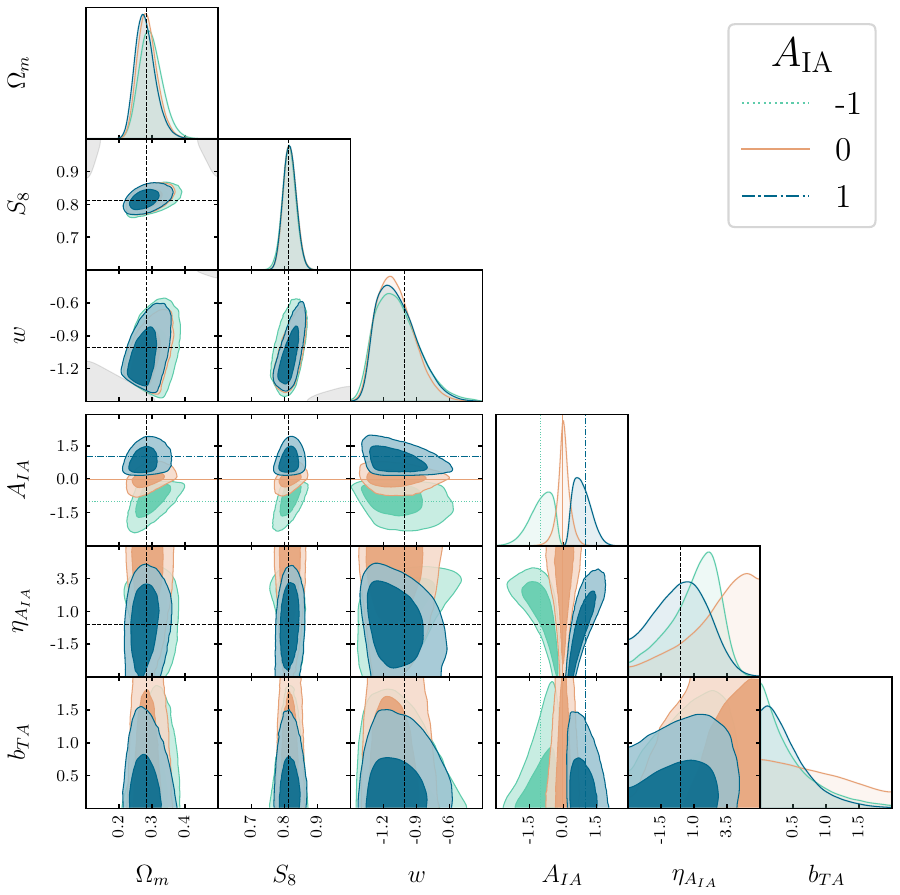}
    \caption{
        Marginal posterior probability densities for \cosmogrid internal mocks of varying intrinsic alignment amplitude \AIA.
        For the intrinsic alignment parameters, the axis limits mark the prior boundaries from \cref{tab:prior_cosmogrid}.
        In all cases, the inference pipeline accurately recovers the true parameters.
    }
    \label{fig:ia_comparison}
\end{figure}

We present results for $\AIA \in \{-1, 0, 1\}$, $\etaIA = 0$, and $\bta = 0$ in \cref{fig:ia_comparison} and find that our inference pipeline accurately recovers the ground truth values in all cases.
The parameters \etaIA and \bta are only weakly constrained and partially hit the prior boundaries.
Everything besides \AIA being fixed, the marginal distributions of the cosmological parameters are only mildly affected.

\end{document}